\documentclass{aa}
\usepackage{graphicx}
\usepackage{longtable}
\usepackage{latexsym}
\usepackage{amssymb}
\usepackage{lscape}
\begin{document}
\title{
UV to radio centimetric spectral energy distributions of
optically-selected late-type galaxies in the Virgo cluster
}

\author{A. Boselli \inst{1}
\and G. Gavazzi \inst{2}
\and G. Sanvito \inst{2}
}

\authorrunning{A.Boselli et al.}
\titlerunning{UV to radio centimetric SED of galaxies in the Virgo cluster}

\offprints{Alessandro Boselli}

\institute{
Laboratoire d'Astrophysique de Marseille, BP 8,
Traverse du Siphon, F-13376 Marseille Cedex 12, France\\
\email {Alessandro.Boselli@oamp.fr}
\and Universit\`a degli Studi di Milano-Bicocca, Dipartimento di Fisica,
Piazza dell'Ateneo Nuovo 1, 20126 Milano, Italy\\
\email {Giuseppe.Gavazzi@mib.infn.it}
\email {Gerry.Sanvito@mib.infn.it}
}

\date{}

\abstract{We present a multifrequency dataset for an optically-selected, 
volume-limited, complete sample of 118 late-type galaxies ($\geq$ S0a)
in the Virgo cluster. The database includes UV, visible, near-IR, mid-IR, 
far-IR, radio continuum photometric data as well as spectroscopic 
data of H$\alpha$, CO and HI lines, homogeneously reduced,
obtained from our own observations or compiled from the literature.\\
Assuming the energy balance between the absorbed stellar light and that
radiated in the IR by dust, we calibarte an empirical attenuation law 
suitable for correcting photometric and spectroscopic data of normal
galaxies.
The data, corrected for internal extinction, 
are used to construct the spectral energy distribution (SED)
of each individual galaxy, and combined to trace the median SED of galaxies
in various classes of morphological type and luminosity. Low-luminosity,
dwarf galaxies have on average bluer stellar continua and higher 
far-IR luminosities per unit galaxy mass than giant, early-type spirals. 
If compared to nearby starburst galaxies such as M82 and Arp 220,
normal spirals have relatively similar observed stellar spectra but
10-100 times lower IR luminosities. The temperature of the cold dust
component increases with the far-IR luminosity, from giant spirals to dwarf
irregulars.
The SED are used to separate the stellar emission from the dust emission
in the mid-IR regime. We show that the contribution of the
stellar emission at 6.75 $\mu$m to the total emission of galaxies is 
generally important, from $\sim$ 80 \% in Sa to $\sim$ 20 \% in Sc.\\
  \keywords{Galaxies: general -- spiral -- ISM -- star
formation}
}

\maketitle
%

\section{Introduction}

An ideal tool for constraining observationally models of galaxy evolution
would consist of a multi-dimensional "data-cube": D$_{imag}(\lambda, Type, Lum, z, Env)$
containing imaging data of complete samples of
galaxies, spanning the broadest possible wavelength ($\lambda$), redshift ($z$), 
morphological type ($Type$) and luminosity ($Lum$) ranges.
Moreover, all environmental conditions ($Env$) should be equally
represented, from the coarsest "field" to the densest cluster's cores.\\ 
Such an ideal data-base is irrealistic. First of all, 
multifrequency images hardly exist, at suitable resolution, even for
galaxies in the Local Group. 
The requirement that the data-cube
consists of "imaging" data  must then be relaxed for the more realistic requirement that it should contain 
"integrated" data, as more commonly available from aperture/CCD photometry. 
Even with these reduced characteristics, very few such data-sets exist
either for high $z$, or for local galaxies. The presently  available samples cover 
a small wavelength window, such as those of Connolly et al. (1995)
or Kinney et al. (1993), or they are biased towards starburst 
and active galaxies (Schmitt et al. 1997), thus they
are not representative of ``normal'' galaxies.\\ 
Within few years from now, however, when SLOAN will reach completion and
the space missions GALEX (UV) and ASTRO-F (FIR) will 
perform their all-sky surveys, large data-sets meeting the above
requirements will be at hands.\\
There is yet a sample which approaches the ideal requirements.
The data-cube we are referring to is an optically selected (complete) one, representative of galaxy
in a broad luminosity range and it is truly multifrequency (from the far-UV to the radio domains). 
It suffers from three limitations: it is local ($z$=0) and it 
represents only late-type galaxies in the densest environment, being composed 
of galaxies in the the Virgo cluster: 
D$_{phot}(UV-radio, Late, Lum, z=0, Virgo)$. It is on this data-base that
the present paper is focused.\\
Skipping through the details of the sample selection and of the available data
that can be found in Sections 2 and 3 of this paper respectively, it is worth spending
some words on what scientific purpouses such data-base is aimed at.\\
Individual galaxies are represented in the data-base under the form of  
Spectral Energy Distributions (SEDs), such as those
presented in Fig. 2 (http://goldmine.mib.infn.it/papers/isosed.html).
SEDs are powerful diagnostic tools for studying the energy balance 
between the principal constituents of galaxies.
From 0.1 to 5 $\rm \mu m$ (UV, Visible, Near-IR) SEDs are dominated by the stellar thermal radiation,
(but include most of the measurable recombination lines providing
the diagnostics of the ISM).
From 5 to 25 $\rm \mu m$ (Mid-IR) the dominant source is the radiation from very small grains of dust,
but the contribution of emission lines (PAH) is relevant.
From 25 to 1000 $\rm \mu m$ (Far-IR, sub-mm radio) the flux of SEDs is due to the thermal radiation from cold
dust (10-100 K). Important diagnostic lines such as the 
[CII] ($\lambda 158 \mu m$) and CO are found in this interval. 
It is here that dust-rich objects peak their flux distributions.
At wavelengths longer than 1 cm (radio) the radiation is non-thermal (synchrotron) by
relativistic cosmic ray electrons and magnetic fields, but the most important ISM diagnostic line,
the 21 cm line of the neutral hydrogen, lies in this domain.
All these components and their complex feedback relations can be studied at once using the SEDs.
First an estimate of the relative fraction of stars in the various age (temperature) classes can be obtained
by fitting populations synthesis models (Bruzual \& Charlot 1993) to the 
stellar continua (see e.g., Gavazzi et al. 2002a).
Once the stellar populations are determined, by studying the ISM emission 
line properties (e.g. the H$\alpha$) one can
learn about the ionization processes in HII regions. From the FIR properties 
we can study the dust heating mechanisms.
Finally from the luminosity of the synchrotron radiation one can study the contribution of the various
stellar populations to the cosmic ray acceleration.\\
Before energy balances can be quantitatively derived, however,
the observed SEDs must be properly corrected for a number of effects that introduce
wavelength dependent distortions to their shape. 
Primarily the SEDs must be rest-framed. 
Galaxies at large redshift require important K corrections. Their cosmic
evolution can be studied by comparing their rest-frame SEDs with those
of normal local galaxies. Hence the importance of obtaining 
template SEDs representative of normal galaxies, unlike those of starburst galaxies such as 
M82 or Arp 220 (see Fig.\ref{fig.3}), often used for such a purpouse.\\
Secondly comes the internal extinction correction. Stellar light is absorbed
and scattered by the dust in a wavelength dependent way. Corrected SEDs can be
derived if the proper amount of extinction is estimated.
The amount of stellar light absorbed in the blue should equal that thermally re-emitted
in the FIR by the dust. Thus the difference of the integral under the stellar continua in
the SEDs before and after the extinction correction gives the energy radiated in the FIR. 
By reversing the argument Buat et al. (2002) derive a robust estimate of 
the internal extinction in normal galaxies.\\
Finally the comparison of SEDs of isolated and cluster galaxies can shed light on
influences of the environment on the various components of galaxies.
Our Virgo sample, spanning a large interval of galactocentric projected distance from M87 
(up to 6 degrees), provides a clue also on this issue.\\
Matter in the present paper is organized as follows
The sample is described in sect. (2); in sect. (3) we give a new prescription for
the determination of the UV, optical and near-IR internal extinction based on
the FIR/UV flux ratio. The adopted extinction law 
is checked in sect. (5.3) using considerations on the energy balance between the emitted far-IR
radiation and the absorbed stellar light.
The SEDs of the sample galaxies are presented in sect. (4),
and analyzed in sect. (5). We construct template SEDs in bins of equal morphological type and luminosity
and compare them to those of starburst galaxies (sect. 5.1). The stellar contribution to the mid-IR emission
of galaxies (sect. 5.2) and the properties of the nonthermal radiation (sect. 5.4)
are also analyzed.  The bolometric properties of the
observed sample are described in sect. (5.5).\\
New optical observations obtained using the 1.2m telescope
of the Observatoire de Haute Provence (OHP), the 0.9m telescope at Kitt Peak
and the 2.5m INT telescope at el Roques de los Muchachos (La Palma)
are given in the appendix. \\
All observations analyzed in the present paper
are contained in a database that has been made available to the international
community via the Word Wide Web site GOLDMine (http://goldmine.mib.infn.it) described in Gavazzi
et al. (2002c).

\section{The sample}

The sample analyzed in this work was extracted from the optically selected 
Virgo Cluster Catalogue (VCC) of Binggeli et al. (1985), which 
is complete to $B_T$~$\leq$~18. Galaxies were selected according to the 
following criteria:

\begin{itemize}
\item B$_T$~$<$~18
\item Hubble type later than S0
\item Classified as cluster member by Binggeli et al. (1985, 1993)
\item lying at a projected angular distance smaller than 2 degrees from M87 (cluster-core)
or greater than 4 degrees from the position of 
maximum projected galaxy density given by Sandage et al. (1985)(cluster-
periphery), but 
excluding galaxies within 1.5 degrees of the M49 sub-cluster. 
\item To limit the spread of distances 
within the sample, galaxies in the M and W clouds, and in the Southern 
Extension ($\delta(1950) < $5$^o$) were also excluded \footnote {Some galaxies
belonging to these three substructures are lying outside the regions delimited as
M and W clouds and Southern Extension in Fig. 1.}.
\end{itemize}

\addtocounter{figure}{0}
\begin{figure*}[!t]
\centerline {}
\vbox{\null\vskip 18.5 cm
\includegraphics{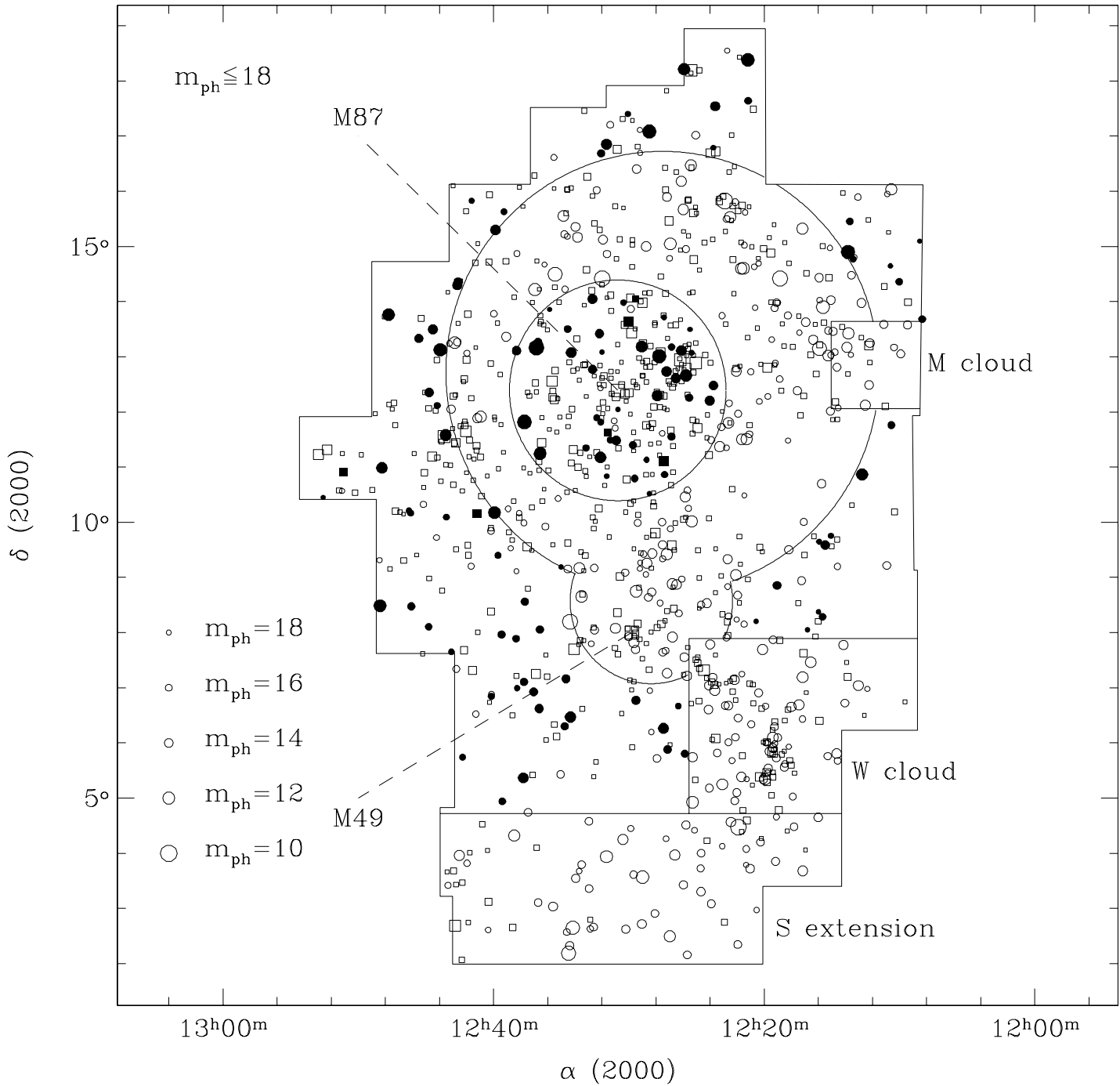}
}
\caption
{Plot of all VCC galaxies classified as members by Binggeli et al. 
(1985) taken from figure 1 of Sandage et al. (1985). The subsample of 
galaxies here analyzed (see section 2) are marked with filled symbols of increasing size 
according to their magnitude. Empty symbols are for Virgo members not included in the
ISO sample; circles for late-type galaxies ($\geq$Sa), squares for early types ($\leq$S0a).
The 2.0 degree radius circle centred on M87 contains the cluster-core subsample.
The inner boundary of the cluster periphery subsample has a radius of 4 
degrees about the position of maximum projected galaxy density.
The 1.5 degree radius circle is centered on the position of maximum projected galaxy density
of M49 subcluster. 
}
\label{fig1}
\end{figure*}

The sky areas from which galaxies were chosen define two
contrasting subsamples to optimise the statistical evaluation
of the cluster environment on observed properties (see Fig. 1). The cluster-core subsample
is composed of 46 galaxies within the X-ray emitting 
``atmosphere'' of M87.
The cluster-periphery subsample includes 72 galaxies in the outskirts.

The resulting sample of 118 galaxies is complete to B$_T\,=\,$18, and
both the cluster-periphery and -core subsamples span the range -21 $<$ $M_B$ 
$<$ -13.
Both subsamples are approximately equally divided between giant
spirals on the one hand and dwarf and irregular galaxies on the other.
The distribution over Hubble type is summarised in Table 1. 

\begin{table}
\caption{Distribution of sample over Hubble type for the cluster-periphery
and cluster-core subsamples.}
\label{Tab1}
\[
\begin{array}{p{0.15\linewidth}cccccc}
\hline
\noalign{\smallskip}
 & S0/a-Sab & Sb-Sc & Scd-Sm & Im & BCD
\\
\noalign{\smallskip}
\hline
\noalign{\smallskip}
periphery &    9  &     16  &     12 &      20  &     15 \\
core      &   16  &     11  &      6 &      10  &      3 \\
Total     &   25  &     27  &     18 &      30  &     18 \\
\noalign{\smallskip}
\hline
\end{array}
\]
\end{table}

\noindent
The parameter of the sample galaxies are given in Table 2, arranged as follows:

\begin {itemize}

\item {Column 1: VCC denomination (Binggeli et al. 1985).}
\item {Column 2: NGC  name.}
\item {Column 3: IC name.}
\item {Column 4: UGC name (Nilson 1973).}
\item {Column 5: CGCG denomination (Zwicky et al. 1961-68)}
\item {Columns 6 and 7: (B2000.0) celestial coordinates, from NED,
with a few arcsec accuracy.}
\item {Column 8: morphological type, from the VCC or from Binggeli et al. (1993).}
\item {Column 9: photographic magnitude from the VCC.}
\item {Columns 10 and 11: major ($a$) and minor ($b$) optical diameters (arcmin) 
determined at the surface brightness of $\rm 25^{th} mag~arcsec^{-2}$. 
For galaxies without a $\rm 25^{th} mag~arcsec^{-2}$ value in the VCC, 
the diameter is computed from the "last visible" isophotal diameter given 
in the VCC using the relation:  $Log~a_{ext}(b_{ext})=0.99 Log~a(b)+0.1$.}
\item {Column 12: heliocentric velocity, in km s$^{-1}$.}
\item {Column 13: distance, in Mpc. Distances to the various 
substructures of Virgo are as given in Gavazzi et al. (1999).}
\item {Column 14: cluster membership as defined in Gavazzi et al. (1999)
\footnote{The Gavazzi et al. (1999) cluster membership criterion, based on the 
3-D distribution of galaxies in the Virgo cluster, is slightly 
different from that given by Binggeli et al. (1985; 1993), used to define the sample.
It is thus not surprising that several objects in Table 2 are listed as members of cluster
 B or of the M and W clouds.}.}
\item {Column 15: projected angular separation from the cluster centre (M87), in degrees.}
\item {Column 16: the model-independent near-IR concentration index parameter $C_{31}$, 
from Gavazzi et al. (2000), defined as  
the ratio between the radii that enclose 75\% and 25\% of the total light. $C_{31}$ is a tracer
of the light distribution within galaxies: values of $C_{31}$ $<$ 3
are for pure exponential discs, $C_{31}$ $>$ 3 for galaxies with bulges.}
\item {Column 17: notes.}

\end{itemize}

\section{The data}

The SED presented in this paper have been constructed using multifrequency data
available in the literature or from our own observations, treated as consistently as possible,
in order to produce an homogeneous data-set.\\ 
The UV data are taken from the FAUST (Lampton et al. 1990) and the FOCA 
(Milliard et al. 1991) experiments. In order to be consistent with our previous works,
we transformed UV magnitudes taken at 1650 \AA~ by Deharveng et al. (1994) 
to 2000 \AA~ assuming a constant colour index UV(2000)=UV(1650)+0.2 mag.
This relation has been obtained by comparing the FAUST 1650 \AA~ with the SCAP (Donas et al. 1987)
2000 \AA~ UV magnitudes of 17 late-type galaxies in the Virgo cluster, observed by both experiments
(Deharveng et al. 1994).
FOCA magnitudes are from Deharveng et al. (2002), and Donas et al., in preparation.
These are total magnitudes, determined by integrating the UV emission up to
the weakest detectable isophote. The estimated error on the UV magnitude is 
0.3 mag in general, but it ranges from 0.2 mag for bright galaxies to 0.5 
mag for weak sources observed in frames with larger than average calibration 
uncertainties.\\
U, B and V  photometry is generally derived from our own CCD measurements consistently with
Gavazzi \& Boselli (1996), as described in the appendix.
When these are not available it is derived from aperture photometry taken from the literature.
The $(U,B,V)$  $D_{25}$ magnitudes, computed at the $\rm 25^{th} mag~arcsec^{-2}$ isophotal $B$ band 
diameter as in Gavazzi \& Boselli (1996),
have $\sim$ 10 \% uncertainty. They are on
average 0.1 mag fainter than the total asymptotic magnitudes.\\
NIR data, from Nicmos3 observations, are taken mostly from Boselli et al. (1997)
and Gavazzi et al. (2001). Magnitudes $(J,H,K)$ 
are determined  consistently with the
optical magnitudes as in Gavazzi \& Boselli (1996). The typical uncertainty in $(J,H,K)$ is 10 \%. As for the 
visible magnitudes, they are on average 0.1 mag fainter than the total asymptotic magnitudes.\\
Mid-IR data, at 6.75 and 15 $\mu$m, are from Boselli et al. (2002c). Flux densities have been
extracted from ISOCAM images by integrating the emission until the weakest detectable isophote.
Even if the mid-IR emission of these galaxies is less extended than in the visible and near-IR bands, 
ISOCAM data provide us with integrated flux denisties representative of the whole galaxy.
The typic uncertainty on the ISOCAM data is $\sim$  30 \%.\\
12, 25, 60 and 100 $\mu$m integrated flux densities from the IRAS survey are taken from 
different sources. The typical uncertainty in the IRAS data is $\sim$ 15 \%.
Alternative Far-IR values at 60 and 100 $\mu$m from ISOPHOT, as well as 170 $\mu$m flux densities, are 
taken from Tuffs et al. (2002), with a typical nominal uncertainty of $\sim$ 10 \%. The comparison
of ISO and IRAS data for the sample galaxies detected in both surveys reveals a systematic difference of 
ISO/IRAS=0.95 and 0.82 at 60 and 100 $\mu$m respectively (Tuffs et al. 2002).\\
We collected radio continuum data at 2.8, 6.3, 12.6 and 21 cm from different sources.
21cm radio continuum data, available for the whole sample, are mostly from the NVSS survey (Condon et al. 1998)
(see Gavazzi \& Boselli 1999). 
All radio continuum data are integrated fluxes. The typical uncertainty is $\sim$ 20 \%.

The photometric data for the whole sample are given in Table 3, arranged as follows:

\begin {itemize}

\item {Column 1: VCC denomination.}
\item {Column 2: UV magnitude at 2000 \AA, uncorrected for galactic and internal extinction.}
\item {Columns 3-8: U, B, V, J, H, K magnitudes in the Johnson system, determined
as described in Gavazzi \& Boselli (1996), uncorrected for galactic and internal extinction.}
\item {Column 9: ISOCAM 6.75 $\mu$m flux density, in mJy.}
\item {Column 10: IRAS 12 $\mu$m flux density, in mJy.}
\item {Column 11: ISOCAM 15 $\mu$m flux density, in mJy.}
\item {Column 12: IRAS 25 $\mu$m flux density, in mJy.}
\item {Column 13: IRAS 60 $\mu$m flux density, in mJy.}
\item {Column 14: ISOPHOT 60 $\mu$m flux density, in mJy.}
\item {Column 15: IRAS 100 $\mu$m flux density, in mJy.}
\item {Column 16: ISOPHOT 100 $\mu$m flux density, in mJy.}
\item {Column 17: ISOPHOT 170 $\mu$m flux density, in mJy.}
\item {Columns 18-21: radio continuum flux densities at 2.8, 6.3, 12.6 and 21 cm, in mJy.}

\end{itemize}

\noindent
All data given in Table 3 are observed quantities. The UV, optical
and near-IR data are uncorrected for dust extinction, the mid-IR data for the contribution of the 
stellar component, the radio continuum data for the
contribution of the nuclear emission. 

References to the photometric data are given in Table 4.

Additional emission line data are given in Table 5, arranged as follows:
\begin {itemize}

\item {Column 1: VCC denomination.}
\item {Column 2: H$\alpha$+[NII] equivalent width, in \AA.}
\item {Column 3: logarithm of the H$\alpha$+[NII] line, in erg cm$^{-2}$ s$^{-1}$.}
\item {Column 4: reference to the H$\alpha$ data.}
\item {Column 5: logarithm of the HI mass, in solar units, defined as $MHI = 2.36~ 10^5 D^2 SHI$,
where $D$ is the distance of the galaxy in Mpc (from Table 2), and $SHI$ is the integrated HI flux,
in Jy km s$^{-1}$.}
\item {Column 6: the HI-deficiency parameter, defined as the ratio of the HI mass to the 
average HI mass of isolated objects of similar morphological type and linear size 
(Haynes \& Giovanelli 1984); it is used
to discriminate between "normal" galaxies and galaxies suffering for
gas depletion due to ram pressure. Galaxies with an HI-deficiency parameter 
$\leq 0.3$ can be treated as unperturbed, isolated galaxies.}
\item {Column 7: the quality of the HI profile: 1 stands for high signal to noise, two-horns profiles,
2 for  high signal to noise gaussian profiles, 3 for average signal to noise profiles, 4 for poor 
quality data and 5 for those objects whose profile is not available in the literature.}
\item {Column 8: HI line width, measured as the average value of the width at 20 and 50 \% 
of the peak, in km s$^{-1}$.}
\item {Column 9: reference to the HI data.}
\item {Column 10: logarithm of the H$_2$ mass, in solar units, defined as in Boselli et al. (2002b),
determined assuming a luminosity dependent $X$ conversion factor between the CO intensity and
the H$_2$ surface density $log X = -0.38 log L_H +24.23$ mol cm$^{-2}$ (K km s$^{-1}$)$^{-1}$.As shown in Boselli et al. (2002b), the use of a luminosity dependent
rather than a metallicity dependent $X$ conversion
factor does not affect the uncertainty in the determination of the molecular hydrogen
mass.}
\item {Column 11: reference to the CO data.}

\end{itemize}

\subsection{The extinction correction}

UV to near-IR data have been corrected for galactic extinction according to 
Burstein \& Heiles (1982). The galactic extinction $A_g(B)$, taken from NED
and listed in Table 7, have been transformed to $A_g(\lambda)$ assuming a 
standard galactic extinction law (see Table 6): 
$A_g(\lambda)$=$c(\lambda)$ $A_g(B)$, where $c(\lambda)$=$k(\lambda)/k(B)$.

\addtocounter{table}{4}
\begin{table}
\caption{Galactic extinction law}
\label{Tab6}
\[
\begin{array}{p{0.1\linewidth}ccc}
\hline
\noalign{\smallskip}
\hline
Filter & \lambda & c(\lambda) \\
\hline
       &   \AA  &            \\
UV     &  2000  & 2.10       \\
U      &  3650  & 1.15       \\
B      &  4400  & 1.00       \\
V      &  5500  & 0.75       \\
J      & 12500  & 0.21       \\
H      & 16500  & 0.14       \\
K'     & 21000  & 0.10       \\
\noalign{\smallskip}
\hline              
\end{array}
\]
\end{table}

The observed stellar radiation of galaxies, from UV to near-IR
wavelengths, is subject to internal extinction (absorption plus 
scattering) by the interstellar dust. In order to quantify the 
emission of the various stellar populations, UV, optical and, to a lesser 
amount, near-IR fluxes must be corrected for dust attenuation.
Furthermore, since dust extinction varies from galaxy to galaxy (according to
their geometrical parameters such as the inclination, their history of star formation 
and metallicity), corrections appropriate to each individual galaxy must be determined.\\
Estimating the dust extinction at different $\lambda$ in external galaxies 
is however very difficult (it has been done only for the Magellanic clouds).
Buat et al. (2002) have shown that, for example, the Calzetti's
law calibrated on the central part of starburst galaxies (Calzetti 2001)
strongly overestimates the extinction in normal, late-type objects.
This difficulty is mainly due to two reasons: a) the extinction strongly
depends on the relative geometry of the emitting stars and of the 
absorbing dust within the disc of galaxies. The young stellar population are 
mostly located along the disc in a thin layer, while the old populations forms a thicker layer.
This point is further complicated by the fact that different dust components 
(very small grains, big grains etc.), which have different opacities to 
the UV, visible or near-IR light, have themselves different geometrical distributions both on 
the large and small
scales. b) it is still uncertain whether the Galactic extinction law
is universal, or if it changes with metallicity and/or with the UV radiation field. Detailed
observations of resolved stars in the Small Magellanic Cloud by Bouchet et 
al. (1985) indicate that the extinction law in the optical domain is not significantly different 
from the Galactic one in galaxies with a UV field $\sim$ 10 times higher and 
a metallicity $\sim$ 10 times lower than those of the Milky Way.
A steeper UV rise and a weaker 2200 \AA ~bump than in the Galactic
extinction law have been however observed in the LMC and SMC (Mathis 1990).\\
While the adoption of the Galactic extinction law for external 
galaxies seems reasonable 
(even though it is questionable for low-luminosity galaxies), no simple analytic
functions describing the geometrical distribution 
of emitting stars and absorbing dust, both on small and large scales, are yet available.\\
The radiative transfer models of Witt \& Gordon (2000) have however shown that
the FIR to UV flux ratio, being mostly independent of the geometry,
of the star formation history (the two radiations are produced by similar stellar 
populations) and of the adopted extinction law, 
is a robust estimator of the dust extinction at UV wavelengths.
Here we will use this method to estimate the 
extinction correction in the UV, the wavelength most affected 
by dust.\\
We propose an internal extinction correction prescription similar 
to that described in Gavazzi et al. (2002a).\\

Our semi-empirical 
determination of $A(UV)$ takes into account the scattered light.
Following Buat et al. (1999), we estimate $A_i(UV)$ from the relation:

\begin{displaymath}
A_i(UV) = 0.466 + Log(FIR/UV)+
\end{displaymath}

\begin{equation}
+0.433 \times (Log(FIR/UV))^2    ~~~~~~~~~~~~~~~~~~~[{\rm mag}]
\end{equation}

\noindent
where 

\begin{displaymath}
FIR=1.26 \times (2.58 \times 10^{12} \times F_{60} +
\end{displaymath}

\begin{equation}
+ 10^{12} \times F_{100}) \times 10^{-26}     ~~~~~~[{\rm Wm^{-2}}]
\end{equation}

\noindent
$F_{60}$ and $F_{100}$ are the IRAS FIR fluxes (in Jy) and 

\begin{equation}
UV=10^{-3} \times 2000*10^{(UV_{mag}+21.175)/-2.5}             ~~~~~[{\rm Wm^{-2}}]
\end{equation}

$A_i(\lambda)$ can be derived from $A_i(UV)$ once an extinction law
and a geometry for the dust and star distribution are assumed. We
adopt the sandwitch model, where a thin layer of dust of thickness $\zeta$
is embedded in a thick layer of stars:

\begin{displaymath}
A_i(\lambda)=-2.5 \cdot log([\frac{1-\zeta(\lambda)}{2}](1+e^{-\tau(\lambda) \cdot sec(i)})+
\end{displaymath}

\begin{equation}
+[\frac{\zeta(\lambda)}{\tau(\lambda) \cdot sec(i)}] \cdot (1-e^{-\tau(\lambda) \cdot sec(i)}))  ~~~~~~~~~~~~~~~~~~[{\rm mag}]
\end{equation}

\noindent
where the dust to stars scale height ratio $\zeta(\lambda)$ depends on 
$\lambda$ (in units of \AA) as:

\begin{equation}
\zeta(\lambda)=1.0867-5.501 \cdot 10^{-5} \cdot \lambda
\end{equation}

\noindent
Relation (5) has been calibrated adopting 
the average between the optically thin and optically thick cases with $\lambda$ dependent
dust to star scale height ratios  given by 
Boselli \& Gavazzi (1994). Observations of some
edge-on nearby galaxies show that it is still unclear whether $\zeta$ 
depends or not on $\lambda$ (Xilouris et al. 1999). As shown in 
Gavazzi et al. (2002a), however, similar values of $A_i(\lambda)$
are obtained in the case of a sandwitch model and of the extreme case of
a slab model ($\zeta$=1), meaning that the high uncertainty on 
$\zeta$ is not reflected on $A_i(\lambda)$.\\
In the case of the UV band ($\lambda$=2000 \AA), $\zeta$=1, and eq. (4)
reduces to a simple slab model. In this case $\tau(UV)$ can be derived by 
inverting eq. (4):

\begin{displaymath}
\tau(UV)=[1/sec(i)] \cdot (0.0259+1.2002 \times A_i(UV)+
\end{displaymath}

\begin{equation}
+1.5543 \times A_i(UV)^2-0.7409 \times A_i(UV)^3 +0.2246 \times A_i(UV)^4)
\end{equation}

\noindent
using the galactic extinction law $k(\lambda)$ (Savage \& Mathis 1979), we than derive: 
\begin{equation}
\tau(\lambda) = \tau(UV) \cdot k(\lambda) / k(UV)
\end{equation}

\noindent
and we compute the complete set of $A_i(\lambda)$ using eq. (4).\\
FIR/UV is available for 44 objects.
If FIR or UV measurements are unavailable we assume the average values $A_i(UV)$ = 1.28; 0.85; 0.68 mag
for Sa-Sbc; Sc-Scd; Sd-Im-BCD galaxies respectively, as determined 
when FIR and UV measurements are available.\\
Once corrected adopting the aformentioned prescription, we checked empirically 
that the SED do not contain a residual
dependence on galaxy inclination. The corrected SEDs of 32 Sc galaxies,
binned in 4 intervals of inclination, and their fit parameters were found 
very consistent one another.
The galactic and internal extinction correction (in magnitude) for the observed galaxies 
are given in Table 7.\\
This empirical attenuation law gives 
a zeroth order estimate of the attenuation in the UV regime, the most affected 
by dust. We stress however that the shape of the corrected spectrum, in particular at 
UV wavelengths, is still uncertain. This is due not only to the lack of 
observational constraints other than the 2000 \AA~ flux, but also to
the large uncertainties on the relative geometrical distributions of dust and stars
and on the extinction law, which might significantely depend on the
UV field and metallicity in this wavelength regime.

\section{The SEDs}

Figure 2 shows the SEDs of the sample galaxies obtained using the data given in Table 3 (only for those
galaxies with at least 2 photometric data points).
UV, optical and near-IR data are corrected for galactic and internal extinction as described 
in the previous section. FIR data at 60 and 100 $\mu$m
are average values between IRAS and ISOPHOT data when both are available. When one of
the two data is an upper limit, we take the detection \footnote{The IRAS data of VCC 17 and 1725
from Almoznino \& Brosch (1998) are inconsistent with the PHOT data of Tuffs et al. (2002)
and with our CAM data, and are thus not used in the construction of these SEDs.}.
To be as consistent as possible with IRAS, ISOPHOT data have been corrected for the average ISOPHOT/IRAS
ratio found by Tuffs et al. (2002) for Virgo galaxies
detected with both instruments, ISOPHOT/IRAS=0.95 and 0.82 at 60 and 100 $\mu$m respectively.\\

\noindent
The morphological type given in Table 2 and the logarithm of the H band luminosity, defined as
$log L_H = 11.36 - 0.4H_T +2logD$ (in solar units), 
where $H_T$ is the total $H$ band magnitude and $D$ is the distance 
to the source (in Mpc), are labeled in Fig. 2 (http://goldmine.mib.infn.it/papers/isosed.html). 
For few objects we derive the H luminosity
from K band measurements assuming an average H-K colour of
0.25 mag (independent of type; see Gavazzi et al. 2000).
A minority of the objects in our sample have an H band magnitude
obtained from aperture photometry, thus with no asymptotic extrapolation.
For these we use the $H$ magnitude determined as in Gavazzi \& Boselli 
(1996) at the optical radius
which is on average 0.1 magnitudes fainter than $H_T$ (Gavazzi et al. 2000).\\
The continuum line in the optical domain gives the integrated spectrum obtained by Gavazzi et al. (2002a).
The two dashed lines at $\lambda$ $<$ 10 $\mu$m are the 
Bruzual \& Charlot stellar population synthesis 
models (GISSEL 2001). 
The upper curves represent the models which best fit the extinction corrected data,
as determined by Gavazzi et al. (2002a). The lower curves represent the same models
attenuated by dust extinction using the inverse relations
of sect. (3.1).
For galaxies with insufficient photometric points for fitting a  model, we adopt the 
Bruzual \& Charlot model that best-fits a template SED of similar morphological type (Fig. 9
in Gavazzi et al. 2002a). To be consistent with Gavazzi et al. (2002a), all models are normalized 
to the $V$ band photometric data when available, or to the $K$ band. 
Given the poor
quality of the fit, models are not shown for the galaxies VCC 1217 and VCC 1313.\\
We have preferred not to give fits in the Mid-IR range for two reasons: 1) because
the very small grains and the carriers of the Aromatic Infrared Bands responsable
for the  mid-IR dust emission are not in thermal
equilibrium with the radiation, but are stochastically heated (mostly) by UV photons (Boselli
et al. 2002c). Thus modified black-body functions cannot be used to fit the mid-IR data.
2) mid-IR spectra obtained with the CVF camera onboard ISO in various galactic and extragalactic 
environments has shown a variety of strong emission lines with fluxes comparable with
the continuum. It is thus difficult to estimate a typic mid-IR spectrum of galaxies.\\
The dashed line in the FIR domain (20-2000 $\mu$m) reprsents a two dust components 
model. Two modified blackbodies $F(\nu)$ $\sim$ $\nu^{\beta}B(\nu)(T_D)$,
with $\beta$=2, one with a fixed warm temperature of $T_w$=47 K (tracing the star forming
regions), the other with a (variable) cold temperature $T_c$ (tracing the cirrus emission), were 
determined consistently with Popescu et al. (2002). The two components 
are calibrated to match the 60 and 170 $\mu$m data respectively. 
For galaxies not observed by PHOT but detected by IRAS at 60 and 100 $\mu$m, we adopted  
a modified blackbody with $T_w$=47 K for the warm component and we assume $T_c$=18 K 
(the average value of Popescu et al. 2002), for the cold component. They are calibrated 
to match the 60 and 100 $\mu$m fluxes respectively.\\
The far-IR to mm domain, from 170 $\mu$m to $\sim$ 1 cm, is totally unexplored.
Submillimetric observation should provide constraints on the cold dust temperature and on 
the total dust mass of the sample galaxies. From $\sim$ 1 mm to 1 cm, data are needed to
estimate the relative contribution of the thermal and synchrotron radio emission.\\
The dashed line in the centimetric domain, given for all galaxies with more than two detections,
represents the power-law regression to the radio continuum data. The best-fit parameters are given in Table 8.\\

\section{Analysis}

Previous analyses, each devoted to a limited spectral domain, have attempted 
to interpret the SEDs of galaxies:
Gavazzi et al. (2002a) for the continuum stellar radiation, Boselli et al. (1998, 2002c and in preparation) for
the mid-IR emission, Popescu et al. (2002) for the FIR emission, and Niklas et al. (1997) for
the radio emission.
In this work, for the first time we analyze the SEDs as determined in the whole spectral range.

\subsection{The template SED}
 
The template SEDs in bins of morphological type and luminosity are obtained as median combinations of 
the normalized (to the K band) SEDs. We used only the detected values and imposed that
at least 2 photometric points were available. 
The resulting extinction corrected template SEDs in different classes of morphological
type and luminosity are shown in Fig. 3 a and c respectively. The observed 
(dust attenuated) SEDs of M82 and Arp220 (from Elbaz et al. 2002), are given for comparison 
in Fig. 3 b and d. The median values of $F(\lambda)/F(K)$ for the templates 
in the 18 bands considered in this work are given in Table 9, while the fitting
models in the visible (corrected and uncorrected for dust extinction) and in the FIR
are given in Table 10, 11 and 12 respectively \footnote {A sample of Table 10 is 
given at the end of the paper; the whole Tables 10, 11 and 12 
are available only in electronic format at http://cdsweb.u-strasbg.fr}.
The values in parenthesis in Table 9 give the total number of objects in each Hubble type  
and wavelength bin that were combined to form the templates. 

\addtocounter{figure}{1}
\begin{figure*}[!hpbm]
\centerline {}
\vbox{\null\vskip 20.0 cm
\includegraphics{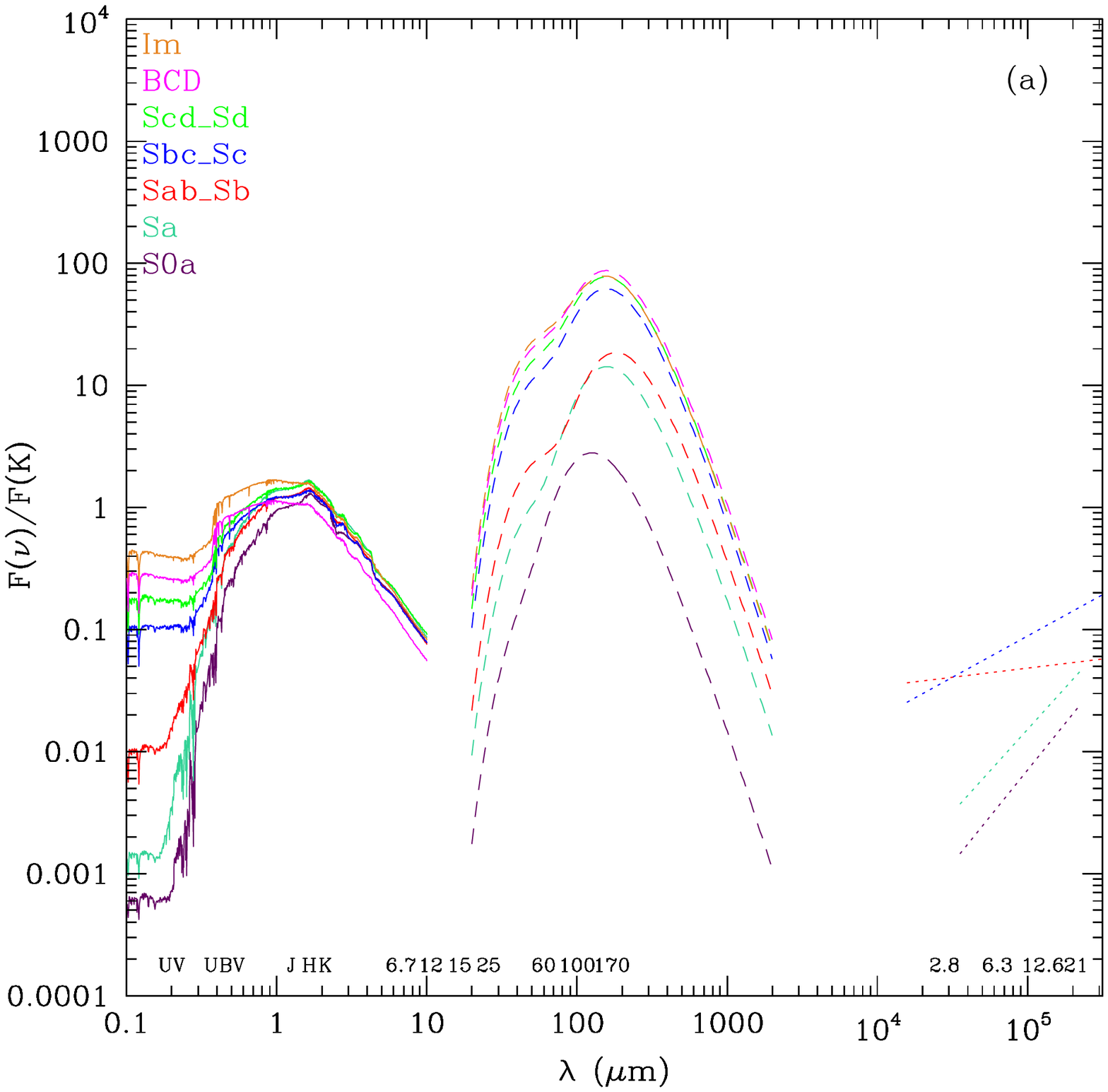}
\includegraphics{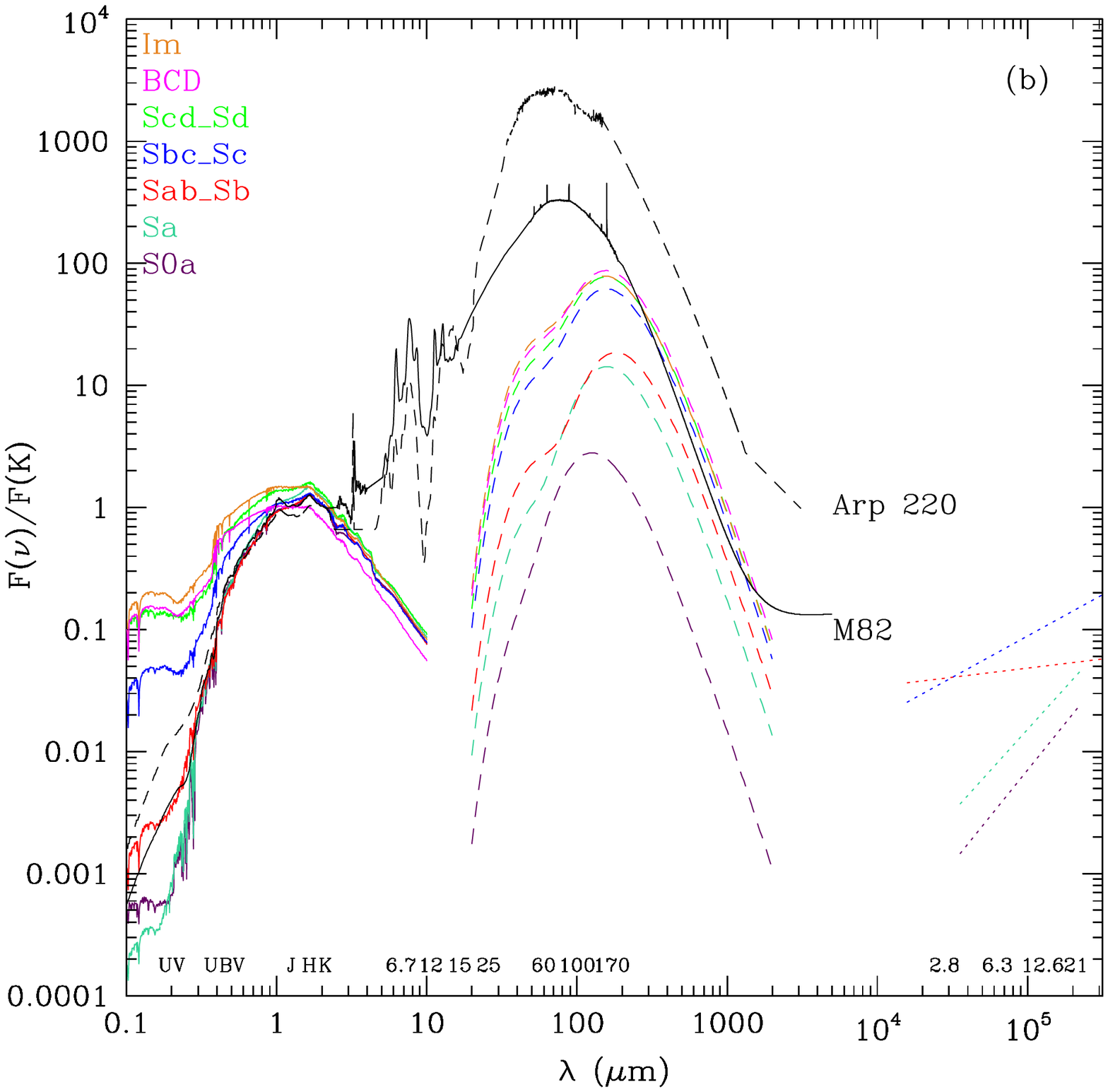}
\includegraphics{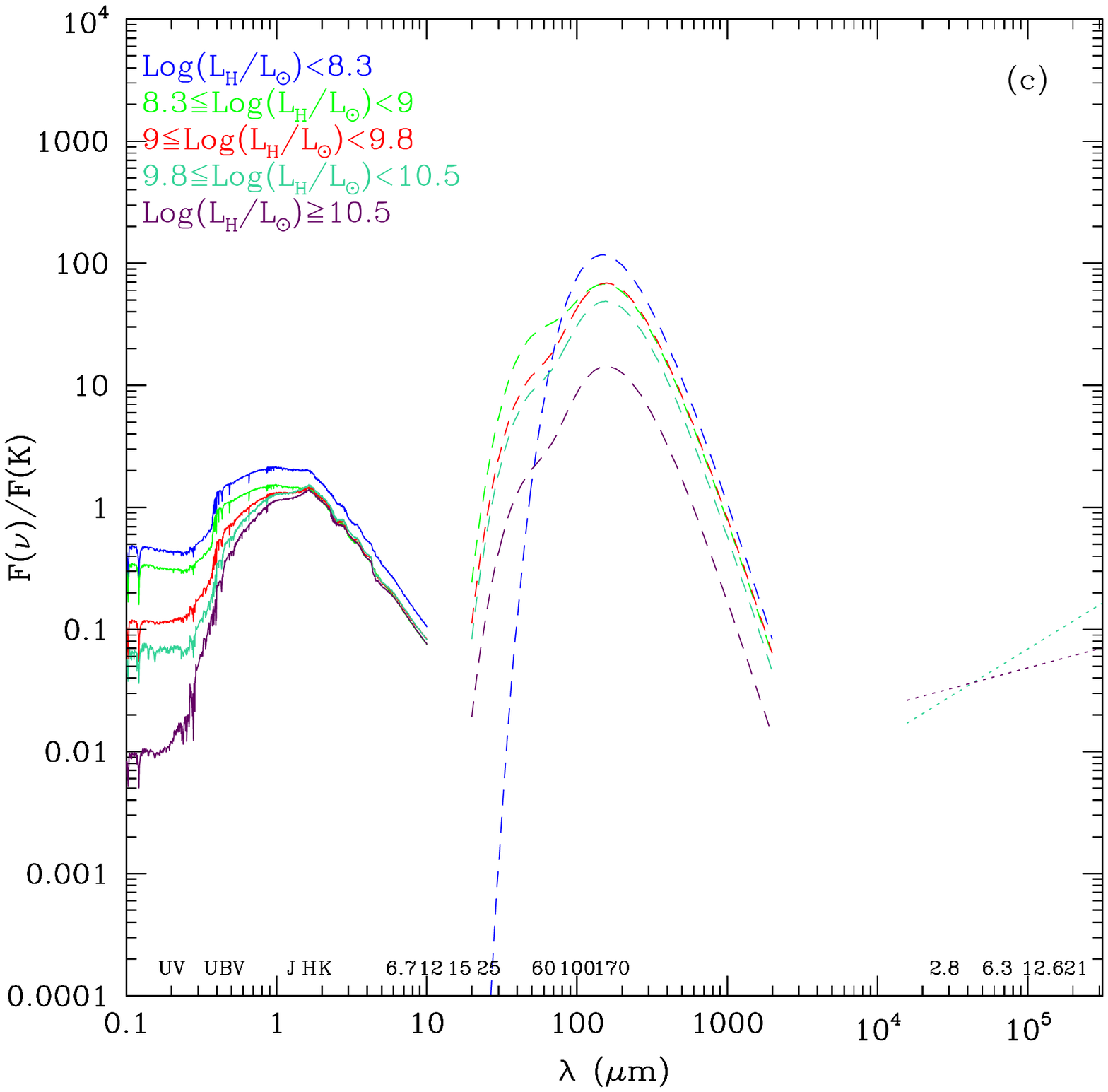}
\includegraphics{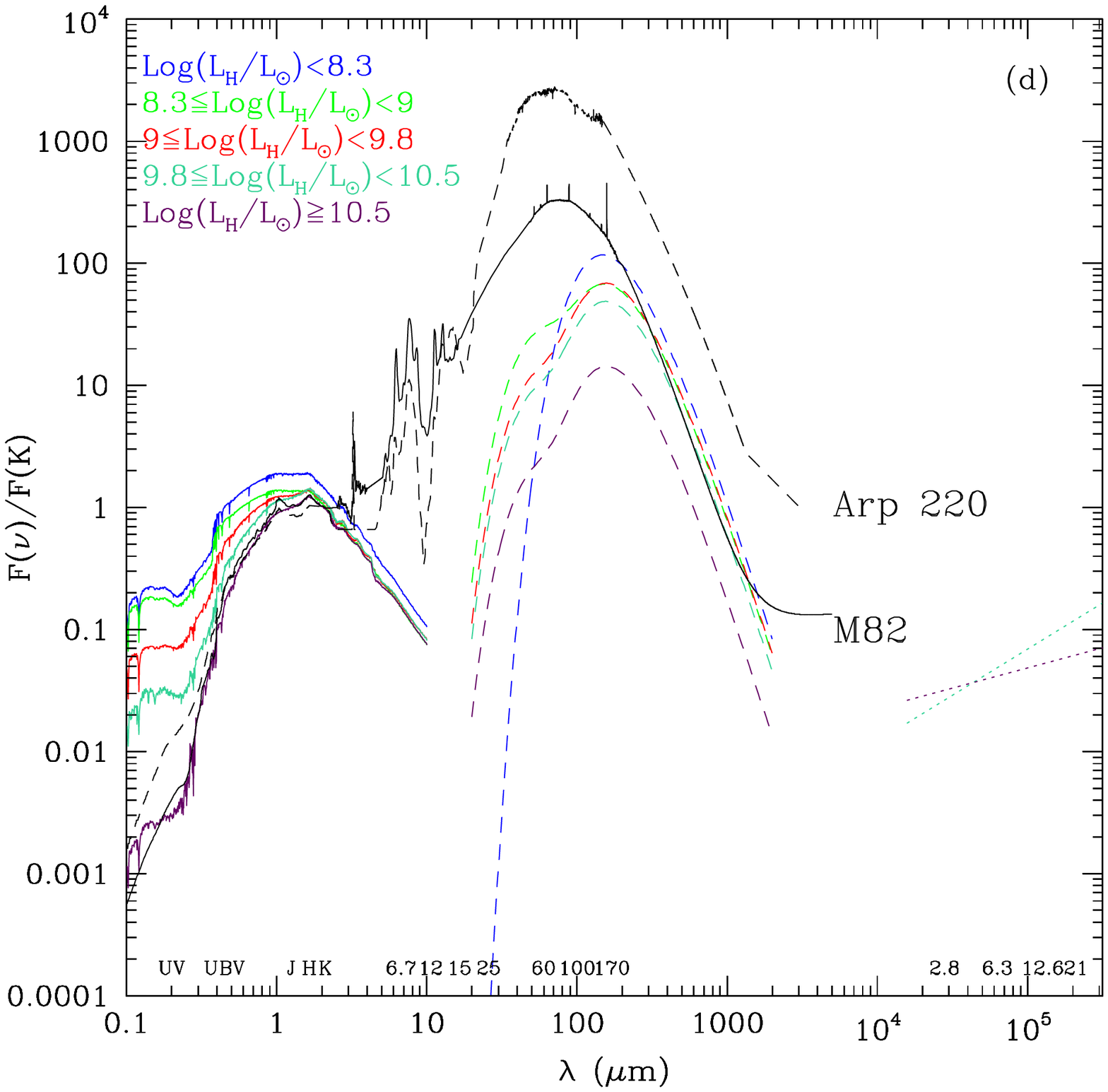}
}
\caption{The extinction corrected (a) and dust attenuated (b) template SEDs in bins of 
morphological type and luminosity (c,d). The dust attenuated SEDs of M82 (continuum
black line) and Arp 220 (dotted black line) are also given for comparison.
}
\label{fig.3}
\end{figure*}

\noindent
By analyzing Fig. 3 we can observe that: a) the relative contribution to the SED of the young
stellar component, emitting in the UV, and of the relatively cold dust emitting at $\sim$ 60-200
$\mu$m increases from early to late-type spirals and/or from high-mass to low-mass objects; 
b) the 60 to 100 $\mu$m flux density ratio increases with the total FIR emission, indicating a
general increase of the big grains dust temperature  
from massive Sa to low-luminosity Scd-Im-BCD and, to a much higher degree, in starburst galaxies.
c) optically selected spirals have UV to near-IR SEDs similar to those of sturburst
galaxies such as M82 or Arp 220, despite the fact that these extreme objects have dust attenuations
several order of magnitudes higher than normal galaxies, $A(UV)$ $\sim$ 1 for
optically selected spirals vs. $A(UV)$$\sim$ 3.5 for M82 (Buat et al. 2002) and $A(UV)$$\geq$ 100 for  
Arp 220 (Haas et al. 2001). 
At the same time 
the far-IR emission of optically-selected, normal galaxies is more than a factor of 10-100 less
important than in sturbust galaxies. \\
It is thus extremely dangerous to use the SEDs of starburst galaxies such as M82 and Arp 220 as
templates of normal late-type galaxies at high redshift, as often done, since these
objects may not be representative of the mean late-type galaxy population even at earlier epochs, 
when star formation was expected to be more active. 

\subsection{The stellar contribution to the mid-IR emission}

The Bruzual \& Charlot models fitted to the data trace the stellar emission 
from 1000 \AA~ to 10 $\mu$m, and can thus be used to estimate the stellar contribution to the
emission of our target galaxies at 6.75 $\mu$m.
The ratio of the total flux (dust plus star) to the stellar flux at 6.75 $\mu$m, 
$[F_{6.75}(d+s)/F_{6.75}(s)]$, determined for all galaxies detected at 6.75 $\mu$m, and with 
available visible or near-IR photometry, is given in Table 8, while the median value for each morphological class 
in Table 13.

\addtocounter{figure}{0}
\begin{figure}[!h]
\centerline {}
\vbox{\null\vskip 8.0 cm
\includegraphics{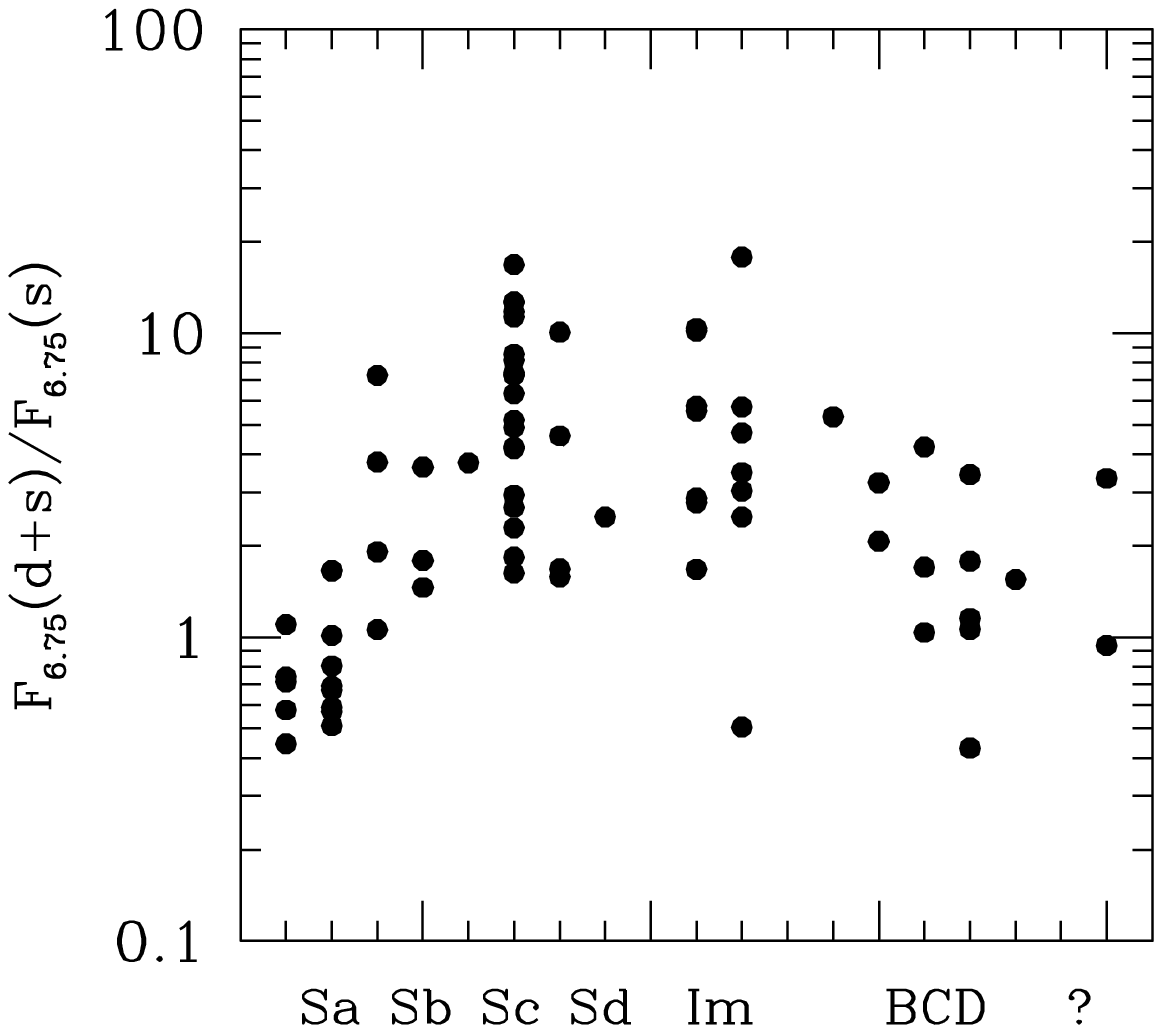}
}
\caption{The relationship between the total flux (dust plus stars) to the stellar flux at 6.75 $\mu$m, 
$[F_{6.75}(d+s)/F_{6.75}(s)]$, and the morphological type.
}
\label{fig.4}
\end{figure}

\noindent
Figure 4 shows the relationship between $[F_{6.75}(d+s)/F_{6.75}(s)]$ and the morphological type.
The stellar contribution to the total  mid-IR emission of galaxies strongly depends on the 
morphological type. In early-types ($\leq$ S0a), the emission at 6.75 $\mu$m is completely
dominated by the photosphere of the cold stellar population (see Table 13).
The average stellar contribution to the 6.75 $\mu$m emission of spiral galaxies is always 
important, ranging from $\sim$ 80 \% in Sa to $\sim$ 20 \% to Sc and Im. In BCD the 
stellar emission contributes on average at $\sim$ 50 \%. Given the low detection rate in irregular
galaxies (Im and BCD), their average $[F_{6.75}(d+s)/F_{6.75}(s)]$ ratios might be biased towards 
objects whose stellar contribution to the mid-IR emission is important, the only ones with detectable
6.75 $\mu$m flux. The decrease of the dust emission observed in BCD and Im galaxies, however, 
could be due either to their low metallicity, or to the
destruction of the carriers of the UIB expected in high UV radiation fields (Boselli et al. 1998).
We do not see any strong relationship between the $[F_{6.75}(d+s)/F_{6.75}(s)]$
ratio and the total $K$ band luminosity or concentration index parameter.
However all galaxies with $C_{31}(K)$ $>$ 4 have their mid-IR
emission at 6.75 $\mu$m dominated by stars. Among the ISOCAM resolved galaxies, these
objects have also a $C_{31}$(6.75$\mu$m) index $>$ 4 (Boselli et al. 2002c), suggesting
that the spatial distribution of the stellar component dominating the mid-IR emission
is similar to that emitting in the near-IR. 

\addtocounter{table}{6}
\begin{table}
\caption{The average stellar contribution to the 6.75 $\mu$m emission for different
morphological classes}
\label{Tab13}
\[
\begin{array}{p{0.3\linewidth}cr}
\hline
\noalign{\smallskip}
\hline
Type    & log[F_{6.75}(d+s)/F_{6.75}(s)]\\
\hline  
S0a     & -0.17  \pm 0.13       \\  
Sa-Sab  & 0.08   \pm 0.33       \\  
Sb-Sbc  & 0.39   \pm 0.18       \\  
Sc      & 0.76   \pm 0.29       \\  
Scd-Sd  & 0.50   \pm 0.30       \\  
Sm-Im   & 0.60   \pm 0.37       \\  
BCD     & 0.27   \pm 0.31       \\  
\noalign{\smallskip}
\hline              
\end{array}
\]
\end{table}

\addtocounter{figure}{0}
\begin{figure}[htbp]
\centerline {}
\vbox{\null\vskip 22.7 cm
\includegraphics{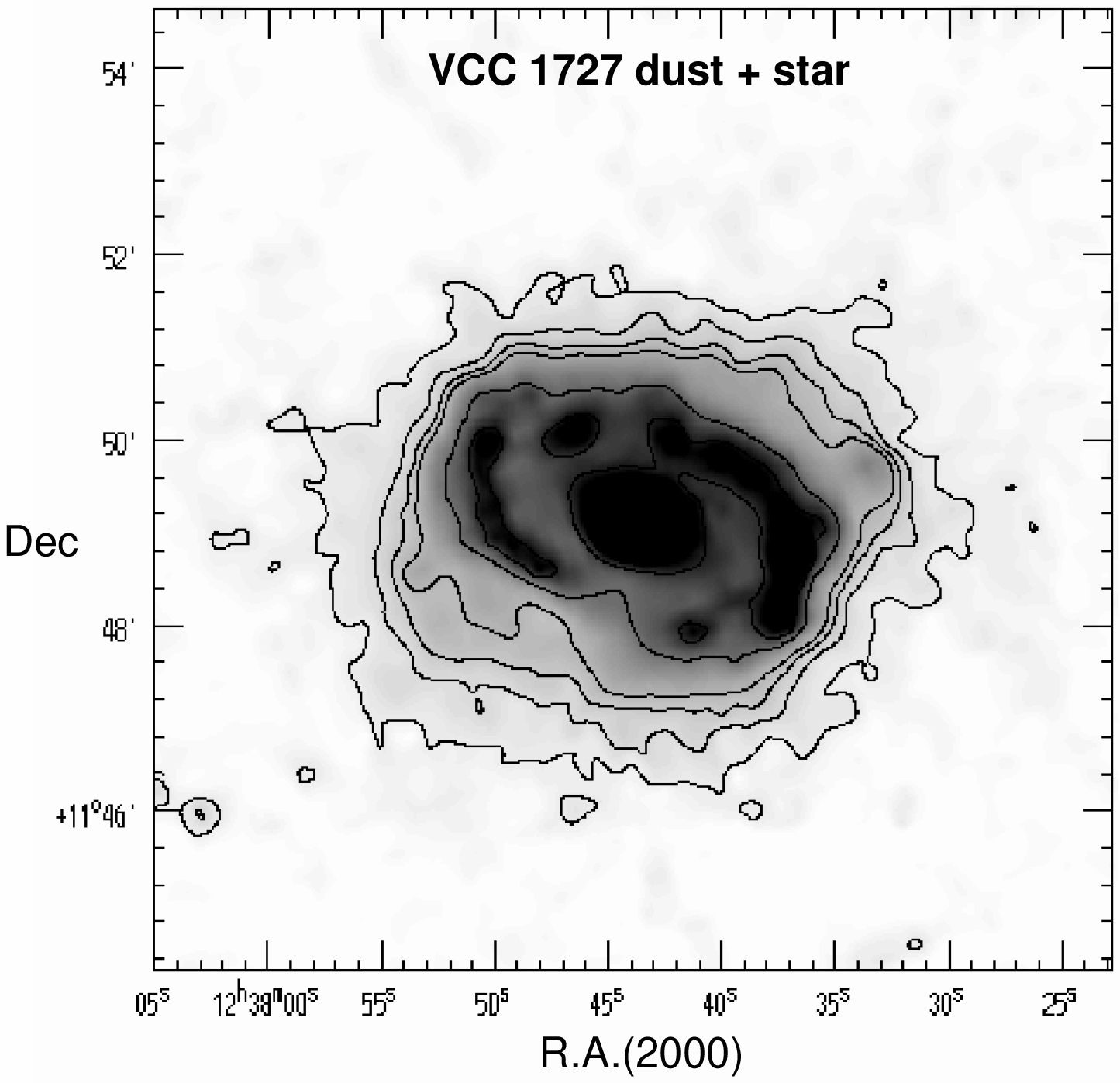}
\includegraphics{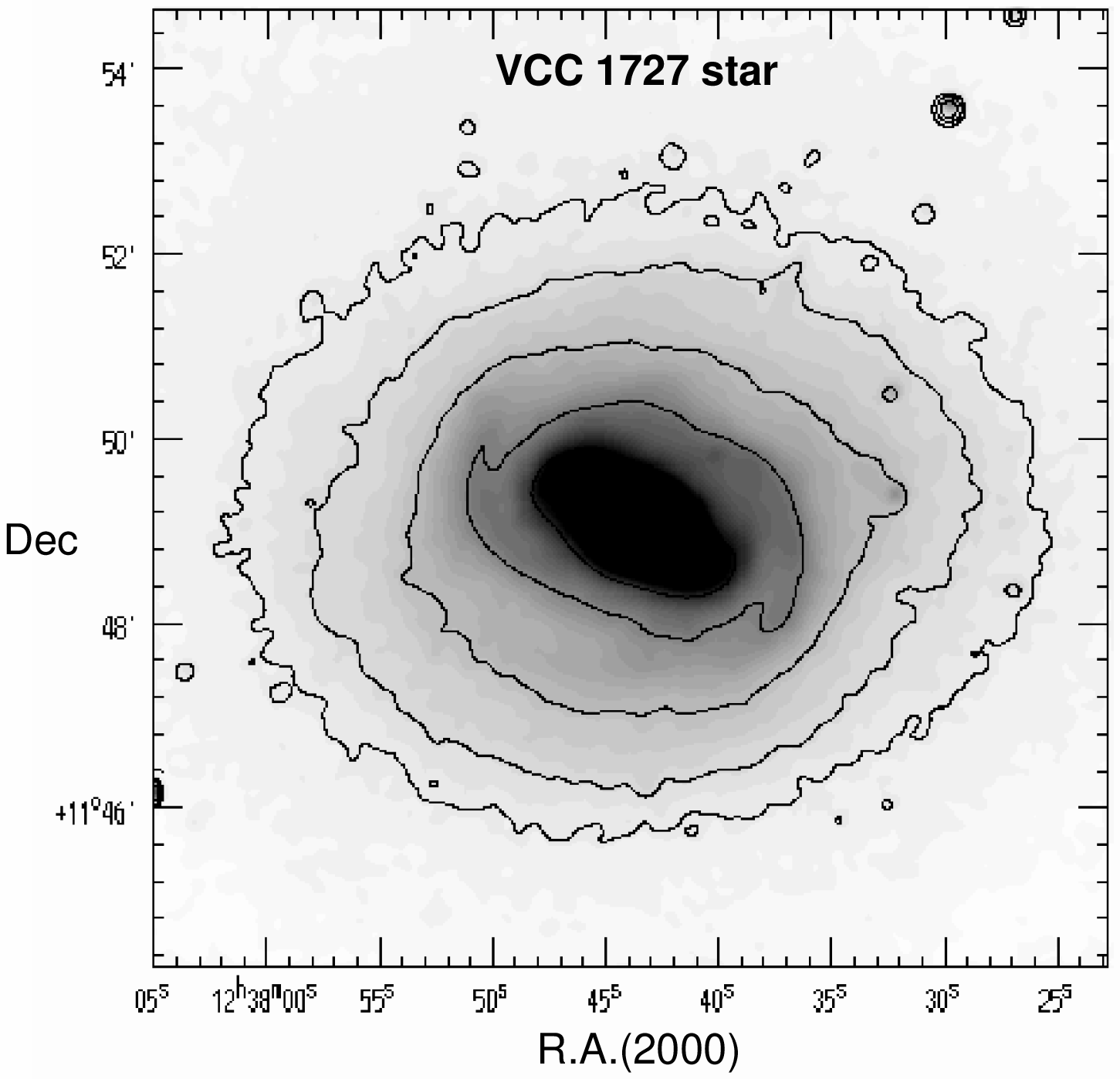}
\includegraphics{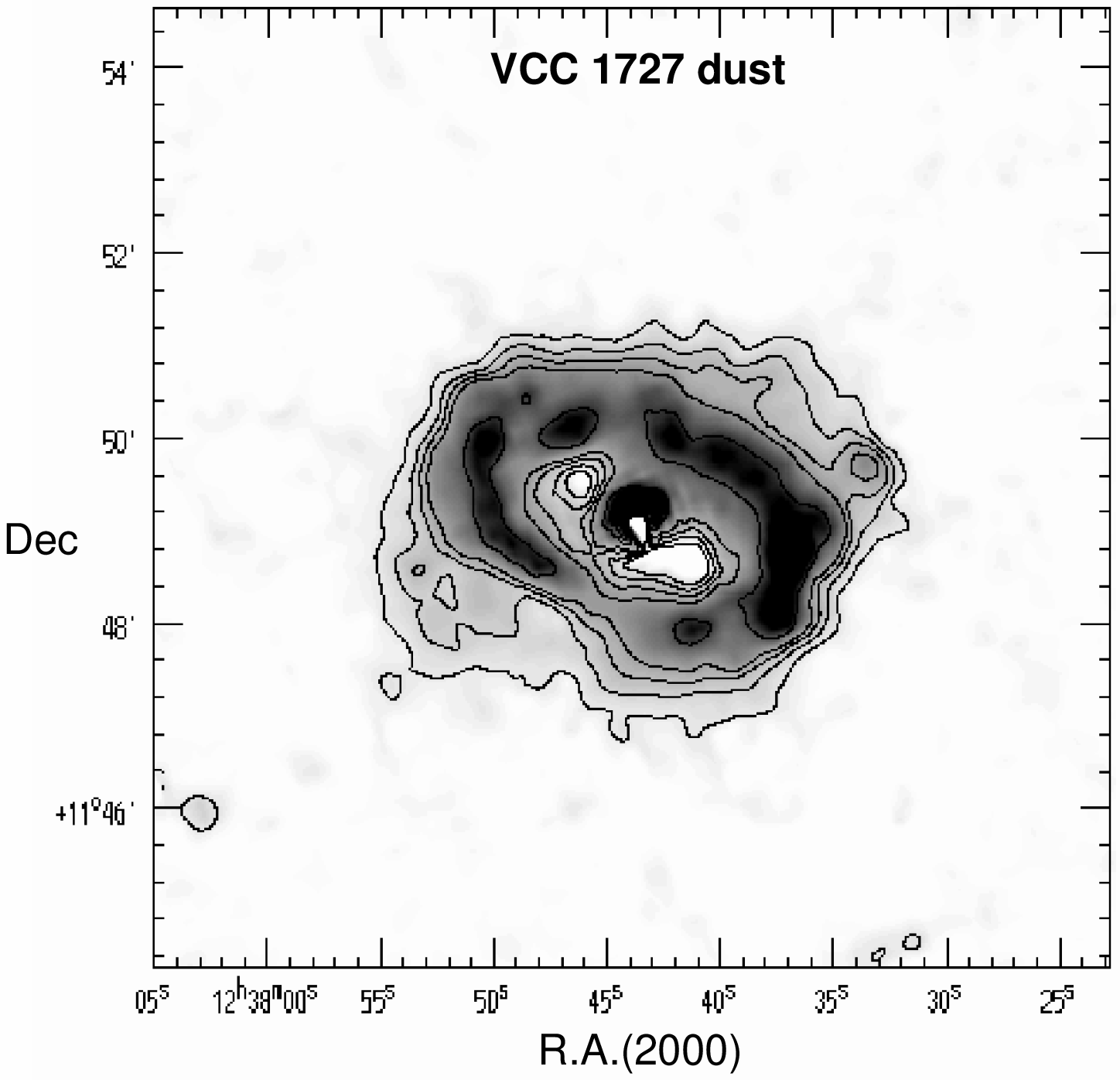}
}
\caption{The gray-level images of the Sab galaxy VCC 1727 at 6.75 $\mu$m:
a) the observed images (dust+star), b) the stellar image (scaled from the $K$ band image),
c) the image of the dust emission, corrected for stellar contribution (dust). The three
images are displayed with the same cuts and on the same scale (6.45 $\times$ 6.45
arcmin).}
\label{fig.5}
\end{figure}

\noindent
In the assumption that the stars dominating the emission at $\sim$ 7 $\mu$m have
a spatial distribution similar to those emitting in the near-IR, we can re-scale our $K$ band 
images (Boselli et al. 1997) using Table 8 and subtract them from the 
ISOCAM LW2 images of Boselli et al. (2002c) to obtain images of the pure dust emission at 6.75 $\mu$m.
We apply this correction, as an exercize to the Sab galaxy VCC 1727 (Fig. 5). The ISOCAM LW2 image at 6.75 $\mu$m
shows a very pronunced nucleus, a clumpy, ring-like structure and a smoothed, diffuse external region.
The emitting dust, on the contrary, is mostly located along the ring-like structure. Most of the
nuclear and part of the diffuse emission in the 6.75 $\mu$m image is stellar.\\
The determination of the stellar contribution to the 12 and 15 $\mu$m emission of galaxies
cannot be easely quantified since the Bruzual \& Charlot models are limited to the spectral
domain $\lambda$ $\leq$ 10 $\mu$m. 
The extrapolation of our fit (Fig. 2 (http://goldmine.mib.infn.it/papers/isosed.html)) indicates that 
the stellar contribution can be important at 15 $\mu$m, even though less than at 6.75 $\mu$m.\\
This result has to be taken in serious consideration when mid-IR deep surveys are used
to estimate the star formation activity of galaxies at high $z$, where rest-frame mid-IR
fluxes might be dominated by the stellar emission.

\subsection{The dust emission}

As extensively discussed in sect. (3.1), in a given galaxy the energy emitted by the various stellar
populations and absorbed by dust must equal the total energy radiated in 
the mid- and far-IR domain. 
However $A(UV)$ was estimated in sect. (3.1) just from $FIR$, which is a
combination of the 60 and 100 $\mu$m fluxes, not from 
the integral of the dust emission as determined on the SEDs. 
It remains to be checked
whether the global extinction A(1000 \AA $<$$\lambda$$<$ 10 $\mu$m), which depends on the 
adopted geometrical model and on the choice of the galactic extinction law, 
is consistent with the observed mid- and far-IR
emission.\\

The energy of the stellar light absorbed by dust is equal to the 
difference between the integrals of the stellar SEDs (i.e. the Bruzual \& Charlot models) 
prior and after the extinction correction. This should equal the energy radiated in the FIR:

\begin{equation}
\int\limits_{20 \mu m}^{2000 \mu m} F(\lambda)d\lambda=
                 \int\limits_{1000 \AA}^{10 \mu m} F_{corr}(\lambda)d\lambda- 
                 \int\limits_{1000 \AA}^{10 \mu m} F_{obs}(\lambda)d\lambda
\end{equation}

\noindent
where the integral on the left is 
performed under the two modified black-body functions
fitted to the data between 20 and 2000 $\mu$m (far-IR). The integrals on the right are performed under the
Bruzual \& Charlot models prior and after the extinction correction. We disregard the dust emission in the 
range  5 -- 20 $\mu$m due to the lack of model fitting in the mid-IR
domain, whose energy contribution to the total should however be small.

\addtocounter{figure}{0}
\begin{figure}[!h]
\centerline {}
\vbox{\null\vskip 8.0 cm
  \includegraphics{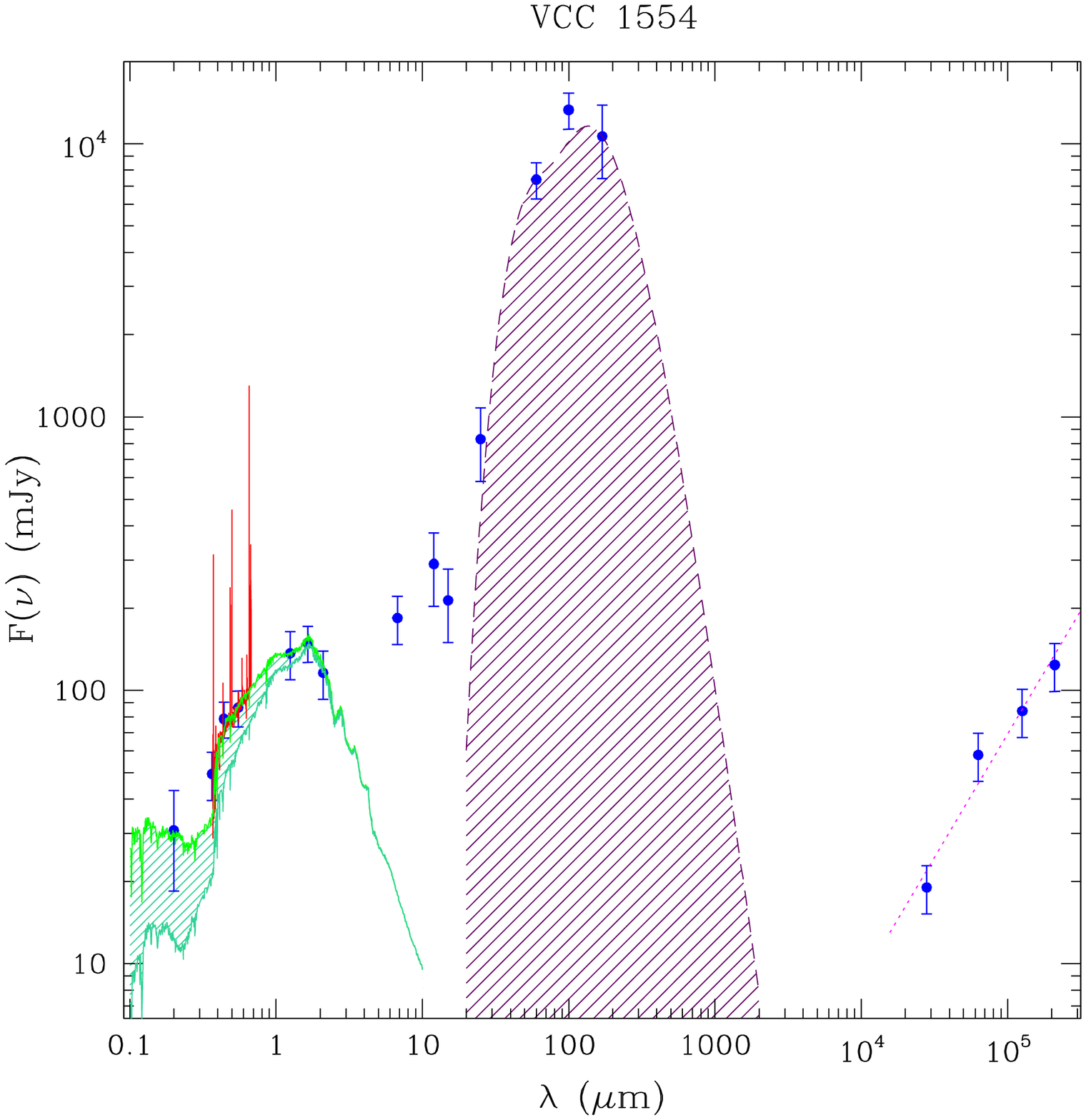}
}
\caption{The SED of the galaxy VCC 1554. The energy of the stellar light absorbed by dust
is indicated by the shaded region in between the Bruzual \&  Chralot extinction 
corrected model (green continuum line) and the observed one (sky-blue continuum line)
in the wavelength range 0.1-10 $\mu$m. The energy re-emitted by dust in the FIR
is shown by the violet shaded region in the wavelength range 20-2000 $\mu$m. The photometric
data are shown by blue dots, the optical spectrum is given in red.      
}
\label{fig.6}
\end{figure}

\noindent
To illustrate our method we give in Fig. 6 the SED of the galaxy VCC 1554. The energy of the 
stellar light absorbed by dust is marked by the shaded region shortward of 10 $\mu$m, 
the energy re-emitted in the FIR by the shaded region between 20 and 2000 $\mu$m.

\addtocounter{figure}{0}
\begin{figure}[!h]
\centerline {}
\vbox{\null\vskip 8.0 cm
\includegraphics{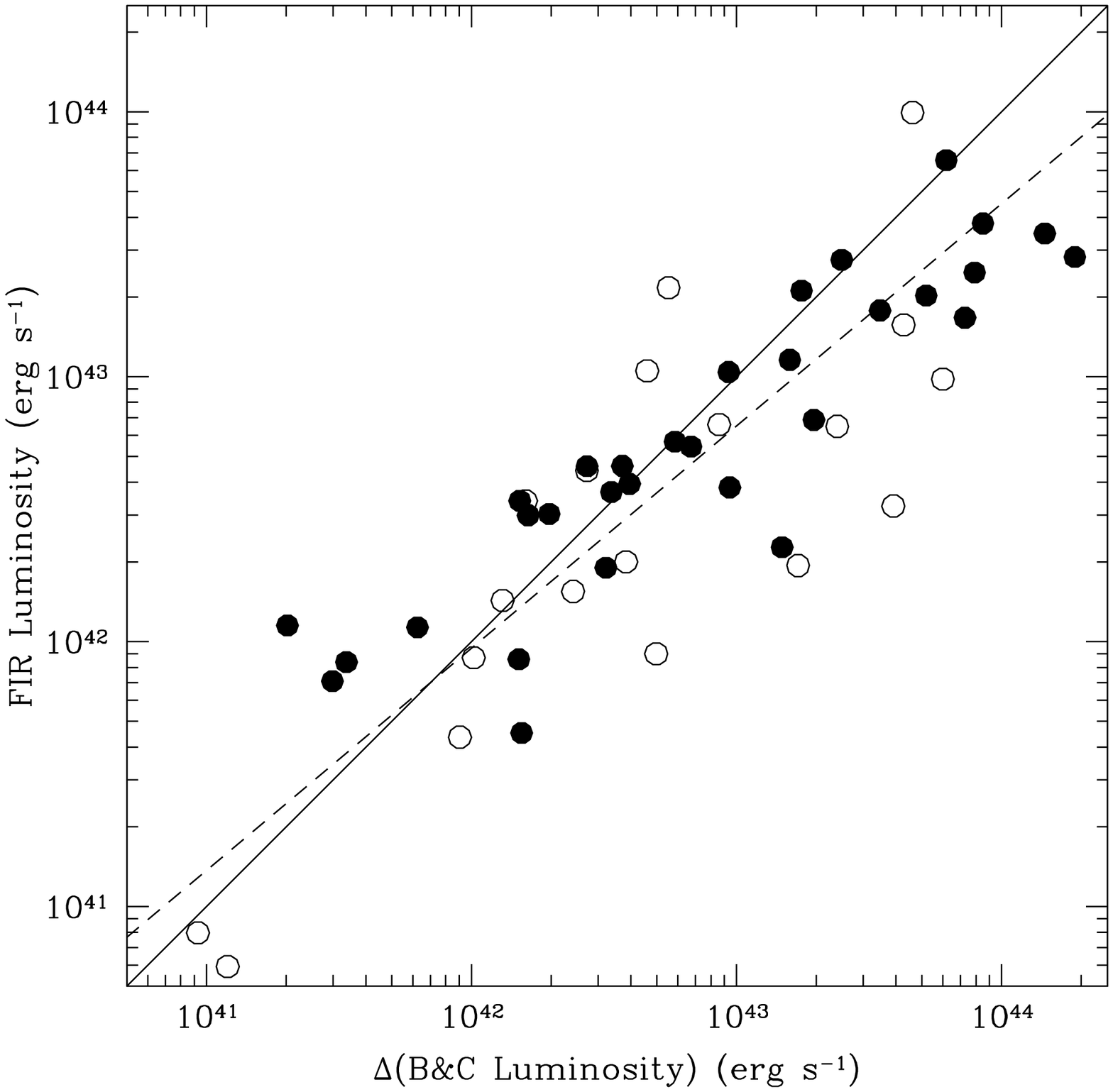}
}
\caption{The relationship between the total energy emitted in the far-IR and that
emitted by stars and absorbed by dust in the range between 1000 \AA ~and 10 $\mu$m (see eq. 8). 
The continuum line is the one to one relation, while the dashed line
is the bysector fit. Filled symbols are
galaxies whose extinction has been determined directly using the observed $FIR/UV$ ratio,
empty symbols are objects without far-IR and/or UV data, whose extinction has
been determined using the average $A(UV)$ for their morphological type class.   
}
\label{fig.7}
\end{figure}

\noindent
Figure 7 shows the relationship between the total energy emitted in the far-IR and that
emitted by stars and absorbed by dust in the range between 1000 \AA ~and 10 $\mu$m (eq. 8). 
The median value of the ratio between the energy 
absorbed by dust and that emitted in the far-IR is 1.27 for the entire sample, 1.03 for those objects 
whose extinction has been determined directly using the observed $FIR/UV$ ratio, as illustrated in Fig. 8.

\addtocounter{figure}{0}
\begin{figure}[!h]
\centerline {}
\vbox{\null\vskip 8.0 cm
\includegraphics{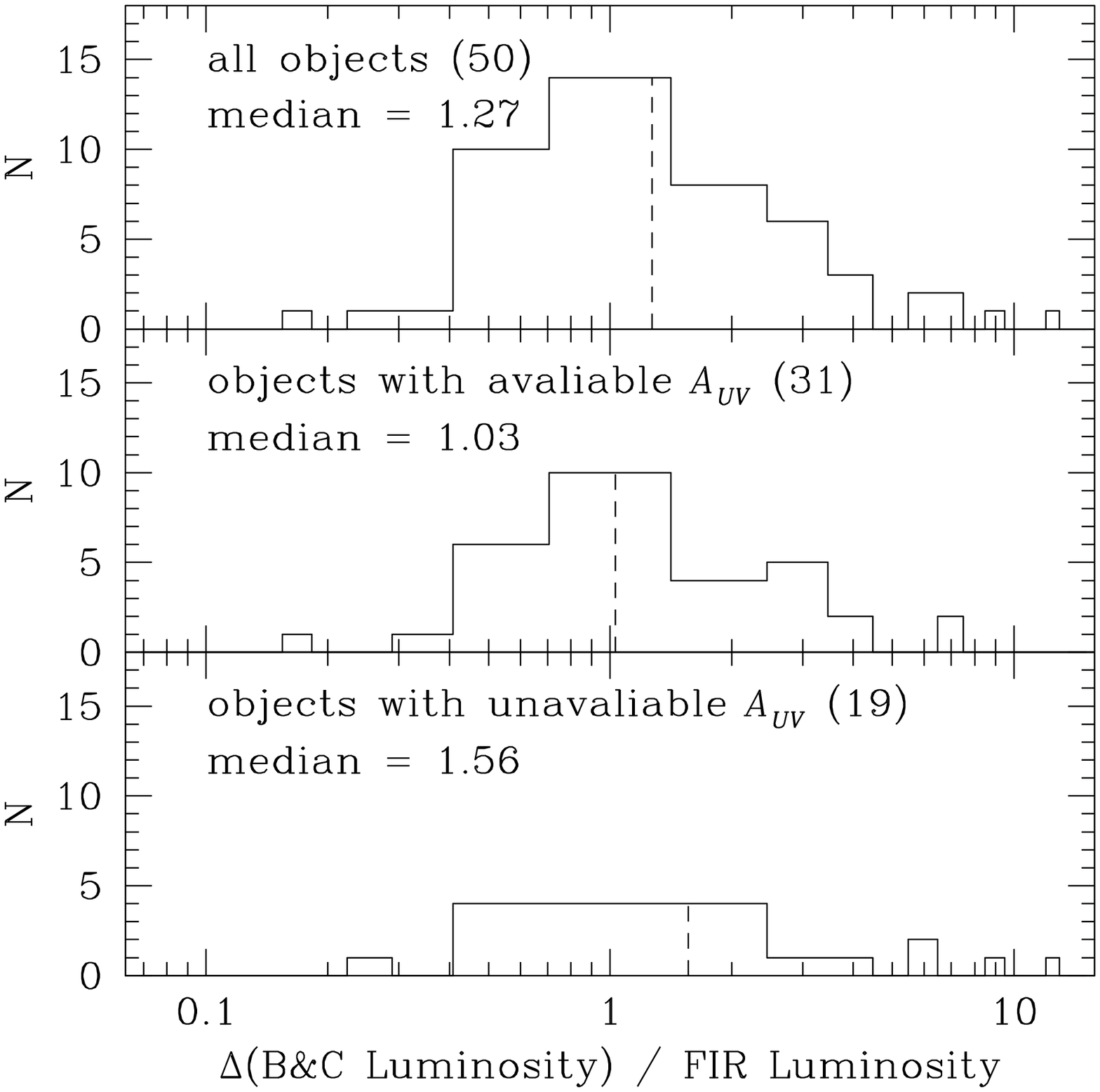}
}
\caption{The distribution of the ratio between the energy absorbed by dust and that 
emitted in the far-IR for the entire sample (a),
for galaxies whose extinction has been determined using the observed $FIR/UV$ ratio (b),
and for those objects without far-IR and/or UV data whose extinction has
been determined using the average value of $A(UV)$ for their morphological class (c).   
}
\label{fig.8}
\end{figure}

\noindent
The almost linear relation between the absorbed star light and the energy emitted by
dust, combined with their ratio close to one,  
leads us to conclude that the prescription 
given in sect. (3.1) to correct stellar SEDs is sufficiently
accurate for optically-selected spiral galaxies, even for objects without 
UV and far-IR data. 

\addtocounter{figure}{0}
\begin{figure}[!h]
\centerline {}
\vbox{\null\vskip 8.0 cm
\includegraphics{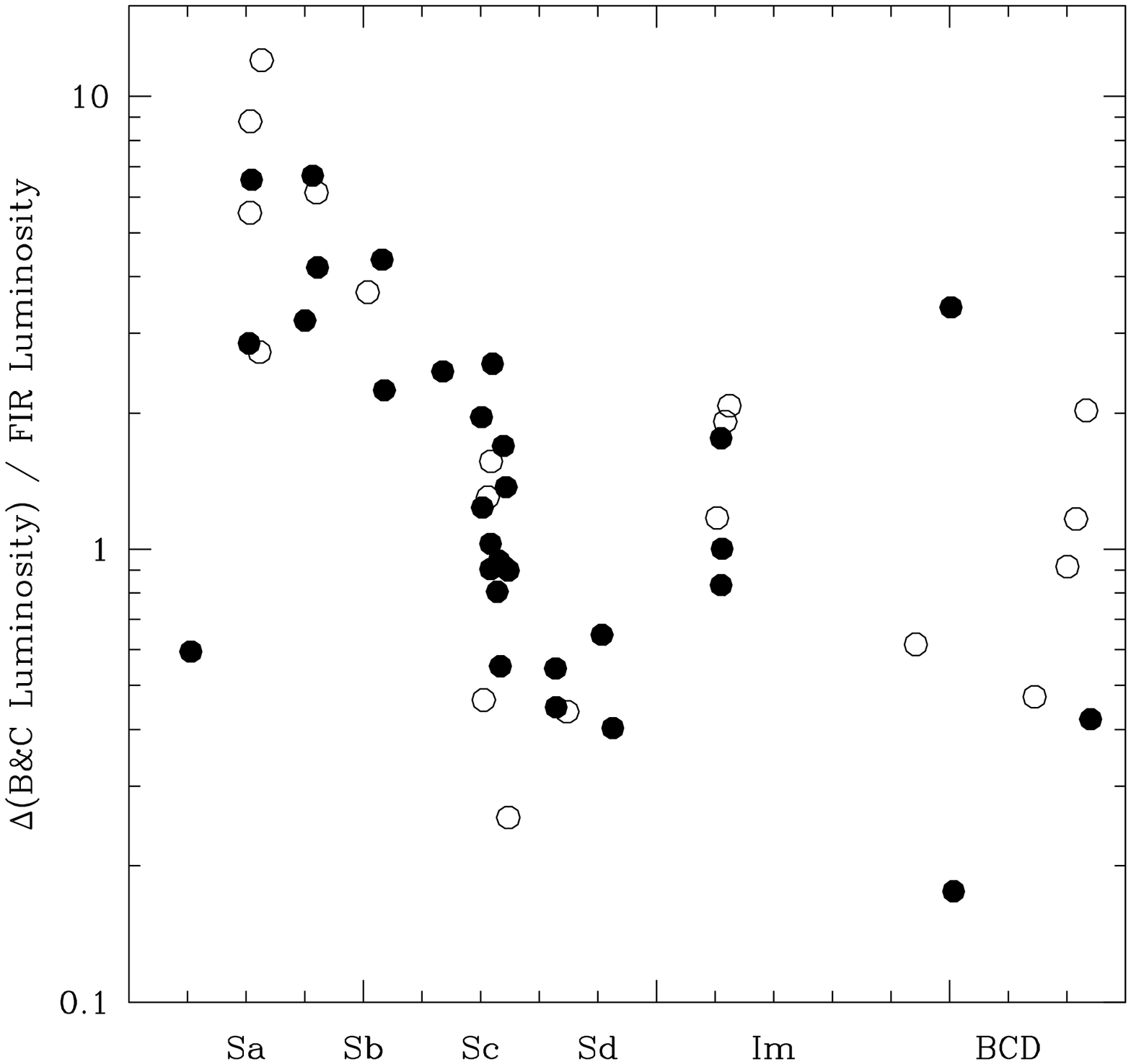}
}
\caption{The ratio between the energy absorbed by dust and that emitted in the 
far-IR as a function of the morphological type. Symbols as in Fig. 7.  
}
\label{fig.9}
\end{figure}

\noindent
The ratio between the energy absorbed by dust and that emitted in the 
far-IR shows however a weak residual trend with morphological type (Fig. 9) 
and luminosity (Fig. 10): it is significantely larger than unity in early-type, massive galaxies.

\addtocounter{figure}{0}
\begin{figure}[!h]
\centerline {}
\vbox{\null\vskip 8.0 cm
\includegraphics{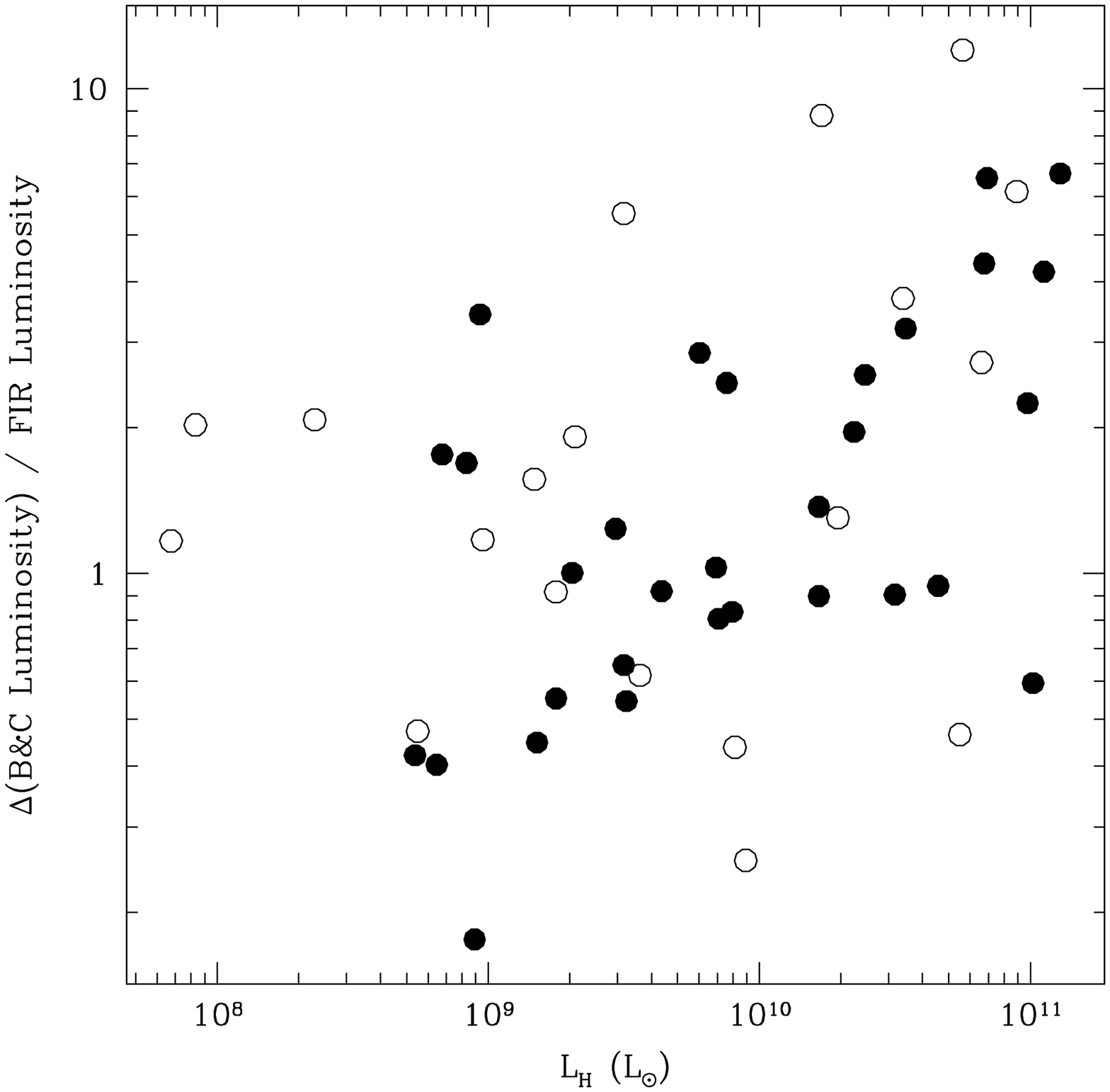}
}
\caption{The ratio between the energy absorbed by dust and that emitted in the 
far-IR as a function of the H band luminosity. Symbols as in Fig. 7.  
}
\label{fig.10}
\end{figure}
\noindent
This increase could be due to an underestimate of the far-IR emission of massive, early-type
galaxies, that could exist if we missed a colder dust component in quiescent objects
with low UV interstellar radiation field.

\noindent
We remind that the extinction values derived using this prescription are significantly
smaller than those obtained using the Calzetti's law, which is probably more accurate for
starburst galaxies (see Gavazzi et al. 2002a and Buat et al. 2002 for a detailed discussion on 
this issue). 

\subsection{The radio emission}

For 25 galaxies detected at more than one frequency in the centimetric domain, we
derive the slope of the radio continuum spectrum by a simple linear fit to the data. 
Excluding galaxies VCC 857, 1110 and 1450 showing 
large inconsistencies in the radio continuum flux densities and 8 additional objects  
with signs of nuclear activity (LINER, Seyfert, see Table 2)
we obtain an average spectral slope $\alpha$=0.76 $\pm$ 0.27, consistent with 
the canonical synchrotron slope $\alpha$=0.8 found by Niklas et al. (1997) by
carefully separating the contribution of the thermal from the synchrotron emission (see Table 8).



\subsection{The bolometric luminosity of optically-selected, late-type galaxies}

By integrating the fit models in the stellar and FIR domain, we 
calculate the (observed) bolometric luminosity of our target galaxies: 

\begin{equation}
        L_{Bol} = \int\limits_{1000 \AA}^{10 \mu m} F_{obs}(\lambda)d\lambda +
\int\limits_{20 \mu m}^{2000 \mu m} F(\lambda)d\lambda 
\end{equation}

As before, we disregard the contribution of UIB 
and of the very small grains in the 5-50 $\mu$m range, thus this esitimate gives
a lower limits to the total bolometric luminosity.

\addtocounter{figure}{0}
\begin{figure}[!h]
\centerline {}
\vbox{\null\vskip 8.0 cm
\includegraphics{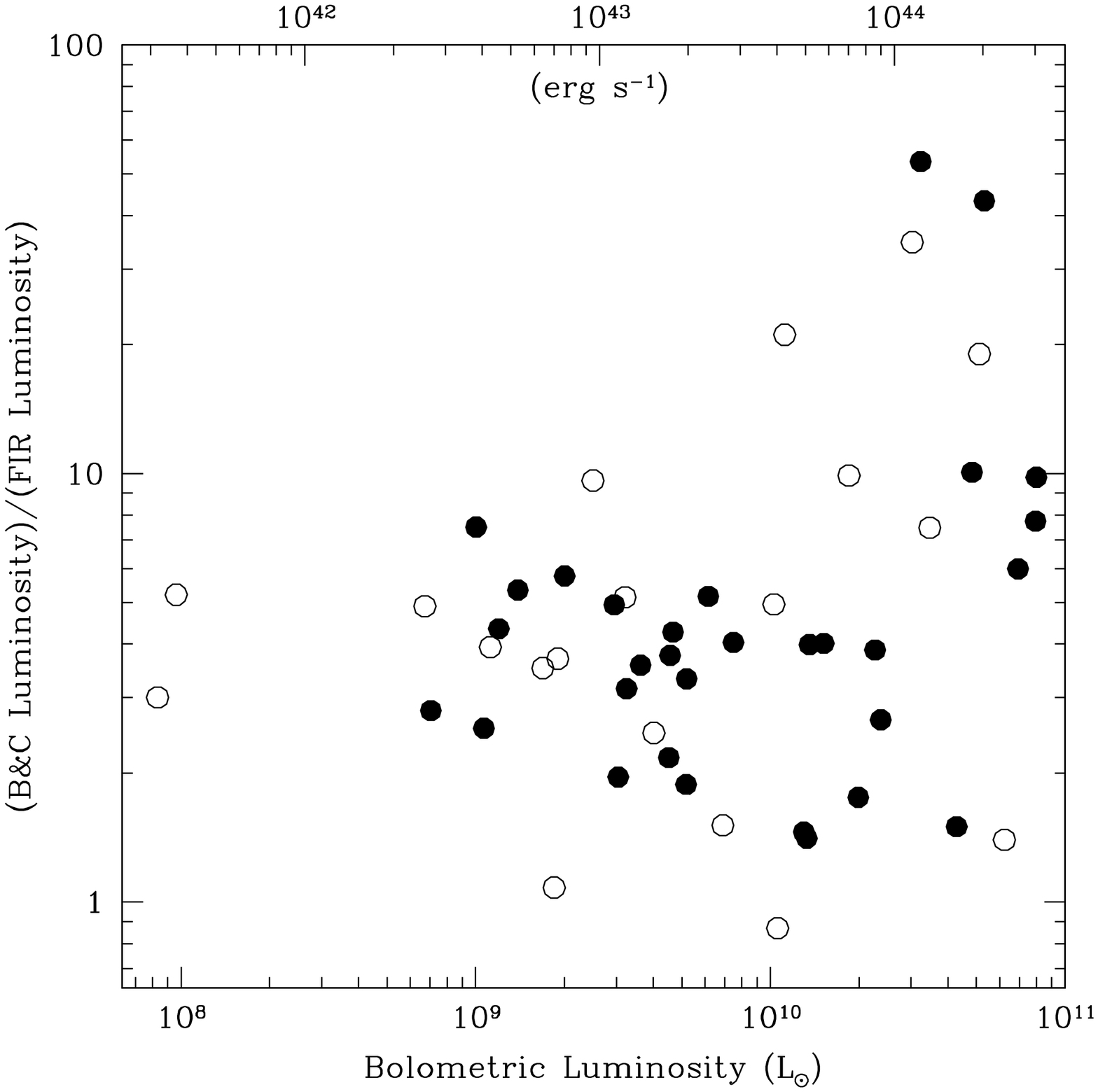}
}
\caption{The relationship of the ratio of the total uncorrected 
stellar luminosity (from the Bruzual \& Charlot model) to the total FIR
luminosity versus the bolometric luminosity of the target galaxies. Symbols are as in Fig. 7.
}
\label{bolo}
\end{figure}

\noindent
Figure \ref{bolo} shows that the bolometric luminosity of optically-selected late-type galaxies
in the range 10$^8$ $\leq$ $L_{bol}$ $\leq$ 10$^{11}$ L$\odot$,
is dominated by the stellar emission. The median value of the ratio between
the energy emitted by stars in the 1000 \AA - 10 $\mu$m range and by dust in the Far-IR 
is 4.0, 
significantly higher than $f_B/f_{FIR} \sim$ 1.6 
found by Soifer et al. (1987) who determined the stellar emission from the B band luminosity alone.
No relation is observed between the stellar to FIR ratio and the bolometric luminosity,
except for an higher dispersion at high luminosity.

\addtocounter{figure}{0}
\begin{figure}[!h]
\centerline {}
\vbox{\null\vskip 8.0 cm
\includegraphics{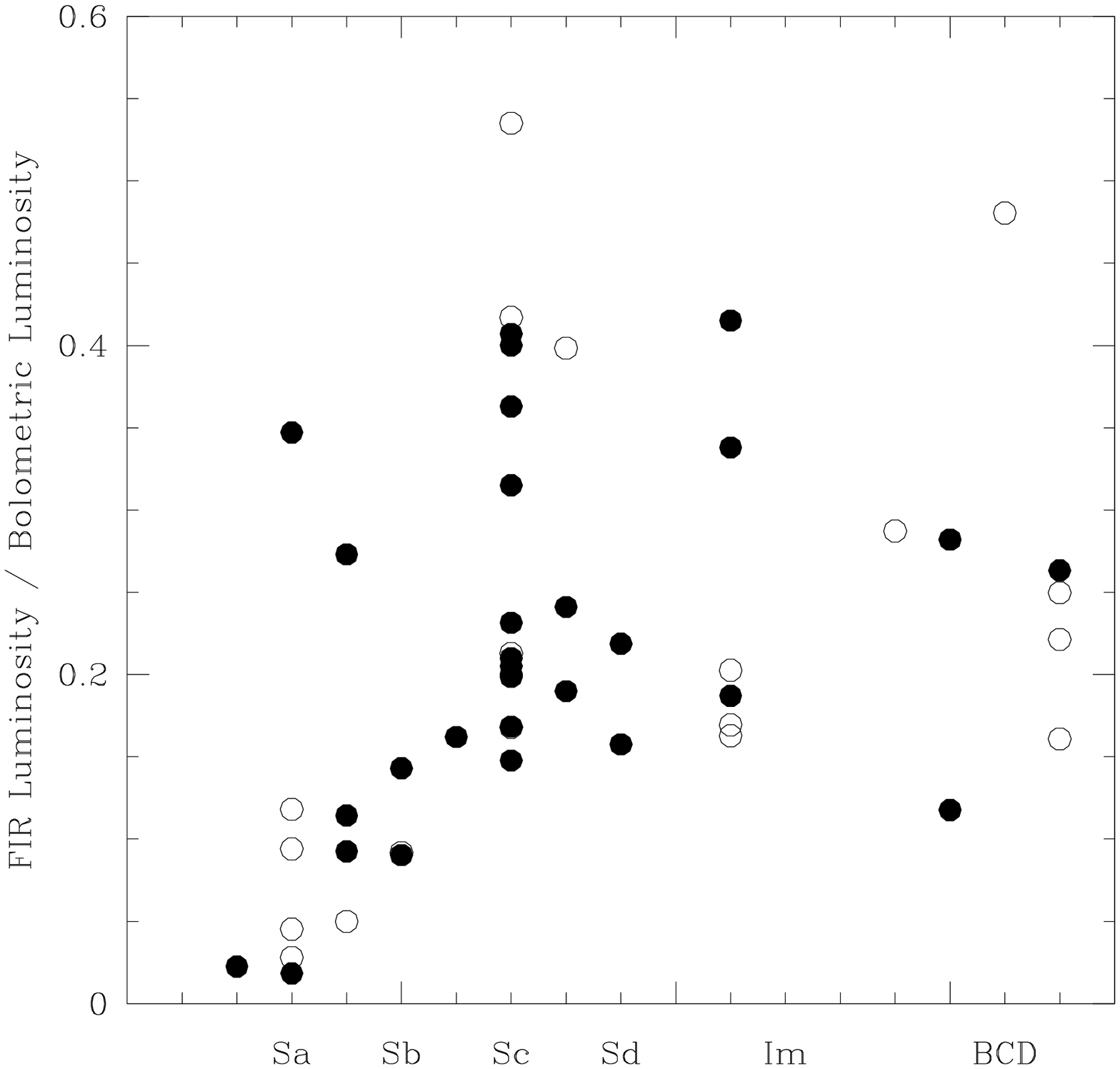}
}
\caption{The relationship between the far-IR to bolometric luminosity ratio and the 
morphological type. Symbols are as in Fig. 7.
}
\label{bolotype}
\end{figure}

Figure \ref{bolotype} shows that the far-IR to bolometric luminosity ratio increases 
from early Sa spirals ($L_{FIR}/L_{bol} \leq 0.1$) to Sc-Sd galaxies ($L_{FIR}/L_{bol} \sim 0.2-0.4$),
consistently with Popescu \& Tuffs (2002). BCDs have values of $L_{FIR}/L_{bol} \sim 0.2$, significantly 
lower than those estimated by Popescu \& Tuffs (2002) ($L_{FIR}/L_{bol} > 0.5$). This 
discrepancy is probably due to a different determination of the stellar contribution to the 
bolometric luminosity. Our complete and homogeneous dataset allowed a more accurate determination of
the total stellar emission than in Popescu \& Tuffs (2002), in particular for BCD galaxies where
the contribution of the UV emission, here determined for several objects using unpublished UV data,
might be dominant.

\section{Summary and Conclusions}

We present a multifrequency dataset comprising an optically-selected, 
volume-limited, complete sample of 118 galaxies in the Virgo cluster. 
The sample includes all late-type ($\geq$ S0a) Virgo A members
in the core of the cluster, with projected distance $\Theta<$ 2 degrees 
from M87, or at the peryphery of the cluster ($\Theta>$ 4 degrees from the position 
of maximum projected galaxy density).\\
The database includes UV, visible, 
near-IR, mid-IR, far-IR, radio continuum photometric data as well 
as spectroscopic data on the H$\alpha$, CO and HI lines.\\
Spectral energy distributions (SEDs) of the individual galaxies, 
as well as templates SEDs in bins of morphological type and 
luminosity are derived. The SEDs are fitted with
stellar population synthesis models providing an estimate of the total stellar radiation, 
with modified black-bodies fitted to the far-IR data giving the energy re-emitted by dust, 
and with power laws representing the synchrotron emission.\\
Assuming the energy balance between the absorbed stellar light and the energy
radiated in the IR by dust, we calibrate an empirical attenuation law 
suitable for correcting photometric and spectroscopic data of normal
galaxies.\\
The analysis of the SED show that low-luminosity,
dwarf galaxies have on average bluer stellar continua and higher 
far-IR luminosities (per unit galaxy mass) than giant, early-type spirals.
Normal spirals have relatively similar observed stellar spectra but
10 to 100 times lower IR luminosities 
than nearby starburst galaxies such as M82 and Arp 220. 
The temperature of the cold dust
component increases with the far-IR luminosity, from giant spirals to dwarf
irregulars and to an higher extent in starburst galaxies.
SEDs of starburst
galaxies should not be used as templates of normal high redshift galaxies.\\
We show that the contribution of the
stellar emission to the 6.75 $\mu$m mid-IR flux is 
generally important, from $\sim$ 80 \% in Sa to $\sim$ 20 \% in Sc.\\


\acknowledgements

We thank J. Donas and D. Pierini for providing us with unpublished UV and near-IR data.
We thank S. Arnouts, V. Buat and M. Sauvage for stimulating discussions, and D. Elbaz for providing us with
the SED of M82 and Arp 220. This research has made use of the NASA/IPAC Extragalactic Database (NED) which is operated 
by the Jet Propulsion Laboratory, California Institute of Technology, under contract with the
National Aeronautics and Space Administration. most of the data presented in this
work are available through the WEB page http://goldmine.mib.infn.it.

\noindent
Appendix: UBV CCD photometry of Virgo galaxies \footnote{The Observatoire 
de Haute Provence (OHP) (France), is
operated by the French CNRS; the INT telescope is operated on the island 
of La Palma by the ING team in the 
Spanish Observatorio del Roque de Los Muchachos of the 
Instituto de Astrofísica de Canarias; KPNO is operated by AURA, Inc. 
under contract to the National Science Foundation.}

The present work is partly based on new CCD optical photometry of 36 galaxies 
obtained at the 1.20m Newton telescope at the Observatoire d'Haute Provance (OHP, France),
at the 0.9m telescope at Kitt Peak and at the 2.5m INT telescope at La Palma.
The OHP and the INT observations were taken during the H$\alpha$ surveys presented
in Boselli \& Gavazzi (2002) and Boselli et al. (2002a) respectively. Details on the
observations and data reduction procedures are found in these papers.
Kitt Peak targets were observed as fillers during an H$\alpha$ survey of isolated
galaxies.\\  
The $f/6$ 1.2m OHP telescope was equipped with 
a thinned TK1024$\times$1024 pixels CCD detector, with a pixel size of 0.69 arcsec
and a field of view of 11.8$\times$11.8 arcminuts.
At the adopted gain, the electron/adu conversion is 3.5 $e^-$/adu, with a readout noise of
8.5 $e^-$.
Thirty galaxies of the present sample were observed during 26 nights in two runs, 
in 1998 and 2000. Fourteen galaxies were imaged in the $V$, 
30 in the $B$ and 1 in the $U$ band. The observations were done in poor seeing 
conditions, ranging from 2 to 4 arcsec.
The typical integration time was 10 minuts in the $V$, 15 in the $B$ and 30
in the $U$ bands.\\
INT $B$ band imaging of 2 galaxies were obtained in 1999 using the  
Wide Field Camera (WFC) attached at the prime focus of the $f/3.29$
2.5m telescope. 
The WFC is composed by a science array of four thinned AR
coated EEV 4K$\times$2K CCDs, plus a fifth acting as
autoguider. The pixel scale at the detectors is 0.33
arcsec~pixel$^{-1}$, which gives a total field of view of about $34\times 34$~
arcmin$^{2}$. The observations were done during photometric conditions, with an 
average seeing of 1.5-2 arcsec and an integration time of 10 minutes.\\
Kitt Peak $B$ band imaging of 5 galaxies were obtained during 4 nights in 1995 
using the 0.9m telescope in the $f/13$ configuration, equipped with a 
T2KA 2048$\times$2048 pixel CCD,
with a pixel size of 0.384 arcsec~pixel$^{-1}$ and a total field of view of
13.1$\times$13.1 arcminuts. At the adopted gain, the 
electron/adu conversion is 2 $e^-$/adu, with a lecture noise of 4 $e^-$. 
The observations were done during non photometric conditions, with an 
average seeing of 1-1.5 arcsec and an integration time of 15 minutes.\\

The observations were calibrated and transformed into the Johnson $UBV$ 
system using standard stars in the catalogue of Landolt (1983). Observations
of the standard stars were repeated every 2 hours. Repeated measurements
gave $<$ 0.10 mag differences, which we assume as the typical uncertainty of the 
photometric result given in this work. Not all frames were obtained in photometric 
conditions. When the zero point was varing by more than 0.05 mag due to cirrus,
we choose to observe only galaxies with available multiaperture photometry in
order to perform the calibration a posteriori.


The data reduction of the CCD images follows a procedure identical to the one 
described in previous papers of the series (Gavazzi et al. 1995), based on the IRAF
STSDAS data reduction packages. To remove the detector response each image is
bias subtracted and devided by the mean of 5 flat field exposures obtained on
the twilight sky. Direct inspection of the frames allows manual cosmic rays
removal and subtraction of contaminating objects, such as nearby stars and galaxies.
The sky background is determined in each frame in concentric object-free annuli 
around the object. The typical uncertainty on the mean background is estimated 
10 \% of the rms in the individual pixels. This represents the dominant source of 
error in low S/N regions. 
The determination of object centroid is performed by fitting gaussian two-dimensional 
profiles to the data, centered on the brightest excess in each object, generally 
corresponding to the nucleus. At the central coordinates determined, a growth curve 
is derived for each object by integrating the counts in concentric circular rings of 
increasing radii. The obtained growth curves, transformed from counts to magnitudes, 
are then compared with the multiaperture photometry available in the literature, in 
order to check our photometric calibration and to obtain a zero point for those  
objects observed in non-photometric conditions. At this stage stars projected within 
the target galaxies were not subtracted since, unless specified, reference aperture 
photometry usually includes them.
Once the accurate zero point is obtained for each galaxy, a similar procedure is repeated 
after subtracting contaminating stars and galaxies.
Following the procedure described in Gavazzi \& Boselli (1996), a magnitude is obtained after 
integrating along circular, concentric annuli up to the isophotal 25 mag arcsec$^{-2}$
$B$ diameter. To improve the photometric accuracy, this procedure is applyed adding our 
measurements with aperture photometry available in the literature.
$UBV$ magnitudes of the target galaxies are given in Table 3.
The estimated error on the magnitude is $\sim$ 10 \%.\\

\addtocounter{table}{-12}
\onecolumn
\tiny
\begin{longtable}{rccccrrcrrrcrrrcc}
\caption{The target galaxies}\\
\hline
\noalign{\smallskip}
\hline
\noalign{\smallskip}
{\rm VCC}&{\rm NGC}& {\rm IC} &{\rm UGC}& {\rm CGCG}&{\rm R.A.(2000)}&{\rm dec}&{\rm type}&$m_{pg}$&
$a$&$b$&$vel$&$Dist$&{\rm memb.}&$\theta$& $C_{31}$&com\\
         &         &            &       &           & h~ m~ s        & $^{o}$~ '~ "&          &mag &    
 ' & ' &km s$^{-1}$&Mpc&          &deg   & &\\
(1) & (2) & (3) & (4) & (5) & (6) & (7) & (8) & (9) & (10) & (11) & (12) & (13) 
& (14) & (15) & (16) & (17)\\
\hline                                                                          
1    &-      &-      &-     &69059  &120820.02  &134100.2  &BCD?    &        14.78&  0.80 & 0.18 & 2267  &32 & M & 5.63 &  3.36 &\\
4    &-      &-      &-     &-      &120830.75  &150548.2  &Im      &        17.50&  0.50 & 0.43 & 589   &32 & M & 6.06 &  -    &\\
17   &-      &3023   &7150  &-      &121001.86  &142142.4  &Im      &        15.20&  0.91 & 0.45 & 819   &32 & M & 5.43 &  2.85 &\\
24   &-      &-      &-     &69070  &121035.65  &114538.5  &BCD     &        14.95&  1.00 & 0.37 & 1289  &32 & M & 4.99 &  6.23 &\\
26   &-      &-      &-     &-      &121040.20  &143848.5  &Im      &        17.50&  0.43 & 0.27 & 2469  &32 & M & 5.39 &  2.35 &\\
66   &4178   &-      &7215  &69088  &121246.27  &105156.0  &SBc(s)  &        11.89&  5.35 & 1.87 & 369   &17 & N & 4.68 &  3.24 &*\\
81   &4186   &-      &7223  &-      &121326.18  &144620.1  &d:Sc    &        15.60&  0.95 & 0.81 & 2075  &17 & N & 4.85 &  3.22 &\\
87   &-      &-      &-     &98106  &121340.91  &152713.2  &Sm      &        15.00&  1.45 & 0.72 & -134  &17 & N & 5.17 &  3.18 &\\
92   &4192   &-      &7231  &98108  &121348.24  &145401.2  &Sb:     &        10.92&  9.78 & 2.60 & -135  &17 & N & 4.84 &  5.04 &*\\
130  &-      &-      &-     &-      &121504.22  & 94513.5  &BCD     &        16.50&  0.63 & 0.25 & 2189  &17 & N & 4.68 &  2.71 &\\
152  &4207   &-      &7268  &69107  &121530.31  & 93508.6  &Scd(on edge)&    13.48&  1.96 & 0.89 & 592   &17 & N & 4.69 &  3.54 &\\
159  &-      &-      &-     &69108  &121541.50  & 81707.7  &Im      &        15.08&  1.04 & 0.52 & 2584  &32 & W & 5.54 &  2.73 &\\
169  &-      &-      &-     &-      &121556.39  & 93855.7  &Im      &        16.50&  0.85 & 0.43 & 2222  &17 & N & 4.57 &  -    &\\
171  &-      &-      &-     &-      &121558.88  & 82225.8  &Im      &        17.40&  0.57 & 0.36 & 875   &32 & W & 5.43 &  -    &\\
207  &-      &-      &-     &-      &121648.07  & 80302.0  &BCD     &        17.20&  0.36 & 0.13 & 2564  &32 & W & 5.55 &  2.63 &\\
318  &-      &776    &7352  &70005  &121903.40  & 85122.7  &SBcd    &        14.01&  1.71 & 1.00 & 2469  &32 & W & 4.57 &  2.93 &\\
425  &-      &-      &-     &-      &122035.90  & 81209.3  &Im:     &        17.30&  0.43 & 0.38 & -     &23 & B & 4.89 &  -    &\\
459  &-      &-      &-     &99022  &122111.46  &173818.5  &BCD     &        14.95&  0.84 & 0.36 & 2108  &17 & A & 5.74 &  3.09 &\\
460  &4293   &-      &7405  &99023  &122112.68  &182256.5  &Sa pec  &        11.20&  5.10 & 2.92 & 921   &17 & A & 6.42 &  3.52 &*\\
655  &4344   &-      &7468  &99037  &122337.45  &173228.5  &S pec,N:/BCD&    13.21&  1.55 & 1.55 & 1147  &17 & A & 5.44 &  2.50 &\\
664  &-      &3258   &7470  &70042  &122344.36  &122842.5  &Sc      &        13.50&  2.60 & 1.87 & -427  &17 & A & 1.73 &  2.55 &\\
666  &-      &-      &-     &-      &122346.13  &164728.5  &Im:     &        16.80&  1.00 & 0.57 & -     &17 & A & 4.72 &  2.80 &\\
692  &4351   &-      &7476  &70045  &122401.37  &121216.6  &Sc(s)   &        12.93&  2.92 & 1.87 & 2324  &17 & A & 1.67 &  2.95 &\\
793  &-      &-      &-     &-      &122521.88  &130423.2  &Im,N?   &        16.74&  0.47 & 0.34 & 1906  &17 & A & 1.50 &  2.25 &\\
802  &-      &-      &-     &-      &122529.01  &132947.3  &BCD     &        17.40&  0.64 & 0.21 & -215  &17 & A & 1.71 &  2.58 &\\
809  &-      &3311   &7510  &70063  &122533.17  &121536.3  &Sc (on edge)&    14.55&  1.45 & 0.36 & -142  &17 & A & 1.30 &  3.24 &\\
836  &4388   &-      &7520  &70068  &122546.60  &123940.4  &Sab     &        11.83&  5.10 & 1.24 & 2515  &17 & A & 1.26 &  4.69 &*\\
848  &-      &-      &-     &42097  &122552.78  & 54829.5  &Im pec/BCD&      14.72&  1.16 & 0.98 & 1537  &23 & B & 6.70 &  2.86 &\\
857  &4394   &-      &7523  &99047  &122555.64  &181249.5  &SBb(sr) &        11.76&  3.60 & 3.60 & 914   &17 & A & 5.94 &  5.64 &*\\
873  &4402   &-      &7528  &70071  &122607.32  &130643.6  &Sc (on edge)&    12.56&  3.95 & 1.16 & 234   &17 & A & 1.36 &  2.95 &\\
890  &-      &-      &-     &-      &122620.85  & 64005.7  &BCD     &        16.00&  0.21 & 0.21 & 1483  &23 & B & 5.83 &  2.64 &\\
912  &4413   &-      &7538  &70076  &122632.16  &123639.8  &SBbc(rs)&        12.97&  2.92 & 1.75 & 105   &17 & A & 1.07 &  3.14 &\\
945  &-      &3355   &7548  &70085  &122651.06  &131032.9  &SBm     &        15.31&  1.29 & 0.57 & -9    &17 & A & 1.25 &  2.67 &\\
950  &-      &3356   &7547  &70084  &122651.38  &113316.9  &Sm      &        14.49&  1.71 & 0.85 & 1098  &17 & A & 1.28 &  2.78 &\\
971  &4423   &-      &7556  &42107  &122708.93  & 55248.1  &Sd (on edge)&    14.28&  3.06 & 0.43 & 1120  &23 & B & 6.57 &  3.44 &\\
984  &4425   &-      &7562  &70091  &122713.30  &124405.1  &SBa     &        12.82&  2.99 & 1.00 & 1883  &17 & A & 0.94 &  4.68 &\\
995  &-      &3371   &7565  &70092  &122721.55  &105155.2  &Sc (on edge)&    15.32&  1.53 & 0.11 & 928   &17 & A & 1.75 &  3.43 &\\
1001 &-      &-      &-     &-      &122724.65  &134300.2  &Im      &        16.60&  0.73 & 0.47 & 338   &17 & A & 1.56 &  2.44 &\\
1002 &4430   &-      &7566  &42111  &122726.37  & 61544.2  &SBc(r)  &        12.48&  3.02 & 2.69 & 1450  &23 & B & 6.19 &  2.73 &\\
1003 &4429   &-      &7568  &70093  &122726.31  &110629.2  &S0/Sa pec&       11.15&  8.12 & 3.52 & 1130  &17 & A & 1.53 &  5.48 &*\\
1043 &4438   &-      &7574  &70097  &122745.52  &130031.4  &Sb (tides)&      10.91&  8.12 & 3.68 & 70    &17 & A & 0.97 &  10.21&*\\
1047 &4440   &-      &7581  &70099  &122753.52  &121735.5  &SBa(sr) &        12.48&  2.01 & 1.71 & 724   &17 & A & 0.72 &  7.42 &\\
1106 &-      &-      &-     &-      &122829.23  &103112.8  &Im:     &        17.50&  0.59 & 0.41 & -     &17 & A & 1.96 &  2.26 &\\
1110 &4450   &-      &7594  &99062  &122829.27  &170506.8  &Sab pec &        10.93&  6.15 & 4.04 & 1954  &17 & A & 4.73 &  4.33 &*\\
1121 &-      &-      &-     &-      &122841.73  &110754.9  &Im?     &        16.48&  0.71 & 0.56 & -     &17 & A & 1.36 &  -    &\\
1158 &4461   &-      &7613  &70115  &122903.01  &131101.1  &Sa      &        12.09&  3.52 & 1.29 & 1919  &17 & A & 0.90 &  7.45 &\\
1189 &-      &3414   &7621  &42129  &122928.83  & 64612.3  &Sc(s)   &        13.70&  1.84 & 1.07 & 597   &17 & S & 5.63 &  2.42 &\\
1196 &4468   &-      &7628  &70122  &122931.25  &140258.3  &S0/Sa   &        13.80&  1.76 & 1.06 & 895   &17 & A & 1.69 &  4.31 &\\
1200 &-      &3416   &-     &70124  &122934.53  &104737.3  &Im      &        15.10&  1.26 & 0.84 & -123  &17 & A & 1.63 &  2.76 &\\
1217 &-      &3418   &7630  &-      &122942.54  &112404.4  &SBm     &        14.59&  1.87 & 1.29 & -     &17 & A & 1.03 &  2.80 &\\
1253 &4477   &-      &7638  &70129  &123002.37  &133810.6  &SB0/SBa &        11.31&  3.60 & 3.60 & 1353  &17 & A & 1.26 &  8.73 &*\\
1257 &-      &-      &-     &-      &123004.68  &172401.6  &Im pec  &        16.50&  1.36 & 0.32 & 2488  &17 & A & 5.01 &  2.55 &\\
1287 &-      &-      &-     &-      &123023.79  &135855.8  &Im      &        16.00&  0.85 & 0.85 & -     &17 & A & 1.59 &  -    &\\
1313 &-      &-      &-     &-      &123048.47  &120242.0  &BCD     &        17.15&  0.45 & 0.20 & 1254  &17 & A & 0.35 &  2.07 &\\
1326 &4491   &-      &7657  &70140  &123057.15  &112859.1  &SBa(s)  &        13.41&  1.89 & 0.94 & 497   &17 & A & 0.91 &  2.87 &\\
1356 &-      &3446   &-     &70142  &123122.92  &112934.3  &Sm/BCD  &        15.55&  1.10 & 0.43 & 1251  &17 & A & 0.91 &  2.71 &\\
1368 &4497   &-      &7665  &70145  &123132.79  &113736.4  &SB0/SBa &        13.12&  2.01 & 0.85 & 1123  &17 & A & 0.78 &  2.81 &\\
1377 &-      &-      &-     &-      &123139.21  &105008.5  &Im:     &        16.87&  0.61 & 0.43 & -     &17 & A & 1.57 &  2.79 &\\
1379 &4498   &-      &7669  &99075  &123139.62  &165107.5  &SBc(s)  &        12.62&  2.85 & 1.53 & 1505  &17 & A & 4.46 &  2.27 &\\
1403 &-      &-      &-     &-      &123159.63  &130459.7  &Im?     &        17.15&  0.71 & 0.43 & -     &17 & A & 0.75 &  -    &\\
1410 &4502   &-      &7677  &99078  &123203.22  &164114.7  &Sm      &        14.57&  1.48 & 0.78 & 1629  &17 & A & 4.31 &  2.80 &\\
1411 &-      &3466   &-     &70150  &123204.83  &114902.7  &pec,N   &        15.72&  0.70 & 0.43 & 911   &17 & A & 0.65 &  2.74 &\\
1412 &4503   &-      &7680  &70149  &123206.13  &111034.8  &Sa      &        12.12&  4.33 & 1.71 & 1342  &17 & A & 1.25 &  5.95 &\\
1419 &4506   &-      &7682  &70152  &123210.46  &132509.8  &Spec(dust)&      13.64&  2.16 & 1.29 & 737   &17 & A & 1.08 &  3.44 &\\
1426 &-      &-      &-     &-      &123222.80  &115338.9  &Im?     &        15.64&  0.80 & 0.80 & 1110  &17 & A & 0.63 &  3.08 &\\
1448 &-      &3475   &7692  &70156  &123240.83  &124613.1  &Im      &        13.87&  2.31 & 1.83 & 2583  &17 & A & 0.59 &  2.88 &\\
1450 &-      &3476   &7695  &70157  &123241.91  &140256.1  &Sc(s)   &        13.29&  2.60 & 2.01 & -173  &17 & A & 1.72 &  2.93 &\\
1486 &-      &3483   &-     &70160  &123309.94  &112049.4  &Spec,N  &        15.30&  1.10 & 0.78 & 129   &17 & A & 1.19 &  7.18 &\\
1552 &4531   &-      &7729  &70175  &123415.77  &130429.1  &Sa pec  &        12.58&  4.24 & 2.42 & 195   &17 & A & 1.08 &  3.00 &\\
1554 &4532   &-      &7726  &42158  &123419.31  & 62807.1  &Sm      &        12.30&  2.60 & 1.00 & 2021  &17 & S & 5.99 &  2.92 &\\
1569 &-      &3520   &-     &70178  &123431.68  &133013.2  &Scd:    &        15.00&  1.07 & 0.71 & 799   &17 & A & 1.43 &  3.16 &\\
1575 &-      &3521   &7736  &42162  &123439.28  & 70938.3  &SBm pec &        13.98&  2.00 & 1.41 & 597   &17 & S & 5.32 &  2.44 &\\
1581 &-      &-      &7739  &42163  &123444.93  & 61807.4  &Sm      &        14.55&  1.46 & 1.16 & 2065  &17 & S & 6.17 &  2.77 &\\
1596 &-      &-      &-     &-      &123500.91  & 91116.5  &Im:     &        17.24&  0.35 & 0.16 & 1286  &17 & S & 3.36 &  -    &\\
\hline
 
\newpage
 
\caption{continue}\\
 
\hline
\noalign{\smallskip}
{\rm VCC}&{\rm NGC}& {\rm IC} &{\rm UGC}& {\rm CGCG}&{\rm R.A.(2000)}&{\rm dec}&{\rm type}&$m_{pg}$&
$a$&$b$&$vel$&$Dist$&{\rm memb.}&$\theta$& $C_{31}$&com\\
         &         &            &       &           & h~ m~ s        & $^{o}$~ '~ "&          &mag &
 ' & ' &km s$^{-1}$&Mpc&          &deg  &   &\\
(1) & (2) & (3) & (4) & (5) & (6) & (7) & (8) & (9) & (10) & (11) & (12) & (13)
& (14) & (15) & (16) & (17)\\
\hline
1644 &-      &-      &-     &-      &123551.82  &135133.1  &Sm      &        17.50&  0.98 & 0.17 & 756   &17 & A & 1.91&  2.93 &\\
1673 &4567   &-      &7777  &70189  &123632.66  &111528.6  &Sc(s)   &        12.08&  2.92 & 1.87 & 2277  &17 & A & 1.80&  2.73 &*\\
1675 &-      &-      &-     &42174  &123634.65  & 80317.6  &Pec     &        14.47&  1.26 & 0.74 & 1795  &17 & S & 4.56&  2.87 &\\
1676 &4568   &-      &7776  &70188  &123634.16  &111419.6  &Sc(s)   &        11.70&  5.10 & 1.75 & 2255  &17 & A & 1.82&  4.27 &*\\
1678 &-      &3576   &7781  &42176  &123637.61  & 63716.6  &SBd     &        13.70&  2.16 & 1.87 & 1073  &17 & S & 5.94&  2.91 &*\\
1686 &-      &3583   &7784  &70191  &123643.57  &131531.7  &Sm      &        13.95&  2.79 & 1.71 & 1122  &17 & A & 1.68&  2.98 &\\
1690 &4569   &-      &7786  &70192  &123649.78  &130945.7  &Sab(s)  &        10.25& 10.73 & 5.35 & -216  &17 & A & 1.65&  4.37 &*\\
1699 &-      &3589   &7790  &42179  &123702.24  & 65530.9  &SBm     &        14.11&  1.55 & 0.83 & 1635  &17 & S & 5.68&  2.64 &\\
1725 &-      &-      &-     &70196  &123741.51  & 83331.3  &Sm/BCD  &        14.51&  1.55 & 0.97 & 1068  &17 & S & 4.19&  2.92 &\\
1726 &-      &-      &7795  &42184  &123745.08  & 70622.4  &Sdm     &        14.54&  1.29 & 1.00 & 61    &17 & S & 5.55&  2.73 &\\
1727 &4579   &-      &7796  &70197  &123743.48  &114904.4  &Sab(s)  &        10.56&  6.29 & 4.87 & 1520  &17 & A & 1.78&  4.51 &*\\
1730 &4580   &-      &7794  &42183  &123748.60  & 52206.4  &Sc/Sa   &        12.61&  2.16 & 1.60 & 1032  &17 & S & 7.23&  2.68 &\\
1750 &-      &-      &-     &-      &123815.48  & 65938.7  &BCD?    &        16.50&  0.31 & 0.16 & -117  &17 & S & 5.70&  2.54 &\\
1757 &4584   &-      &7803  &70199  &123817.79  &130635.8  &Sa(s)pec&        13.60&  1.87 & 1.00 & 1783  &17 & A & 1.96&  3.53 &\\
1758 &-      &-      &7802  &42186  &123820.81  & 75328.8  &Sc (on edge)&    14.99&  1.71 & 0.27 & 1788  &17 & S & 4.87&  3.47 &\\
1784 &-      &-      &-     &-      &123913.81  &153749.4  &Im      &        15.84&  0.79 & 0.63 & 57    &17 & E & 3.83&  2.80 &\\
1789 &-      &-      &-     &42192  &123921.34  & 45619.5  &Im      &        15.07&  1.10 & 0.62 & 1619  &17 & S & 7.74&  2.46 &\\
1791 &-      &3617   &7822  &42194  &123924.55  & 75752.5  &SBm/BCD &        14.67&  1.29 & 0.64 & 2079  &17 & S & 4.90&  2.86 &\\
1804 &-      &-      &-     &-      &123940.25  & 92355.7  &Im/BCD  &        15.63&  0.75 & 0.30 & 1898  &17 & E & 3.70&  4.33 &\\
1811 &4595   &-      &7826  &99106  &123951.63  &151753.9  &Sc(s)   &        12.92&  2.16 & 1.42 & 632   &17 & E & 3.64&  2.71 &\\
1813 &4596   &-      &7828  &70206  &123955.88  &101034.9  &SBa     &        11.51&  4.76 & 4.04 & 1834  &17 & E & 3.14&  5.44 &\\
1822 &-      &-      &-     &-      &124010.14  & 65050.1  &Im      &        15.60&  0.63 & 0.25 & 1012  &17 & S & 6.00&  2.79 &\\
1869 &4608   &-      &7842  &70214  &124113.52  &100922.9  &SB0/a   &        12.05&  4.30 & 3.42 & 1864  &17 & E & 3.39&  9.68 &\\
1885 &-      &-      &-     &-      &124137.57  &154933.2  &Im      &        16.41&  1.16 & 0.57 & -     &17 & E & 4.32&  2.80 &\\
1918 &-      &-      &-     &-      &124218.10  & 54421.7  &Im      &        15.80&  1.03 & 0.36 & 980   &17 & S & 7.23&  2.81 &\\
1929 &4633   &-      &7874  &99111  &124237.12  &142122.0  &Scd(s)  &        13.77&  2.48 & 1.07 & 291   &17 & E & 3.48&  3.14 &\\
1932 &4634   &-      &7875  &99112  &124240.83  &141746.0  &Sc (on edge)&    13.19&  2.92 & 0.87 & 116   &17 & E & 3.45&  3.08 &\\
1952 &-      &-      &-     &-      &124306.86  & 73858.4  &Im      &        16.00&  0.71 & 0.35 & 1308  &17 & E & 5.62&  2.93 &\\
1970 &-      &-      &-     &71013  &124329.11  &100534.7  &Im,N?   &        15.80&  0.71 & 0.50 & 1325  &17 & E & 3.86&  2.94 &\\
1972 &4647   &-      &7896  &71015  &124332.28  &113454.7  &Sc(rs)  &        12.03&  2.60 & 2.16 & 1422  &17 & E & 3.21&  3.06 &*\\
1987 &4654   &-      &7902  &71019  &124356.71  &130734.0  &SBc(rs) &        11.14&  4.99 & 2.60 & 1039  &17 & E & 3.28&  2.93 &\\
1992 &-      &-      &7906  &-      &124410.02  &120659.2  &Im      &        15.50&  0.81 & 0.51 & 1003  &17 & E & 3.27&  2.80 &\\
1999 &4659   &-      &7915  &71024  &124429.38  &132953.5  &Sa      &        13.08&  1.99 & 1.25 & 267   &17 & E & 3.51&  6.03 &\\
2006 &-      &3718   &7920  &71026  &124445.93  &122111.7  &Amorphous&       13.68&  2.60 & 0.71 & 844   &17 & E & 3.40&  3.19 &\\
2007 &-      &3716   &-     &43016  &124447.50  & 80629.7  &Im/BCD: &        15.20&  0.78 & 0.41 & 1857  &17 & E & 5.49&  2.70 &\\
2023 &-      &3742   &7932  &71032  &124531.55  &131951.3  &SBc(s)  &        13.86&  2.01 & 1.00 & 958   &17 & E & 3.70&  3.09 &\\
2033 &-      &-      &-     &71033  &124604.76  & 82830.8  &BCD     &        14.65&  0.73 & 0.73 & 1486  &17 & E & 5.42&  3.70 &\\
2034 &-      &-      &-     &-      &124607.96  &100948.8  &Im      &        15.82&  0.78 & 0.52 & 1500  &17 & E & 4.36&  2.46 &\\
2037 &-      &-      &-     &-      &124615.15  &101224.9  &Im/BCD  &        15.92&  0.88 & 0.38 & 1142  &17 & E & 4.37&  2.92 &\\
2058 &4689   &-      &7965  &71043  &124745.39  &134548.3  &Sc(s)   &        11.55&  5.86 & 4.44 & 1620  &17 & E & 4.34&  2.80 &\\
2066 &4694   &-      &7969  &71044  &124815.05  &105906.7  &Amorphous&       12.19&  3.20 & 1.16 & 1181  &17 & E & 4.49&  4.21 &*\\
2070 &4698   &-      &7970  &71045  &124822.96  & 82913.8  &Sa      &        11.53&  5.67 & 2.84 & 1008  &17 & E & 5.82&  5.78 &*\\
2087 &4733   &-      &7997  &71054  &125106.81  &105444.3  &SB0/a   &        12.63&  1.96 & 1.96 & 908   &17 & E & 5.18&  2.73 &\\
2094 &-      &-      &-     &-      &125235.75  &102648.7  &Im:     &        17.80&  0.37 & 0.37 & -     &17 & E & 5.68&  -    &\\
\hline
\end{longtable}
\normalsize
Notes on morphological type, from NED: 

VCC 66: HII;
VCC 92: M98: HII and Seyfert;
VCC 460: LINER;
VCC 836: Seyfert2;
VCC 857: LINER;
VCC 1003: HII LINER;
VCC 1043: LINER, tidally interacting with VCC 1030;
VCC 1110: LINER;
VCC 1253: Seyfert 2;
VCC 1673: interacting with VCC 1676?;
VCC 1676: interacting with VCC 1673?;
VCC 1678: HII;
VCC 1690: M90: LINER, Seyfert;
VCC 1727: M58; LINER, Seyfert 1.9;
VCC 1972: interacting with VCC 1978 (M60)?;
VCC 2066: HII;
VCC 2070: Seyfert 2;
\twocolumn

\addtocounter{table}{0}
\onecolumn
\begin{landscape}
\tiny
\begin{longtable}{rrrrrrrrrrrrrrrrrrrrr}
\caption{The photometric data}\\
\hline
\noalign{\smallskip}
\hline
{\rm VCC}&{\rm UV}& {\rm U} &{\rm B} & {\rm V}&{\rm J}&{\rm H}&{\rm K}& {\rm C6.75} & {\rm I12} & {\rm C15} & {\rm I25} & {\rm I60} & {\rm P60} & {\rm I100} & {\rm P100} & {\rm P170} & {\rm r2.8} & {\rm r6.3} & {\rm r12.6} & {\rm r21}\\  
$\lambda$ & 2000\AA &3650\AA&4400\AA &5500\AA &1.25$\mu$m&1.65$\mu$m&2.1$\mu$m&6.75$\mu$m & 12$\mu$m&15$\mu$m&25$\mu$m&60$\mu$m&60$\mu$m&100$\mu$m&
100$\mu$m&170$\mu$m&2.8cm& 6.3cm &12.6cm&21cm\\
{\rm units}&mag & mag& mag& mag& mag& mag& mag& mJy &mJy & mJy &mJy &mJy &mJy &mJy &mJy &mJy &mJy &mJy &mJy &mJy \\
(1) & (2) & (3) & (4) & (5) & (6) & (7) & (8) & (9) & (10) & (11) & (12) & (13) & (14) & (15) & (16) & (17) & (18) & (19) & (20) & (21) \\
\hline                            
1   & -     &  -    &15.90 &15.13&13.58 & 12.81 & 12.52 &    1.15   &    -     &  2.86    &$<$340   & $<$390   &   100    & $<$840   &   140     &  330     &    -     &     -   &     -    & $<$1800\\
4   & -     &  -    & -    & -   & -    &  -    & 15.32 & $<$1.31   &    -     &$<$1.94   &$<$340   & $<$390   &     -    & $<$840   &     -     &    -     &    -     &     -   &     -    & $<$1800\\
17  & -     &  -    &16.28 &15.89&14.81 & 14.30 & 14.15 &    0.98   &    -     &  1.76    &  130    &   130    & $<$40    &   460    & $<$30     &$<$70     &    -     &     -   &     -    & $<$1800\\
24  & -     &  -    &15.79 &15.19& -    &  -    & 12.82 &    0.37   &$<$80     &  0.82    &$<$110   & $<$120   & $<$40    & $<$300   & $<$40     &  120     &    -     &     -   &     -    & $<$1800\\
26  & -     &  -    & -    & -   & -    &  -    & 15.96 & $<$0.71   &    -     &$<$1.05   &$<$340   & $<$390   &     -    & $<$840   &     -     &    -     &    -     &     -   &     -    & $<$1800\\
66  &12.00  & 11.91 &11.98 &11.39&9.89  & 9.14  &  8.92 &  229.59   &  110     &  192.69  &$<$140   &   2110   &   2470   &   8080   &   5090    &  11270   &  6000    &  29000  &   13000  &   26200\\
81  & -     &  -    &15.85 &15.30& -    &  -    & 13.21 & $<$5.86   &    -     &$<$6.94   &$<$340   & $<$390   & $<$40    & $<$840   & $<$30     &  770     &    -     &     -   &     -    & $<$1800\\
87  & -     & 15.10 &15.27 &14.80& -    &  -    & 13.08 &    2.18   &    -     &  0.74    &$<$340   & $<$390   &   100    & $<$840   &   150     &  380     &    -     &     -   &     -    & $<$1800\\
92  &11.17  & 11.17 &10.73 &9.83 &7.72  & 6.85  &  6.59 &  900.15   &  1100    &  692.42  &  1460   &   8110   &   4700   &   23070  &   11460   &  40290   &  18000   &  33000  &   37000  &   73300\\
130 & -     &  -    & -    & -   & -    &  -    & 14.71 & $<$0.96   &$<$120    &$<$1.42   &$<$170   & $<$110   & $<$50    & $<$280   &   70      &  80      &    -     &     -   &     -    & $<$1800\\
152 & -     & 13.78 &13.56 &12.72& -    & 9.75  &  9.49 &  173.99   &  230     &  145.13  &  240    &   3080   &   1870   &   7470   &   5480    &  8380    &    -     &     -   &   11000  &   19800\\
159 & -     &  -    &16.02 &15.66& -    &  -    & 14.09 & $<$3.30   &    -     &$<$4.88   &$<$340   & $<$390   & $<$40    & $<$840   & $<$40     &  160     &    -     &     -   &     -    & $<$1800\\
169 & -     &  -    & -    & -   & -    &  -    &  -    & $<$2.23   &    -     &$<$3.3    &$<$340   & $<$390   & $<$30    & $<$840   & $<$50     &$<$50     &    -     &     -   &     -    & $<$1800\\
171 & -     &  -    & -    & -   & -    &  -    &  -    & $<$1.25   &    -     &$<$1.85   &$<$340   & $<$390   &     -    & $<$840   &     -     &    -     &    -     &     -   &     -    & $<$1800\\
207 & -     &  -    & -    & -   & -    &  -    & 14.88 & $<$0.25   &$<$120    &$<$0.42   &$<$160   & $<$140   &     -    & $<$340   &     -     &    -     &    -     &     -   &     -    & $<$1800\\
318 &13.47  & 14.26 &14.45 &14.09& -    &  -    & 11.95 &    3.08   &    -     &  5.16    &$<$740   &   240    &   130    &   620    &   350     &  960     &    -     &     -   &     -    & $<$1800\\
425 & -     &  -    & -    & -   & -    &  -    &  -    & $<$1.00   &    -     &$<$1.33   &$<$340   & $<$390   &     -    & $<$840   &     -     &    -     &    -     &     -   &     -    & $<$1800\\
459 &13.49  &  -    & -    & -   & -    &  -    & 12.60 &    2.87   &$<$100    &  2.96    &$<$130   &   240    &   130    &   540    &   380     &  500     &    -     &     -   &     -    & $<$1800\\
460 & -     & 12.01 &11.45 &10.50&8.39  & 7.49  &  7.33 &  195.11   &  180     &  186.36  &  510    &   4580   &   3290   &   10390  &   9110    &  11250   &  6000    &  11000  &   20000  &   19100\\
655 & -     & 13.62 &13.59 &12.93& -    &  -    & 10.29 &   42.05   &$<$140    &  15.17   &  140    &   470    &   420    &   1890   &   1110    &  5360    &    -     &     -   &   1000   & $<$1800\\
664 &13.15  & 13.25 &13.60 &13.16& -    &  -    & 11.29 &    7.09   &$<$70     &  15.17   &  140    &   600    &   750    &   1030   &   770     &  970     &    -     &     -   &     -    & $<$1800\\
666 & -     &  -    & -    & -   & -    &  -    & 14.46 & $<$4.34   &    -     &$<$6.17   &$<$340   & $<$390   & $<$30    & $<$840   & $<$30     &$<$40     &    -     &     -   &     -    & $<$1800\\
692 &13.05  & 13.07 &12.99 &12.52& -    & 10.34 & 10.19 &   33.64   &$<$90     &  25.22   &$<$180   &   710    &   540    &   2010   &   1430    &  3910    &    -     &  2000   & $<$4000  & $<$1800\\
793 & -     & 17.27 &17.26 &16.89&15.55 & 14.97 & 15.16 &$<$99      &    -     &$<$99     &$<$340   & $<$390   &     -    & $<$840   &     -     &    -     &    -     &     -   &     -    & $<$1800\\
802 & -     & 17.15 &17.61 & -   & -    &  -    & 14.81 &$<$99      &$<$100    &$<$99     &$<$170   & $<$130   &     -    & $<$620   &     -     &    -     &    -     &     -   &     -    & $<$1800\\
809 &15.14  & 15.14 &15.11 &14.43& -    &  -    & 12.03 &    5.97   &    -     &  3.57    &$<$500   & $<$470   &     -    & $<$1100  &     -     &    -     &    -     &     -   &     -    & $<$1800\\
836 &12.56  & 12.00 &11.86 &11.11&9.32  & 8.37  &  7.92 &  528.11   &  1060    &  1064.56 &  3420   &   10050  &   7030   &   17400  &   14220   &  11630   &  36000   &  84000  &   129000 &   119400\\
848 & -     & 15.01 &15.18 &14.76& -    &  -    & 12.91 &    1.20   &    -     &$<$10.25  &$<$340   & $<$390   & $<$60    & $<$840   &   50      &  1410    &    -     &     -   &     -    & $<$1800\\
857 & -     & 12.28 &11.92 &11.09& -    &  -    &  8.02 &  114.95   &  150     &  98.3    &  150    &   960    &   650    &   4020   &   2830    &  7760    &    -     &  2000   & $<$4000  &     700\\
873 &13.72  & 13.02 &12.64 &11.80& -    & 8.67  &  8.39 &  500.01   &  790     &  525.45  &  640    &   5430   &   3820   &   17480  &   8610    &  17720   &  12000   &  21000  &   50000  &   59500\\
890 & -     &  -    & -    & -   & -    &  -    & 14.61 & $<$0.27   &$<$90     &$<$0.32   &$<$190   & $<$150   & $<$40    & $<$360   & $<$30     &  120     &    -     &     -   &     -    & $<$1800\\
912 &13.40  & 13.01 &12.97 &12.34& -    & 9.73  &  9.53 &   60.94   &  140     &  54.72   &  180    &   1000   &   830    &   3100   &   2340    &  2820    &    -     &$<$1000  &   6000   & $<$1800\\
945 &14.32  & 15.24 &15.47 &15.15& -    &  -    & 13.45 &$<$99      &    -     &$<$99     &$<$340   & $<$390   &     -    & $<$840   &     -     &    -     &    -     &     -   &     -    & $<$1800\\
950 &14.94  & 15.60 &15.76 &15.32& -    &  -    & 13.85 &$<$99      &    -     &$<$99     &$<$340   & $<$390   &     -    & $<$840   &     -     &    -     &    -     &     -   &     -    & $<$1800\\
971 &13.32  & 14.06 &14.17 &13.61& -    &  -    & 11.32 &    8.21   &$<$90     &  6.57    &$<$140   &   470    &   290    &   1100   &   830     &  1380    &    -     &     -   &     -    &    3600\\
984 &16.29  & 13.27 &12.86 &11.95&10.08 & 9.36  &  8.95 &   19.56   &$<$120    &  9.69    &$<$170   & $<$180   & $<$40    & $<$340   & $<$40     &$<$170    &    -     &$<$1000  & $<$4000  & $<$1800\\
995 & -     & 15.21 &15.39 &14.78& -    &  -    & 12.61 &    1.40   &    -     &  3.02    &$<$340   & $<$390   &     -    & $<$840   &     -     &    -     &    -     &     -   &     -    & $<$1800\\
1001& -     &  -    & -    & -   & -    &  -    & 14.79 &$<$99      &    -     &$<$99     &$<$340   & $<$390   &     -    & $<$840   & $<$20     &  230     &    -     &     -   &     -    & $<$1800\\
1002&12.66  &  -    &12.74 &12.09&10.54 & 9.63  &  9.33 &  119.35   &  130     &  72.17   &  230    &   1150   &   940    &   3780   &   3360    &  5300    &    -     &  2000   &   4000   &    7900\\
1003&14.98  & 11.51 &10.95 &9.96 &7.69  & 6.90  &  6.63 &  267.86   &  310     &  192.87  &  220    &   1510   &   870    &   4130   &   3940    &  3120    &    -     &  1000   & $<$4000  & $<$1800\\
1043&12.58  & 11.64 &11.08 &10.07&8.13  & 7.35  &  6.89 &  259.53   &  210     &  247.26  &  170    &   3760   &   2360   &   11270  &   7670    &  16720   &  44000   &  97000  &   109000 &   148900\\
1047&16.27  &  -    &12.85 &11.88&10.02 & 9.05  &  8.92 &$<$99      &$<$150    &  5.09    &$<$160   &   210    & $<$40    & $<$280   & $<$40     &$<$210    &$<$1000   &     -   &   27000  & $<$4000\\
1106& -     &  -    & -    & -   & -    &  -    & 15.00 &$<$99      &    -     &$<$99     &$<$340   & $<$390   &     -    & $<$840   &     -     &    -     &    -     &     -   &     -    & $<$1800\\
1110& -     & 11.48 &10.96 &10.07&7.74  & 7.03  &  6.74 &  236.00   &  150     &  252.78  &  170    &   1800   &   1790   &   7910   &   4080    &  10010   &  14000   &  28000  &   11000  &   10200\\
1121& -     &  -    & -    & -   & -    &  -    & 15.22 &$<$99      &    -     &$<$99     &$<$340   & $<$390   &     -    & $<$840   & $<$20     &$<$140    &    -     &     -   &     -    & $<$1800\\
1158&15.26  & 12.72 &12.19 &11.23&8.91  & 8.22  &  7.99 &   44.15   &$<$120    &  16.45   &$<$180   & $<$120   & $<$40    & $<$310   & $<$50     &$<$140    &    -     &$<$1000  & $<$4000  & $<$1800\\
1189&13.35  & 13.90 &14.02 &13.56& -    &  -    & 11.32 &   12.89   &$<$60     &  5.8     &$<$170   &   230    &   180    &   730    &   490     &  1160    &    -     &     -   &     -    & $<$1800\\
1196& -     & 14.32 &13.92 &13.07&11.03 & 10.38 & 10.17 &    6.03   &    -     &  1.82    &$<$100   & $<$180   & $<$40    & $<$420   & $<$40     &  1240    &    -     &     -   &     -    & $<$1800\\
1200& -     & 15.33 &15.56 &15.06& -    & 13.35 & 13.01 &$<$99      &    -     &$<$99     &$<$340   & $<$390   &     -    & $<$840   &     -     &    -     &    -     &     -   &     -    &   13000\\
1217&15.49  &  -    &14.52 &14.17&13.26 & 12.72 & 12.55 &    3.65   &    -     &$<$21.75  &$<$340   & $<$390   & $<$40    & $<$840   & $<$30     &$<$70     &    -     &     -   &     -    & $<$1800\\
1253& -     & 12.11 &11.52 &10.57&8.26  & 7.51  &  7.25 &   96.85   &$<$100    &  28.74   &$<$170   &   540    &   140    &   1180   &   810     &  1130    &$<$500    &  5000   &   8000   & $<$1800\\
1257& -     &  -    & -    & -   & -    &  -    & 14.08 & $<$2.65   &    -     &  1.75    &$<$340   & $<$390   &     -    & $<$840   &     -     &    -     &    -     &     -   &     -    & $<$1800\\
1287& -     &  -    & -    & -   & -    &  -    &  -    &$<$99      &    -     &$<$99     &$<$340   & $<$390   &     -    & $<$840   &     -     &    -     &    -     &     -   &     -    & $<$1800\\
1313&15.35  & 16.67 &17.30 &16.96& -    & 15.58 &  -    &$<$99      &$<$150    &$<$99     &$<$110   & $<$140   &     -    & $<$350   &     -     &    -     &    -     &     -   &     -    & $<$19800\\
1326&15.01  & 13.74 &13.52 &12.69& -    & 10.07 &  9.95 &   27.00   &$<$90     &  70.1    &  420    &   2770   &   2360   &   3490   &   2890    &  2980    &  5000    &     -   & $<$4000  & $<$1800\\
1356&14.65  & 15.35 &15.56 &15.11&-     & -     & 12.84 &  -        &-         & -        &$<$340   &$<$390    &    -     & $<$840   &    -      &   -      &    -     &  -      &  -       & $<$1800\\
\hline                                                                                                                                                     
\newpage                                                                        
                                              
\caption{continue}\\                                                            
                                              
\hline                                                                                                                                             
{\rm VCC}&{\rm UV}& {\rm U} &{\rm B} & {\rm V}&{\rm J}&{\rm H}&{\rm K}& {\rm C6.75} & {\rm I12} & {\rm C15} & {\rm I25} & {\rm I60} & {\rm P60} & {\rm I100} & {\rm P100} & {\rm P170} & {\rm r2.8} & {\rm r6.3} & {\rm r12.6} & {\rm r21}\\  
$\lambda$ & 2000\AA &3650\AA&4400\AA &5500\AA &1.25$\mu$m&1.65$\mu$m&2.1$\mu$m&6.75$\mu$m & 12$\mu$m&15$\mu$m&25$\mu$m&60$\mu$m&60$\mu$m&100$\mu$m&
100$\mu$m&170$\mu$m&2.8cm& 6.3cm &12.6cm&21cm\\
{\rm units}&mag & mag& mag& mag& mag& mag& mag& mJy &mJy & mJy &mJy &mJy &mJy &mJy &mJy &mJy &mJy &mJy &mJy &mJy \\
(1) & (2) & (3) & (4) & (5) & (6) & (7) & (8) & (9) & (10) & (11) & (12) & (13) & (14) & (15) & (16) & (17) & (18) & (19) & (20) & (21) \\
\hline                                                                                                                                             
1368&17.22  & 13.76 &13.44 &12.56& -    & 9.88  &  9.70 &    8.62   &$<$120    &  3.59    &$<$130   & $<$240   & $<$150   & $<$890   & $<$30     &  200     &    -     &     -   &   9000   & $<$4000\\
1377& -     &  -    &17.23 &16.49& -    &  -    & 14.06 &$<$99      &    -     &$<$99     &$<$340   & $<$390   &     -    & $<$840   &     -     &    -     &    -     &     -   &     -    & $<$1800\\
1379&12.31  & 12.71 &12.76 &12.17& -    & 9.97  &  9.74 &   95.88   &  150     &  62.02   &  90     &   1200   &   1140   &   3700   &   3270    &  4980    &  1000    &     -   &     -    &    4600\\
1403& -     &  -    & -    & -   & -    &  -    &  -    &$<$99      &    -     &$<$99     &$<$340   & $<$390   &     -    & $<$840   &     -     &    -     &    -     &     -   &     -    & $<$4000\\
1410& -     & 14.46 &14.54 &14.05& -    &  -    & 11.93 &    9.68   &    -     &  4.51    &$<$190   &   230    &   220    &   620    &   360     &  690     &    -     &     -   &     -    & $<$1800\\
1411&16.10  &  -    &15.99 &15.49& -    &  -    & 13.61 &$<$99      &    -     &$<$99     &$<$340   & $<$390   &     -    & $<$840   &     -     &    -     &    -     &     -   &     -    & $<$4000\\
1412& -     & 12.79 &12.18 &11.18&8.91  & 8.18  &  7.88 &   47.83   &$<$150    &  13.47   &$<$140   & $<$150   & $<$60    & $<$390   & $<$70     &  710     &    -     &$<$1000  &   8000   & $<$1800\\
1419& -     & 14.10 &13.82 &13.00& -    & 10.38 & 10.32 &   16.31   &    -     &  12.5    &$<$240   &   150    &   60     &   640    &   280     &  440     &    -     &     -   &     -    & $<$1800\\
1426& -     &  -    &16.43 &15.73& -    &  -    & 13.48 &$<$99      &    -     &$<$99     &$<$340   & $<$390   &     -    & $<$840   &     -     &    -     &    -     &     -   &     -    & $<$4000\\
1448& -     & 14.70 &14.55 &13.87&12.40 & 11.68 & 11.50 &$<$99      &$<$100    &$<$99     &$<$180   & $<$200   &     -    & $<$340   &     -     &    -     &    -     &     -   &     -    &    6300\\
1450&12.53  &  -    &13.43 &12.98& -    & 10.82 & 10.55 &   72.46   &  190     &  59.61   &  350    &   1850   &   740    &   3100   &   2990    &  3960    &  10000   &  24000  &   15000  &   10100\\
1486& -     & 15.02 &14.99 &14.31& -    &  -    & 11.54 &$<$99      &    -     &$<$99     &$<$340   & $<$390   &     -    & $<$840   &     -     &    -     &    -     &     -   &     -    & $<$1800\\
1552& -     & 12.85 &12.52 &11.65&9.66  & 8.91  &  8.75 &   40.15   &$<$100    &  27.73   &$<$140   &   360    &   290    &   1720   &   1080    &  1940    &$<$1000   &     -   &   6000   & $<$1800\\
1554&11.34  & 12.01 &12.35 &11.96&10.39 & 9.67  &  9.39 &  183.96   &  290     &  213.47  &  830    &   8930   &   5560   &   15530  &   9070    &  10650   &  19000   &  58000  &   84000  &   123800\\
1569& -     & 15.73 &15.86 &15.39& -    &  -    & 13.50 &    0.80   &    -     &$<$5.48   &$<$340   & $<$390   & $<$50    & $<$840   & $<$50     &$<$60     &    -     &     -   &     -    & $<$1800\\
1575&13.80  & 13.76 &13.79 &13.13& -    &  -    & 10.54 &   47.48   &$<$90     &  46.02   &$<$150   &   1030   &   1110   &   2300   &   2080    &  2740    &    -     &     -   &     -    &    4200\\
1581& -     & 15.08 &15.08 &14.51& -    &  -    & 12.59 & $<$10.32  &    -     &$<$12.22  &$<$340   & $<$390   & $<$30    & $<$840   & $<$30     &  330     &    -     &     -   &     -    & $<$1800\\
1596& -     &  -    & -    & -   & -    &  -    &  -    & $<$0.30   &    -     &$<$0.5    &$<$340   & $<$390   &     -    & $<$840   &     -     &    -     &    -     &     -   &     -    & $<$1800\\
1644& -     &  -    & -    & -   & -    &  -    & 15.19 & $<$99     &    -     &$<$99     &$<$340   & $<$390   &     -    & $<$840   &     -     &    -     &    -     &     -   &     -    & $<$1800\\
1673&12.29  &  -    &11.17 &10.49&9.32  & 8.60  &  7.93 &  324.33   &    -     &  319.63  &$<$240   & $<$390   &     -    & $<$840   &     -     &    -     &    -     &     -   & $<$54000 &   10500\\
1675& -     &  -    & -    & -   & -    &  -    & 12.30 & $<$4.97   &    -     &$<$8.41   &$<$340   & $<$390   &   50     & $<$840   & $<$30     &  160     &    -     &     -   &     -    & $<$1800\\
1676& -     &  -    &11.17 &10.25& -    &  -    &  7.34 &  973.91   &  2000    &  1050.86 &  2580   &   20360  &     -    &   56810  &     -     &    -     &  29000   &  65000  &   75000  &   124200\\
1678&13.26  & 14.32 &14.47 &13.96& -    &  -    & 12.02 & $<$21.54  &$<$100    &$<$29.13  &$<$120   &   300    &   80     &   520    &   430     &  680     &    -     &     -   &     -    & $<$1800\\
1686& -     & 13.00 &13.46 &13.03& -    &  -    & 11.22 &   25.19   &$<$120    &  19.42   &$<$180   &   540    &   450    &   1720   &   1130    &  1490    &    -     &     -   &     -    & $<$1800\\
1690&11.67  & 10.34 &10.10 &9.32 &7.54  & 6.81  &  6.66 &  830.16   &  1310    &  972.71  &  2070   &   10080  &   6200   &   26600  &   16000   &  29160   &  30000   &  40000  &   63000  &   72500\\
1699&13.43  & 14.38 &14.46 &14.04& -    &  -    & 12.19 &    4.15   &    -     &  14.04   &$<$340   & $<$390   &   270    & $<$840   &   390     &  490     &    -     &     -   &     -    & $<$1800\\
1725&13.47  & 14.26 &14.61 &14.23&13.13 & 12.51 & 12.26 &    2.32   &  60      &  2.7     &  100    & $<$180   &   50     &   350    &   300     &  480     &    -     &     -   &     -    & $<$1800\\
1726&13.83  & 14.93 &15.36 &15.13& -    &  -    & 13.38 & $<$5.90   &    -     &$<$9.3    &$<$340   & $<$390   & $<$50    & $<$840   & $<$50     &  240     &    -     &     -   &     -    & $<$1800\\
1727&12.60  & 11.05 &10.51 &9.64 &7.48  & 6.72  &  6.42 &  658.15   &  1110    &  646.18  &  760    &   5850   &   4160   &   20860  &   12340   &  29190   &  82000   &  99000  &   95000  &   97400\\
1730& -     & 13.01 &12.78 &11.94&9.86  & 8.84  &  8.79 &   99.75   &  260     &  97.4    &  540    &   1460   &   1520   &   4820   &   3640    &  5460    &$<$400    &  3000   & $<$4000  & $<$1800\\
1750& -     &  -    & -    & -   & -    &  -    & 14.43 &    0.28   &$<$130    &$<$0.54   &$<$170   & $<$140   & $<$30    & $<$210   &   40      &  80      &    -     &     -   &     -    & $<$1800\\
1757& -     & 14.08 &13.90 &13.14& -    &  -    & 10.52 & $<$9.97   &$<$100    &$<$16.86  &$<$140   &   240    &   100    & $<$640   &   500     &  790     &    -     &     -   &     -    & $<$1800\\
1758& -     & 14.98 &15.00 &14.37& -    &  -    & 11.85 &    5.92   &    -     &  0.98    &$<$340   & $<$390   &     -    & $<$840   &     -     &    -     &    -     &     -   &     -    & $<$1800\\
1784& -     &  -    & -    & -   & -    &  -    & 14.74 & $<$3.03   &    -     &$<$6.28   &$<$340   & $<$390   &     -    & $<$840   &     -     &    -     &    -     &     -   &     -    &    3500\\
1789& -     & 16.00 &15.91 &15.19&13.23 & 12.60 & 12.71 &    2.40   &    -     &$<$7.38   &$<$340   & $<$390   &     -    & $<$840   &     -     &    -     &    -     &     -   &     -    & $<$1800\\
1791&13.07  & 14.41 &14.67 &14.37& -    &  -    & 12.49 &    2.88   &    -     &  3.97    &$<$230   &   270    &     -    &   630    &     -     &    -     &    -     &     -   &     -    &    3000\\
1804& -     & 16.33 &16.30 &15.73& -    &  -    & 13.44 &    0.41   &    -     &$<$2.43   &$<$340   & $<$390   &     -    & $<$840   &     -     &    -     &    -     &     -   &     -    & $<$1800\\
1811&12.81  & 13.11 &13.11 &12.56& -    & 10.22 & 10.04 &   74.49   &  100     &  52.59   &  180    &   900    &     -    &   2670   &     -     &    -     &  1000    &  7000   &   5000   &    7100\\
1813&13.78  & 12.01 &11.48 &10.53&8.17  & 7.44  &  7.19 &  126.86   &  120     &  40.01   &$<$130   &   490    &     -    &   1280   &     -     &    -     &    -     &$<$1000  &   4000   & $<$1800\\
1822& -     & 16.63 &16.85 &16.49& -    &  -    & 14.36 &    0.72   &    -     &$<$1.14   &$<$340   & $<$390   &     -    & $<$840   &     -     &    -     &    -     &     -   &     -    & $<$1800\\
1869& -     &  -    &12.11 &11.16&8.77  & 8.09  &  7.86 &$<$99      &$<$120    &$<$99     &$<$180   & $<$150   &     -    & $<$340   &     -     &    -     &    -     &$<$1000  &   2000   &    2800\\
1885& -     &  -    & -    & -   & -    &  -    & 14.06 & $<$3.53   &    -     &$<$5.96   &$<$340   & $<$390   &     -    & $<$840   &     -     &    -     &    -     &     -   &     -    & $<$1800\\
1918& -     &  -    & -    & -   & -    &  -    & 14.47 &    0.87   &    -     &  1       &$<$340   & $<$390   &     -    & $<$840   &     -     &    -     &    -     &     -   &     -    & $<$1800\\
1929&12.87  & 13.63 &13.77 &13.19& -    &  -    & 10.61 &   29.28   &$<$100    &  35.14   &$<$130   &   500    &     -    &   1810   &     -     &    -     &    -     &     -   &     -    & $<$1800\\
1932& -     & 13.25 &13.18 &12.45&10.52 & 9.67  &  9.25 &  290.08   &  400     &  265.79  &  480    &   4130   &     -    &   12650  &     -     &    -     &  6000    &  20000  &   20000  &   34000\\
1952& -     &  -    & -    & -   & -    &  -    & 14.68 & $<$1.51   &    -     &$<$1.79   &$<$240   & $<$150   &     -    &   250    &     -     &    -     &    -     &     -   &     -    & $<$1800\\
1970& -     &  -    & -    & -   & -    &  -    & 13.75 & $<$2.16   &    -     &$<$3.2    &$<$340   & $<$390   &     -    & $<$840   &     -     &    -     &    -     &     -   &     -    & $<$1800\\
1972&11.88  & 12.36 &12.02 &11.34&10.14 & 8.74  &  8.58 &  501.09   &  980     &  493.38  &  780    &   5350   &     -    &   16040  &     -     &    -     &  6000    &  38000  &   26000  &   56300\\
1987&11.23  & 11.37 &11.31 &10.60&8.71  & 7.86  &  7.57 & 1051.86   &  1190    &  1101.42 &  1910   &   13930  &     -    &   37160  &     -     &    -     &  29000   &  51000  &   59000  &   125300\\
1992& -     & 16.08 &16.58 &16.15&15.48 & 14.88 & 14.54 &    0.58   &    -     &$<$3.72   &$<$340   & $<$390   &     -    & $<$840   &     -     &    -     &    -     &     -   &     -    & $<$1800\\
1999& -     & 13.57 &13.20 &12.34&10.43 & 9.62  &  9.35 &   10.06   &  120     &  3.98    &$<$80    & $<$140   &     -    & $<$570   &     -     &    -     &    -     &     -   & $<$4000  & $<$1800\\
2006& -     &  -    & -    & -   & -    &  -    & 11.15 &    2.93   &$<$120    &  1.49    &$<$140   & $<$150   &     -    & $<$340   &     -     &    -     &    -     &     -   &     -    & $<$1800\\
2007& -     &  -    & -    & -   & -    &  -    & 13.35 &    1.80   &    -     &$<$3.46   &$<$340   & $<$390   &     -    & $<$840   &     -     &    -     &    -     &     -   &     -    & $<$1800\\
2023& -     & 14.02 &14.05 &13.56& -    &  -    & 11.41 &    5.46   &    -     &  5.45    &$<$100   &   250    &     -    &   940    &     -     &    -     &    -     &     -   &     -    & $<$1800\\
2033& -     & 15.37 &15.60 &15.08& -    &  -    & 13.03 &    0.59   &$<$110    &  1.17    &$<$170   &   200    &     -    & $<$350   &     -     &    -     &    -     &     -   &     -    & $<$1800\\
\hline                                                                                                                                                     
\newpage                                                                        
                                              
\caption{continue}\\                                                            
                                              
\hline                                                                                                                                             
{\rm VCC}&{\rm UV}& {\rm U} &{\rm B} & {\rm V}&{\rm J}&{\rm H}&{\rm K}& {\rm C6.75} & {\rm I12} & {\rm C15} & {\rm I25} & {\rm I60} & {\rm P60} & {\rm I100} & {\rm P100} & {\rm P170} & {\rm r2.8} & {\rm r6.3} & {\rm r12.6} & {\rm r21}\\  
$\lambda$ & 2000\AA &3650\AA&4400\AA &5500\AA &1.25$\mu$m&1.65$\mu$m&2.1$\mu$m&6.75$\mu$m & 12$\mu$m&15$\mu$m&25$\mu$m&60$\mu$m&60$\mu$m&100$\mu$m&
100$\mu$m&170$\mu$m&2.8cm& 6.3cm &12.6cm&21cm\\
{\rm units}&mag & mag& mag& mag& mag& mag& mag& mJy &mJy & mJy &mJy &mJy &mJy &mJy &mJy &mJy &mJy &mJy &mJy &mJy \\
(1) & (2) & (3) & (4) & (5) & (6) & (7) & (8) & (9) & (10) & (11) & (12) & (13) & (14) & (15) & (16) & (17) & (18) & (19) & (20) & (21) \\
\hline  
2034& -     &  -    &16.24 &15.78& -    &  -    & 13.37 &    0.19   &    -     &$<$3.66   &$<$340   & $<$390   &     -    & $<$840   &     -     &    -     &    -     &     -   &     -    & $<$1800\\
2037& -     & 16.12 &16.20 &15.79& -    &  -    & 13.49 & $<$1.78   &    -     &$<$2.11   &$<$340   & $<$390   &     -    & $<$840   &     -     &    -     &    -     &     -   &     -    & $<$1800\\
2058&12.62  &  -    &11.73 &10.98& -    & 8.43  &  7.88 &  316.50   &  380     &  352.68  &  400    &   3250   &     -    &   10490  &     -     &    -     &  3000    &  8000   &   7000   &   14300\\
2066& -     & 12.53 &12.31 &11.63&9.86  & 9.16  &  8.91 &   68.00   &  130     &  68.13   &  170    &   1170   &     -    &   2680   &     -     &    -     &  4000    &  3000   &   4000   &    4100\\
2070& -     & 11.99 &11.48 &10.56&8.25  & 7.57  &  7.31 &   88.49   &  280     &  66.58   &$<$460   &   630    &     -    &   1890   &     -     &    -     &    -     &  2000   &   2000   & $<$1800\\
2087& -     & 13.34 &12.98 &12.11&9.99  & 9.24  &  9.03 &   11.46   &$<$120    &  7.06    &$<$180   & $<$130   &     -    & $<$280   &     -     &    -     &  1000    &     -   & $<$4000  & $<$1800\\
2094& -     &  -    & -    & -   & -    &  -    & 16.44 &    0.44   &    -     &$<$1.23   &$<$340   & $<$390   &     -    & $<$840   &     -     &    -     &    -     &     -   &     -    & $<$1800\\
\hline
\end{longtable}
\normalsize
Note to Table 3:

VCC 1379: the 116000 mJy flux at 12.6 cm of Dressel \& Condon is contaminated by a background quasar, visible in the NVSS 20cm map.
\twocolumn
\end{landscape}

\addtocounter{table}{0}
\onecolumn
\scriptsize
\begin{longtable}{rcccccccccc}
\caption{References for the photometric data}\\
\hline
\noalign{\smallskip}
\hline
\noalign{\smallskip}
{\rm VCC}&{\rm UV}& {\rm optical} &{\rm near-IR}& {\rm ISOCAM}&{\rm IRAS}&{\rm ISOPHOT}&{\rm radio 2.8cm}&{\rm radio 6.3cm}&{\rm radio
12.6cm}&{\rm radio 21cm}\\
(1) & (2) & (3) & (4) & (5) & (6) & (7) & (8) & (9) & (10) & (11) \\
\hline            
1       &-&     1         & 1    &  1    &   1   & 1   &   -   &   -       &-  &1     \\%
4       &-&     -         & 3    &  1    &   1   & -   &   -   &   -       &-  &1     \\%
17      &-&     2         & 1    &  1    &   9   & 1   &   -   &   -       &-  &1     \\%
24      &-&     1         & 1    &  1    &   2   & 1   &   -   &   -       &-  &1     \\%
26      &-&     -         & 3    &  1    &   1   & -   &   -   &   -       &-  &1     \\%
66      &1&     1         & 1    &  1    &   4   & 1   &   1   &   1       &-  &1     \\%
81      &-&     1,2       & 1    &  1    &   1   & 1   &   -   &   -       &-  &1     \\%
87      &-&     3,4       & 1    &  1    &   1   & 1   &   -   &   -       &-  &1     \\%
92      &1&     1         & 1    &  1    &   3   & 1   &   1   &   1       &1  &3     \\%
130     &-&     -         & 3    &  1    &   2   & 1   &   -   &   -       &-  &1     \\%
152     &-&     1         & 1    &  1    &   6   & 1   &   -   &   -       &-  &1     \\%
159     &-&     2         & 1    &  1    &   1   & 1   &   -   &   -       &-  &1     \\%
169     &-&     -         & -    &  1    &   1   & 1   &   -   &   -       &-  &1     \\%
171     &-&     -         & -    &  1    &   1   & -   &   -   &   -       &-  &1     \\%
207     &-&     -         & 3    &  1    &   2   & -   &   -   &   -       &-  &1     \\%
318     &1&     3         & 1    &  1    &   10  & 1   &   -   &   -       &-  &1     \\%
425     &-&     -         & -    &  1    &   1   & -   &   -   &   -       &-  &1     \\%
459     &1&     -         & 1    &  1    &   2,7 & 1   &   -   &   -       &-  &1     \\%
460     &-&     1         & 1    &  1    &   4   & 1   &   1   &   1       &1  &1     \\%
655     &-&     4         & 1    &  1    &   4   & 1   &   -   &   -       &-  &1     \\%
664     &1&     7         & 1    &  1    &   2,4 & 1   &   -   &   -       &-  &1     \\%
666     &-&     -         & 3    &  1    &   1   & 1   &   -   &   -       &-  &1     \\%
692     &3&     5         & 1    &  1    &   4   & 1   &   -   &   1       &-  &1     \\%
793     &-&     1         & 3    &  1    &   1   & -   &   -   &   -       &-  &1     \\%
802     &-&     4         & 3    &  -    &   2   & -   &   -   &   -       &-  &1     \\%
809     &3&     3         & 1    &  1    &   10  & -   &   -   &   -       &-  &1     \\%
836     &3&     1         & 1    &  1    &   3   & 1   &   1   &   1       &1  &1     \\%
848     &-&     4         & 1    &  1    &   1   & 1   &   -   &   -       &-  &1     \\%
857     &-&     1         & 1    &  1    &   4   & 1   &   -   &   1       &-  &4     \\%
873     &3&     1         & 1    &  1    &   3   & 1   &   1   &   1       &1  &2     \\%
890     &-&     -         & 1    &  1    &   2   & 1   &   -   &   -       &-  &1     \\%
912     &3&     3,7       & 1    &  1    &   2   & 1   &   -   &   1       &-  &1     \\%
945     &3&     4,6       & 1    &  -    &   1   & -   &   -   &   -       &-  &1     \\%
950     &3&     3,4       & 1    &  -    &   1   & -   &   -   &   -       &-  &1     \\%
971     &1&     3         & 1    &  1    &   4   & 1   &   -   &   -       &-  &1     \\%
984     &3&     1         & 1    &  1    &   2   & 1   &   -   &   1       &-  &1     \\%
995     &-&     2         & 1    &  1    &   1   & -   &   -   &   -       &-  &1     \\%
1001    &-&     -         & 3    &  -    &   1   & 1   &   -   &   -       &-  &1     \\%
1002    &1&     1         & 1    &  1    &   2   & 1   &   -   &   1       &-  &1     \\%
1003    &3&     1         & 1    &  1    &   6   & 1   &   -   &   1       &-  &1     \\%
1043    &3&     1         & 1    &  1    &   4   & 1   &   1   &   1       &1  &2     \\%
1047    &3&     7         & 1,2  &  1    &   5   & 1   &   1   &   -       &-  &1     \\%
1106    &-&     -         & 3    &  -    &   1   & -   &   -   &   -       &-  &1     \\%
1110    &-&     1         & 1    &  1    &   2   & 1   &   1   &   1       &-  &1     \\%
1121    &-&     -         & 3    &  -    &   1   & 1   &   -   &   -       &-  &1     \\%
1158    &2&     1         & 1    &  1    &   2   & 1   &   -   &   1       &-  &1     \\%
1189    &1&     3         & 1    &  1    &   4   & 1   &   -   &   -       &-  &1     \\%
1196    &-&     3,7,8     & 1    &  1    &   5   & 1   &   -   &   -       &-  &1     \\%
1200    &-&     4         & 1    &  -    &   1   & -   &   -   &   -       &-  &1     \\%
1217    &3&     9         & 1    &  1    &   1   & 1   &   -   &   -       &-  &1     \\%
1253    &-&     1         & 1    &  1    &   4   & 1   &   1   &   1       &-  &1     \\%
1257    &-&     -         & 3    &  1    &   1   & -   &   -   &   -       &-  &1     \\%
1287    &-&     -         & -    &  -    &   1   & -   &   -   &   -       &-  &1     \\%
1313    &3&     4         & 4    &  -    &   2   & -   &   -   &   -       &-  &1     \\%
1326    &3&     3,7       & 1    &  1    &   4   & 1   &   1   &   -       &-  &1     \\%
1356    &3&     3         & 1    &  -    &   1   & -   &   -   &   -       &-  &1     \\%
1368    &3&     3,7       & 1    &  1    &   5   & 1   &   -   &   -       &-  &1     \\%
1377    &-&     2         & 1    &  -    &   1   & -   &   -   &   -       &-  &1     \\%
1379    &1&     5         & 1    &  1    &   2   & 1   &   1   &   -       &-  &1     \\%
1403    &-&     -         & -    &  -    &   1   & -   &   -   &   -       &-  &1     \\%
1410    &-&     1         & 1    &  1    &   7   & 1   &   -   &   -       &-  &1     \\%
1411    &3&     2         & 1    &  -    &   1   & -   &   -   &   -       &-  &1     \\%
1412    &-&     1         & 1    &  1    &   2,8 & 1   &   1   &   -       &-  &1     \\%
1419    &-&     3         & 1    &  1    &   8   & 1   &   -   &   -       &-  &1     \\%
1426    &-&     2         & 1    &  -    &   1   & -   &   -   &   -       &-  &1     \\%
1448    &-&     4,7       & 1    &  -    &   2   & -   &   -   &   -       &-  &1     \\%
1450    &1&     7         & 1    &  1    &   2   & 1   &   1   &   1       &-  &1     \\%
1486    &-&     3,4       & 1    &  -    &   1   & -   &   -   &   -       &-  &1     \\%
1552    &-&     1         & 1    &  1    &   2,4 & 1   &   1   &   -       &-  &1     \\%
1554    &1&     1         & 1    &  1    &   3   & 1   &   1   &   1       &1  &1     \\%
1569    &-&     3         & 1    &  1    &   1   & 1   &   -   &   -       &-  &1     \\%
1575    &1&     1         & 1    &  1    &   4   & 1   &   -   &   -       &-  &1     \\%
1581    &-&     4,7       & 1    &  1    &   1   & 1   &   -   &   -       &-  &1     \\%
1596    &-&     -         & -    &  1    &   1   & -   &   -   &   -       &-  &1     \\%
\hline                                                                          
                                              
\newpage                                                                        
                                              
\caption{continue}\\                                                            
\hline
\noalign{\smallskip}
{\rm VCC}&{\rm UV}& {\rm optical} &{\rm near-IR}& {\rm ISOCAM}&{\rm IRAS}&{\rm ISOPHOT}&{\rm radio 2.8cm}&{\rm radio 6.3cm}&{\rm radio
12.6cm}&{\rm radio 21cm}\\
(1) & (2) & (3) & (4) & (5) & (6) & (7) & (8) & (9) & (10) & (11) \\
\hline                                                 
1644    &-&     -         & 3    &  -    &   1   & -   &   -   &   -       &-  &1     \\%
1673    &1&     1         & 1    &  1    &   1   & 1   &   -   &   -       &-  &1     \\%
1675    &-&     -         & 1    &  1    &   1   & 1   &   -   &   -       &-  &1     \\%
1676    &-&     1         & 1    &  1    &   3   & 1   &   1   &   1       &1  &1     \\%
1678    &1&     7         & 1    &  1    &   2   & 1   &   -   &   -       &-  &1     \\%
1686    &-&     1         & -    &  1    &   2   & 1   &   -   &   -       &-  &1     \\%
1690    &1&     1         & 1    &  1    &   3   & 1   &   1   &   1       &1  &1     \\%
1699    &1&     3         & 1    &  1    &   1   & 1   &   -   &   -       &-  &1     \\%
1725    &1&     3,4,10    & 1    &  1    &   9   & 1   &   -   &   -       &-  &1     \\%
1726    &1&     4         & 1    &  1    &   1   & 1   &   -   &   -       &-  &1     \\%
1727    &1&     1         & 1    &  1    &   3   & 1   &   1   &   1       &1  &1     \\%
1730    &-&     3,7       & 1    &  1    &   6   & 1   &   1   &   1       &-  &1     \\%
1750    &-&     -         & 3    &  1    &   2   & 1   &   -   &   -       &-  &1     \\%
1757    &-&     3,7       & 1    &  1    &   4,6 & 1   &   -   &   -       &-  &1     \\%
1758    &-&     3         & 1    &  1    &   1   & -   &   -   &   -       &-  &1     \\%
1784    &-&     -         & 3    &  1    &   1   & -   &   -   &   -       &-  &1     \\%
1789    &-&     4         & 1    &  1    &   1   & -   &   -   &   -       &-  &1     \\%
1791    &1&     3,4,6     & 1    &  1    &   7   & -   &   -   &   -       &-  &1     \\%
1804    &-&     2         & 1    &  1    &   1   & -   &   -   &   -       &-  &1     \\%
1811    &1&     3,11      & 1    &  1    &   2   & -   &   1   &   1       &-  &1     \\%
1813    &1&     1         & 1    &  1    &   7   & -   &   -   &   1       &-  &1     \\%
1822    &-&     2         & 1    &  1    &   1   & -   &   -   &   -       &-  &1     \\%
1869    &-&     1         & 1    &  -    &   2   & -   &   -   &   1       &-  &1     \\%
1885    &-&     -         & 1    &  1    &   1   & -   &   -   &   -       &-  &1     \\%
1918    &-&     -         & 1    &  1    &   1   & -   &   -   &   -       &-  &1     \\%
1929    &1&     1         & 1    &  1    &   2   & -   &   -   &   -       &-  &1     \\%
1932    &-&     1         & 1    &  1    &   6   & -   &   1   &   1       &1  &1     \\%
1952    &-&     -         & 1    &  1    &   9   & -   &   -   &   -       &-  &1     \\%
1970    &-&     -         & 1    &  1    &   1   & -   &   -   &   -       &-  &1     \\%
1972    &1&     7         & 1    &  1    &   3   & -   &   1   &   1       &1  &1     \\%
1987    &1&     1         & 1    &  1    &   3   & -   &   1   &   1       &1  &1     \\%
1992    &-&     2         & 1    &  1    &   1   & -   &   -   &   -       &-  &1     \\%
1999    &-&     3,11      & 1    &  1    &   4   & -   &   -   &   -       &-  &1     \\%
2006    &-&     -         & 1    &  1    &   2   & -   &   -   &   -       &-  &1     \\%
2007    &-&     -         & 1    &  1    &   1   & -   &   -   &   -       &-  &1     \\%
2023    &-&     3         & 1    &  1    &   7   & -   &   -   &   -       &-  &1     \\%
2033    &-&     4         & 1    &  1    &   2   & -   &   -   &   -       &-  &1     \\%
2034    &-&     2         & 1    &  1    &   1   & -   &   -   &   -       &-  &1     \\%
2037    &-&     2         & 1    &  1    &   1   & -   &   -   &   -       &-  &1     \\%
2058    &1&     1         & 1    &  1    &   6   & -   &   1   &   1       &-  &1     \\%
2066    &-&     1         & 1    &  1    &   4   & -   &   1   &   1       &-  &1     \\%
2070    &-&     1         & 1    &  1    &   2   & -   &   -   &   1       &-  &1     \\%
2087    &-&     12,13     & 1    &  1    &   2,5 & -   &   1   &   -       &-  &1     \\%
2094    &-&     -         & 3    &  1    &   1   & -   &   -   &   -       &-  &1     \\%

\hline
\end{longtable}
\normalsize

References:

UV data: 

1: Deharveng et al. (1994)              
2: Deharveng et al. (2002)              
3: Donas et al. (private communication) 

Optical data:

1: this work                            
2: Gavazzi et al. (2001)                %
3: Schroeder \& Visvanathan (1996)      
4: Gallagher \& Hunter (1986)           
5: Gavazzi et al. (1994)                
6: de Vaucouleurs et al. (1981)         
7: Longo et al. (1983-1985)             
8: Prugniel \& Heraudeau P. (1998)      
9: Bothun et al. (1986)                 
10: Takamiya et al. (1995)              
11: Boselli \& Gavazzi (1994)           
12: Burstein et al. (1987)              
13: Frueh et al. (1996)                 

Near-IR:

1: Boselli et al. (1997)                
2: Gavazzi et al. (2001)                
3: Pierini, private communication       
4: Gavazzi et al., in preparation       

ISOCAM data:

1: Boselli et al. (2002c)               

IRAS data:

1: Lonsdale et al. (1985)               
2: Helou et al. (1988)                  
3: Soifer et al. (1989)                 
4: Thuan \& Sauvage (1992)              
5: Isobe \& Feigelson (1992)            
6: Rush et al. (1993)                   
7: Young et al. (1996)                  
8: Tuffs, private communication         
9: Almoznino \& Brosch (1998)           
10: Magri (1994)                        

ISOPHOT data:

Tuffs et al. (2002)                     

Radio continuum data:

2.8cm:

1: Niklas et al. (1995)                 

6.3cm:

1: Niklas et al. (1995)                 

12.6cm:

1: Dressel \& Condon (1978)             

21cm:

1: Gavazzi \& Boselli (1999)            
2: Kotanyi et al. (1980)                
3: Condon et al. (1990)                 
4: Condon et al. (1987)                 

\twocolumn

\addtocounter{table}{0}
\onecolumn
\scriptsize
\begin{longtable}{rcccrrcccrcc}
\caption{The emission lines data}\\
\hline
\noalign{\smallskip}
\hline
\noalign{\smallskip}
{\rm VCC}&$H\alpha+[NII]EW$&$F(H\alpha+[NII])$& ref &$MHI$&$defHI$&qual&$WHI$& ref& $MH_2$ & ref  \\
         &   \AA        &erg cm$^{-2}$ s$^{-1}$&     &M$\odot$& &    &km s$^{-1}$& & M$\odot$&  \\
(1) & (2) & (3) & (4) & (5) & (6) & (7) & (8) & (9) & (10) & (11)  \\
\hline     
1     & 12   &  -13.51  &  1    &    -      & -      & -  & -     &  -     &   -      & -     \\
4     & -    &   -      & -     &    8.25   & 0.00   & 2  & 41    &  1     &   -      & -     \\
17    & 53   &  -12.83  &  6    &    8.78   & -0.04  & 2  & 58    &  1     &   -      & -     \\
24    & 3    &  -14.03  &  1    &    8.93   & -0.11  & 1  & 218   &  1     &   -      & -     \\
26    & -    &   -      & -     &    8.38   & -0.26  & 3  & 47    &  1     &   -      & -     \\
66    & 33   &  -11.72  &  6    &    9.81   & -0.20  & 1  & 269   &  9     &   7.75   & 3     \\
81    & 21   &  -13.26  &  1    &    8.63   & -0.45  & 1  & 103   &  3     &   -      & -     \\
87    & 20   &  -12.94  &  6    &    8.32   & 0.29   & 2  & 109   &  1     &$<$8.13   & 4     \\
92    & 9    &  -11.53  &  1    &    9.76   & 0.33   & 1  & 469   &  11    &   8.75   & 1     \\
130   & -    &   -      & -     &    7.86   & 0.06   & 1  & 119   &  1     &   -      & -     \\
152   & 9    &  -12.63  &  6    &    8.61   & 0.24   & 1  & 216   &  5     &   8.15   & 3     \\
159   & 19   &  -13.21  &  6    &    8.55   & 0.30   & 2  & 67    &  1     &   -      & -     \\
169   & 7    &  -14.36  &  4    &    8.52   & -0.35  & 2  & 35    &  1     &   -      & -     \\
171   & -    &   -      & -     &    7.16   & 1.20   & 4  & 29    &  2     &   -      & -     \\
207   & -    &   -      & -     &    8.25   & -0.25  & 2  & 81    &  2     &   -      & -     \\
318   & 51   &  -12.51  &  1    &    9.39   & -0.13  & 1  & 178   &  5     &$<$8.34   & 4     \\
425   & -    &   -      & -     & $<$7.53   & 0.33   & -  & -     &  2     &   -      & -     \\
459   & 48   &  -12.66  &  2    &    8.22   & -0.07  & 2  & 127   &  1     &$<$7.72   & 4     \\
460   & 8    &  -11.92  &  6    &    7.62   & 1.85   & 3  & 274   &  6     &   8.43   & 1     \\
655   & 6    &  -12.78  &  6    &    7.91   & 0.75   & 2  & 71    &  8     &   7.99   & 3     \\
664   & 101  &  -12.12  &  1    &    8.40   & 0.62   & 2  & 103   &  7     &$<$7.79   & 3     \\
666   & -    &   -      & -     & $<$7.60   & 0.71   & -  & -     &  1     &   -      & -     \\
692   & 16   &  -12.47  &  5    &    8.46   & 0.66   & 1  & 119   &  7     &$<$7.60   & 3     \\
793   & 2    &  -14.41  &  2    &    7.65   & 0.05   & 2  & 41    &  1     &   -      & -     \\
802   & 37   &  -13.54  &  2    & $<$7.66   & 0.28   & -  & -     &  1     &   -      & -     \\
809   & -    &   -      &  -    &    8.39   & 0.15   & 2  & 173   &  3     &   -      & -     \\
836   & 15   &  -11.67  &  1    &    8.78   & 0.69   & 1  & 394   &  7     &   8.29   & 3     \\
848   & 26   &  -13.05  &  2    &    8.85   & -0.18  & 2  & 141   &  1     &   -      & -     \\
857   & 12   &  -11.97  &  1    &    8.51   & 0.86   & 1  & 174   &  7     &   8.30   & 1     \\
873   & 16   &  -11.97  &  2    &    8.74   & 0.63   & 1  & 269   &  4     &   8.96   & 1     \\
890   & -    &   -      & -     &    7.33   & -0.05  & 2  & 66    &  2     &   -      & -     \\
912   & 13   &  -12.56  &  6    &    8.26   & 0.99   & 1  & 157   &  7     &   8.12   & 3     \\
945   & -    &   -      & -     &    8.21   & 0.31   & 3  & 37    &  1     &   -      & -     \\
950   & 23   &  -13.45  &  3    &    8.81   & -0.07  & 2  & 74    &  1     &   -      & -     \\
971   & 29   &  -12.52  &  3    &    9.26   & 0.20   & 2  & 171   &  2     &   8.86   & 3     \\
984   & 1    &  -13.17  &  1    & $<$7.31   & 1.78   & -  & -     &  8     &   -      & -     \\
995   & 32   &  -12.98  &  6    &    8.92   & -0.33  & 1  & 162   &  3     &   -      & -     \\
1001  & -1   &   -      &  6    &    7.49   & 0.57   & 3  & 37    &  1     &   -      & -     \\
1002  & 9    &  -12.11  &  6    &    8.92   & 0.47   & 1  & 152   &  4     &   8.24   & 3     \\
1003  & 5    &  -11.72  &  1    & $<$7.44   & 2.30   & -  & -     &  8     &   7.39   & 7     \\
1043  & 7    &  -11.77  &  6    &    8.62   & 1.33   & 2  & 253   &  8     &   8.98   & 5     \\
1047  & -1   &   -      &  6    & $<$7.44   & 1.37   & -  & -     &  8     &   8.25   & 2     \\
1106  & -    &   -      & -     & $<$7.19   & 0.68   & -  & -     &  1     &   -      & -     \\
1110  & 2    &  -12.30  &  5    &    8.65   & 0.95   & 1  & 319   &  4     &   8.32   & 1     \\
1121  & -    &   -      & -     & $<$7.63   & 0.39   & -  & -     &  1     &   -      & -     \\
1158  & -1   &   -      &  6    & $<$7.13   & 2.07   & -  & -     &  6     &   -      & -     \\
1189  & 21   &  -12.67  &  6    &    8.39   & 0.34   & 3  & 126   &  5     &$<$7.86   & 3     \\
1196  & -    &   -      & -     &    -      & -      & -  & -     &  -     &   -      & -     \\
1200  & -    &  -13.61  & -     & $<$7.39   & 1.11   & -  & -     &  1     &   -      & -     \\
1217  & -    &   -      & -     & $<$7.09   & 1.73   & -  & -     &  2     &   -      & -     \\
1253  & -    &   -      & -     & $<$7.31   & 1.79   & -  & -     &  8     &   -      & -     \\
1257  & -    &   -      & -     &    8.41   & 0.14   & 1  & 157   &  1     &   -      & -     \\
1287  & -    &   -      & -     & $<$7.63   & 0.54   & -  & -     &  1     &   -      & -     \\
1313  & 351  &  -12.84  &  2    &    7.84   & -0.20  & 2  & 107   &  1     &   -      & -     \\
1326  & -1   &   -      &  6    & $<$7.24   & 1.52   & -  & -     &  6     &   7.98   & 3     \\
1356  & 43   &  -13.05  &  1    &    8.34   & 0.04   & 1  & 168   &  3     &   -      & -     \\
1368  & -    &   -      & -     & $<$7.38   & 1.28   & -  & -     &  8     &   -      & -     \\
1377  & -    &   -      & -     & $<$7.53   & 0.37   & -  & -     &  1     &   -      & -     \\
1379  & 36   &  -11.92  &  3    &    8.95   & 0.15   & 1  & 200   &  4     &   7.84   & 3     \\
1403  & -    &   -      & -     & $<$7.57   & 0.45   & -  & -     &  1     &   -      & -     \\
1410  & 35   &  -12.71  &  1    &    8.20   & 0.42   & 1  & 181   &  1     &   -      & -     \\
1411  & 2    &  -14.37  &  3    &    7.96   & 0.51   & 3  & 57    &  1     &   -      & -     \\
1412  & 2    &  -12.52  &  1    & $<$7.02   & 2.33   & -  & -     &  6     &$<$7.76   & 4     \\
1419  & 5    &  -13.06  &  6    & $<$7.27   & 1.59   & -  & -     &  3     &   -      & -     \\
1426  & 6    &  -13.93  &  3    & $<$7.33   & 0.79   & -  & -     &  1     &   -      & -     \\
1448  & -    &   -      & -     &    8.90   & -0.14  & 5  & 70    &  13    &   -      & -     \\
1450  & 69   &  -11.89  &  1    &    8.47   & 0.54   & 1  & 130   &  10    &   7.78   & 3     \\
1486  & 12   &  -13.17  &  6    &    7.66   & 0.72   & 2  & 124   &  5     &   -      & -     \\
1552  & 2    &  -12.98  &  1    & $<$7.16   & 2.18   & -  & -     &  6     &   7.62   & 3     \\
1554  & 75   &  -11.40  &  6    &    9.46   & -0.37  & 1  & 185   &  4     &   8.02   & 4     \\
1569  & 14   &  -13.45  &  6    &    7.47   & 0.90   & 2  & 109   &  3     &   -      & -     \\
1575  & 13   &  -12.73  &  1    &    7.94   & 0.93   & 2  & 113   &  5     &   8.62   & 3     \\
1581  & 6    &  -13.46  &  3    &    8.64   & -0.03  & 2  & 97    &  1     &   -      & -     \\
1596  & -    &   -      & -     &    7.34   & 0.12   & 4  & 56    &  1     &   -      & -     \\
\hline
 
\newpage
 
\caption{continue}\\
 
\hline
\noalign{\smallskip}
{\rm VCC}&$H\alpha+[NII]EW$&$F(H\alpha+[NII])$& ref &$MHI$&$defHI$&qual&$WHI$& ref& $MH_2$ & ref \\
         &   \AA        &erg cm$^{-2}$ s$^{-1}$&     &M$\odot$& &    &km s$^{-1}$& & M$\odot$&  \\
(1) & (2) & (3) & (4) & (5) & (6) & (7) & (8) & (9) & (10) & (11) \\
\hline  
1644  & -    &   -      & -     &    8.15   & 0.14   & 1  & 93    &  1     &   -      & -      \\
1673  & 15   &  -12.00  &  1    &    8.69   & 0.43   & 3  & 214   &  12    &   8.70   & 1      \\
1675  & 4    &  -13.85  &  1    &    7.45   & 1.35   & 3  & 41    &  1     &   -      & -      \\
1676  & 19   &  -11.74  &  1    &    8.99   & 0.58   & 3  & 341   &  12    &   8.99   & 1      \\
1678  & 55   &  -12.56  &  6    &    9.00   & -0.06  & 2  & 57    &  5     &   -      & -      \\
1686  & 44   &  -12.17  &  1    &    8.35   & 0.79   & 1  & 116   &  1     &$<$7.77   & 4      \\
1690  & 2    &  -12.02  &  1    &    8.93   & 1.07   & 1  & 370   &  11    &   9.01   & 1      \\
1699  & 24   &  -12.85  &  3    &    8.62   & 0.04   & 2  & 109   &  1     &   -      & -      \\
1725  & 50   &  -12.62  &  2    &    8.11   & 0.55   & 2  & 93    &  1     &   -      & -      \\
1726  & -    &  -12.80  & -     &    8.52   & 0.00   & 2  & 85    &  1     &   -      & -      \\
1727  & 4    &  -11.49  &  6    &    8.79   & 0.83   & 1  & 374   &  7     &   9.08   & 6      \\
1730  & 4    &  -12.62  &  1    &    7.83   & 1.03   & 3  & 190   &  4     &   8.04   & 3      \\
1750  & -    &   -      & -     &    7.35   & -0.01  & 4  & 73    &  1     &   -      & -      \\
1757  & 7    &  -12.98  &  3    &    7.38   & 1.38   & 5  & -     &  3     &   -      & -      \\
1758  & 17   &  -13.07  &  1    &    8.28   & 0.39   & 1  & 172   &  5     &   -      & -      \\
1784  & 2    &   -      &  4    &    7.34   & 0.78   & 4  & 37    &  1     &   -      & -      \\
1789  & 16   &  -13.25  &  3    &    7.86   & 0.53   & 1  & 103   &  1     &   -      & -      \\
1791  & 72   &  -12.42  &  3    &    8.63   & -0.12  & 2  & 120   &  1     &   -      & -      \\
1804  & 3    &  -14.37  &  2    &    7.23   & 0.84   & 5  & 88    &  1     &   -      & -      \\
1811  & 16   &  -12.50  &  6    &    8.63   & 0.23   & 1  & 150   &  4     &   8.20   & 3      \\
1813  & -1   &   -      &  6    & $<$7.19   & 2.23   & -  & -     &  6     &$<$7.39   & 3      \\
1822  & -1   &   -      &  4    &    7.64   & 0.29   & 2  & 34    &  1     &   -      & -      \\
1869  & -    &   -      & -     & $<$7.44   & 1.81   & -  & -     &  8     &   -      & -      \\
1885  & -    &   -      & -     & $<$7.33   & 1.09   & -  & -     &  1     &   -      & -      \\
1918  & 15   &  -13.90  &  3    &    8.16   & 0.17   & 2  & 77    &  1     &   -      & -      \\
1929  & 13   &  -12.75  &  6    &    8.70   & 0.35   & 1  & 190   &  10    &$<$7.68   & 4      \\
1932  & 16   &  -12.32  &  6    &    8.66   & 0.45   & 1  & 301   &  6     &   8.46   & 3      \\
1952  & 32   &  -13.62  &  4    &    8.21   & -0.19  & 2  & 65    &  1     &   -      & -      \\
1970  & -    &   -      & -     & $<$7.49   & 0.53   & -  & -     &  1     &   -      & -      \\
1972  & 16   &  -11.71  &  6    &    8.75   & 0.27   & 1  & 203   &  4     &   8.69   & 1      \\
1987  & 31   &  -11.34  &  6    &    9.85   & -0.29  & 1  & 308   &  7     &   8.66   & 1      \\
1992  & 24   &  -13.21  &  4    &    8.35   & -0.22  & 1  & 110   &  1     &   -      & -      \\
1999  & -1   &   -      &  6    & $<$7.19   & 1.61   & -  & -     &  6     &   -      & -      \\
2006  & -1   &   -      &  6    &    7.94   & 0.91   & 2  & 78    &  3     &   -      & -      \\
2007  & 10   &  -13.77  &  2    &    7.37   & 0.74   & 3  & 77    &  1     &   -      & -      \\
2023  & 27   &  -12.56  &  3    &    8.85   & -0.05  & 1  & 186   &  3     &$<$7.93   & 4      \\
2033  & 13   &  -13.32  &  2    &    7.45   & 0.61   & 3  & 41    &  1     &   -      & -      \\
2034  & 3    &  -14.28  &  6    &    7.83   & 0.27   & 3  & 62    &  1     &   -      & -      \\
2037  & 16   &  -13.82  &  6    &    7.39   & 0.81   & 3  & 44    &  1     &   -      & -      \\
2058  & 14   &  -11.85  &  6    &    8.79   & 0.90   & 1  & 195   &  4     &   8.70   & 1      \\
2066  & 6    &  -12.47  &  6    &    8.36   & 0.66   & 2  & 106   &  1     &   7.71   & 3      \\
2070  & 2    &  -11.98  &  6    &    9.54   & 0.01   & 1  & 432   &  4     &$<$7.45   & 5      \\
2087  & -    &   -      & -     & $<$7.38   & 1.26   & -  & -     &  8     &   -      & -      \\
2094  & -    &   -      & -     & $<$7.09   & 0.40   & -  & -     &  2     &   -      & -      \\
\hline
\end{longtable}
\normalsize

References:

H$\alpha$+[NII]:

1: Boselli \& Gavazzi (2002); 
2: Boselli et al. (2002a); 
3: Gavazzi et al. (2002b); 
4: Heller et al. (1999); 
5: Koopmann et al. (2001) (equivalent width from ref. 6);
6: Boselli et al., in preparation 


HI:

1: Hoffman et al. (1987); 
2: Hoffman et al. (1989a); 
3: Haynes \& Giovanelli (1986);  
4: Helou et al. (1984); 
5: Hoffman et al. (1989b); 
6: Magri (1994); 
7: Helou et al. (1981); 
8: Giovanardi et al. (1983); 
9: Warmels (1986); 
10: Schneider et al. (1990); 
11: Huchtmeier et al. (1989); 
12: Helou et al. (1982); 
13: Bottinelli et al. (1990); 

CO: 

1: Kenney \& Young (1988); 
2: Stark et al. (1986); 
3: Boselli et al. (1995); 
4: Boselli et al. (2002b); 
5: Combes et al. (1988); 
6: Boselli et al., in preparation; 
7: Sage \& Wrobel (1989); 
\twocolumn

\addtocounter{table}{1}
\onecolumn
\scriptsize
\begin{longtable}{rcccccccc}
\caption{The galactic and internal extinction corrections}\\
\hline
\noalign{\smallskip}
\hline
\noalign{\smallskip}
{\rm VCC}& AB & A(UV) & A(U) & A(B) & A(V) & A(J) & A(H) & A(K) \\ 
(1) & (2) & (3) & (4) & (5) & (6) & (7) & (8) & (9) \\
\hline                                              
1   &  0.02&-   & -   & 0.35& 0.25& 0.07  &  0.05   & 0.03\\
4   &  0.12&-   & -   &  -  &  -  &  -    &   -     & 0.03\\
17  &  0.08&-   & -   & 0.41& 0.31& 0.07  &  0.05   & 0.03\\
24  &  0.01&-   & -   & 0.34& 0.25&  -    &   -     & 0.03\\
26  &  0.13&-   & -   &  -  &  -  &  -    &   -     & 0.03\\
66  &  0.00&0.68& 0.39& 0.33& 0.25& 0.07  &  0.05   & 0.04\\
81  &  0.14&-   & -   & 0.58& 0.43&  -    &   -     & 0.05\\
87  &  0.08&-   & 0.47& 0.41& 0.31&  -    &   -     & 0.03\\
92  &  0.14&1.20& 0.70& 0.60& 0.46& 0.11  &  0.07   & 0.05\\
130 &  0.00&-   & -   &  -  &  -  &  -    &   -     & 0.03\\
152 &  0.00&-   & 0.50& 0.43& 0.33&  -    &  0.07   & 0.05\\
159 &  0.00&-   & -   & 0.33& 0.24&  -    &   -     & 0.03\\
169 &  0.00&-   & -   &  -  &  -  &  -    &   -     &  -  \\
171 &  0.00&-   & -   &  -  &  -  &  -    &   -     &  -  \\
207 &  0.00&-   & -   &  -  &  -  &  -    &   -     & 0.03\\
318 &  0.00&0.24& 0.14& 0.11& 0.07&  -    &   -     & 0.01\\
425 &  0.00&-   & -   &  -  &  -  &  -    &   -     &  -  \\
459 &  0.04&0.31& -   &  -  &  -  &  -    &   -     & 0.01\\
460 &  0.07&-   & 0.88& 0.77& 0.59& 0.16  &  0.11   & 0.08\\
655 &  0.03&-   & 0.42& 0.36& 0.27&  -    &   -     & 0.03\\
664 &  0.07&0.56& 0.29& 0.25& 0.19&  -    &   -     & 0.02\\
666 &  0.03&-   & -   &  -  &  -  &  -    &   -     & 0.03\\
692 &  0.01&0.56& 0.33& 0.28& 0.20&  -    &  0.04   & 0.03\\
793 &  0.12&-   & 0.53& 0.45& 0.33& 0.07  &  0.05   & 0.03\\
802 &  0.11&-   & 0.51& 0.44&  -  &  -    &   -     & 0.03\\
809 &  0.05&0.95& 0.55& 0.48& 0.37&  -    &   -     & 0.05\\
836 &  0.11&2.00& 1.28& 1.13& 0.88& 0.25  &  0.17   & 0.12\\
848 &  0.00&-   & 0.39& 0.33& 0.24&  -    &   -     & 0.03\\
857 &  0.03&-   & 0.83& 0.72& 0.56&  -    &   -     & 0.08\\
873 &  0.12&2.66& 1.73& 1.54& 1.25&  -    &  0.29   & 0.21\\
890 &  0.00&-   & -   &  -  &  -  &  -    &   -     & 0.03\\
912 &  0.10&1.12& 0.66& 0.57& 0.43&  -    &  0.07   & 0.05\\
945 &  0.12&0.93& 0.52& 0.45& 0.34&  -    &   -     & 0.03\\
950 &  0.03&0.74& 0.42& 0.36& 0.27&  -    &   -     & 0.03\\
971 &  0.00&0.43& 0.23& 0.2 & 0.14&  -    &   -     & 0.02\\
984 &  0.10&1.49& 0.92& 0.81& 0.62& 0.16  &  0.11   & 0.08\\
995 &  0.10&-   & 0.62& 0.53& 0.40&  -    &   -     & 0.05\\
1001&  0.08&-   & -   &  -  &  -  &  -    &   -     & 0.03\\
1002&  0.00&0.64& -   & 0.32& 0.23& 0.07  &  0.05   & 0.03\\
1003&  0.05&0.10& 0.07& 0.07& 0.04&  -    &   -     &  -  \\
1043&  0.09&1.45& 0.88& 0.76& 0.59& 0.16  &  0.11   & 0.08\\
1047&  0.08&1.45& -   & 0.77& 0.60& 0.16  &  0.11   & 0.08\\
1106&  0.01&-   & -   &  -  &  -  &  -    &   -     & 0.03\\
1110&  0.03&-   & 0.84& 0.73& 0.56& 0.16  &  0.11   & 0.08\\
1121&  0.04&-   & -   &  -  &  -  &  -    &   -     & 0.03\\
1158&  0.07&1.43& 0.89& 0.78& 0.59& 0.16  &  0.11   & 0.08\\
1189&  0.00&0.23& 0.12& 0.10& 0.07&  -    &   -     & 0.01\\
1196&  0.08&-   & 0.10& 0.09& 0.06&  -    &   -     &  -  \\
1200&  0.03&-   & 0.42& 0.36& 0.28&  -    &  0.05   & 0.03\\
1217&  0.03&0.74& -   & 0.36& 0.28& 0.07  &  0.05   & 0.03\\
1253&  0.02&-   & 0.04& 0.04& 0.02&  -    &   -     &  -  \\
1257&  0.02&-   & -   &  -  &  -  &  -    &   -     & 0.03\\
1287&  0.08&-   & -   &  -  &  -  &  -    &   -     &  -  \\
1313&  0.09&0.87& 0.49& 0.42& 0.31&  -    &  0.05   &  -  \\
1326&  0.02&2.54& 1.69& 1.51& 1.24&  -    &  0.31   & 0.22\\
1356&  0.02&0.72& 0.41& 0.35& 0.26&  -    &   -     & 0.03\\
1368&  0.04&0.08& 0.06& 0.06& 0.04&  -    &   -     &  -  \\
1377&  0.02&-   & -   & 0.35& 0.27&  -    &   -     & 0.03\\
1379&  0.03&0.55& 0.31& 0.26& 0.19&  -    &  0.03   & 0.02\\
1403&  0.07&-   & -   &  -  &  -  &  -    &   -     &  -  \\
1410&  0.03&-   & 0.43& 0.36& 0.27&  -    &   -     & 0.03\\
1411&  0.05&0.78& -   & 0.38& 0.28&  -    &   -     & 0.03\\
1412&  0.01&-   & 0.81& 0.71& 0.55& 0.16  &  0.11   & 0.08\\
1419&  0.01&-   & 0.81& 0.70& 0.54&  -    &  0.11   & 0.08\\
1426&  0.08&-   & -   & 0.41& 0.31&  -    &   -     & 0.03\\
1448&  0.08&-   & 0.12& 0.12& 0.07&  -    &   -     &  -  \\
1450&  0.11&0.89& -   & 0.43& 0.33&  -    &  0.05   & 0.03\\
1486&  0.01&-   & 0.80& 0.70& 0.54&  -    &   -     & 0.08\\
1552&  0.08&-   & 0.88& 0.77& 0.60& 0.16  &  0.11   & 0.08\\
1554&  0.00&0.92& 0.56& 0.47& 0.35& 0.11  &  0.07   & 0.05\\
1569&  0.08&-   & 0.60& 0.51& 0.39&  -    &   -     & 0.05\\
1575&  0.00&1.05& 0.64& 0.55& 0.42&  -    &   -     & 0.06\\
1581&  0.00&-   & 0.40& 0.33& 0.24&  -    &   -     & 0.03\\
1596&  0.00&-   & -   &  -  &  -  &  -    &   -     &  -  \\
\hline
 
\newpage
 
\caption{continue}\\
 
\hline
\noalign{\smallskip}
{\rm VCC}& AB & A(UV) & A(U) & A(B) & A(V) & A(J) & A(H) & A(K) \\ 
(1) & (2) & (3) & (4) & (5) & (6) & (7) & (8) & (9) \\
\hline   
1644&  0.10&-   & -   &  -  &  -  &  -    &   -     & 0.03\\
1673&  0.01&0.87& -   & 0.45& 0.33& 0.10  &  0.07   & 0.05\\
1675&  0.00&-   & -   &  -  &  -  &  -    &   -     & 0.03\\
1676&  0.01&-   & -   & 0.45& 0.33&  -    &   -     & 0.05\\
1678&  0.00&0.19& 0.10& 0.08& 0.06&  -    &   -     & 0.01\\
1686&  0.09&-   & 0.49& 0.42& 0.31&  -    &   -     & 0.03\\
1690&  0.09&1.50& 0.91& 0.79& 0.62& 0.17  &  0.11   & 0.08\\
1699&  0.00&0.68& 0.39& 0.33& 0.24&  -    &   -     & 0.03\\
1725&  0.00&0.68& 0.39& 0.33& 0.25& 0.07  &  0.05   & 0.03\\
1726&  0.00&0.68& 0.38& 0.33& 0.25&  -    &   -     & 0.03\\
1727&  0.14&1.95& 1.23& 1.07& 0.84& 0.23  &  0.16   & 0.11\\
1730&  0.00&-   & 0.51& 0.43& 0.33& 0.10  &  0.07   & 0.05\\
1750&  0.00&-   & -   &  -  &  -  &  -    &   -     & 0.03\\
1757&  0.09&-   & 0.91& 0.80& 0.61&  -    &   -     & 0.08\\
1758&  0.00&-   & 0.51& 0.44& 0.33&  -    &   -     & 0.05\\
1784&  0.02&-   & -   &  -  &  -  &  -    &   -     & 0.03\\
1789&  0.00&-   & 0.39& 0.33& 0.24& 0.07  &  0.05   & 0.03\\
1791&  0.00&0.15& 0.08& 0.06& 0.03&  -    &   -     & 0.01\\
1804&  0.00&-   & 0.39& 0.33& 0.24&  -    &   -     & 0.03\\
1811&  0.03&0.62& 0.35& 0.30& 0.22&  -    &  0.04   & 0.03\\
1813&  0.00&0.67& 0.39& 0.35& 0.25& 0.07  &  0.05   & 0.03\\
1822&  0.00&-   & 0.39& 0.33& 0.25&  -    &   -     & 0.03\\
1869&  0.00&-   & -   & 0.03& 0.01&  -    &   -     &  -  \\
1885&  0.04&-   & -   &  -  &  -  &  -    &   -     & 0.03\\
1918&  0.02&-   & -   &  -  &  -  &  -    &   -     & 0.03\\
1929&  0.03&0.44& 0.23& 0.2 & 0.15&  -    &   -     & 0.02\\
1932&  0.03&-   & 0.53& 0.46& 0.35& 0.10  &  0.07   & 0.05\\
1952&  0.00&-   & -   &  -  &  -  &  -    &   -     & 0.03\\
1970&  0.00&-   & -   &  -  &  -  &  -    &   -     & 0.03\\
1972&  0.04&1.15& 0.70& 0.61& 0.46& 0.13  &  0.09   & 0.06\\
1987&  0.06&1.39& 0.85& 0.74& 0.57& 0.16  &  0.11   & 0.08\\
1992&  0.00&-   & 0.39& 0.33& 0.25& 0.07  &  0.05   & 0.03\\
1999&  0.04&-   & 0.84& 0.73& 0.57& 0.16  &  0.11   & 0.08\\
2006&  0.04&-   & -   &  -  &  -  &  -    &   -     &  -  \\
2007&  0.00&-   & -   &  -  &  -  &  -    &   -     & 0.03\\
2023&  0.06&-   & 0.58& 0.49& 0.37&  -    &   -     & 0.05\\
2033&  0.00&-   & 0.39& 0.33& 0.24&  -    &   -     & 0.03\\
2034&  0.00&-   & -   & 0.33& 0.24&  -    &   -     & 0.03\\
2037&  0.00&-   & 0.39& 0.33& 0.25&  -    &   -     & 0.03\\
2058&  0.05&1.32& -   & 0.70& 0.54&  -    &  0.10   & 0.07\\
2066&  0.00&-   & 0.01& 0.02& 0.01&  -    &   -     &  -  \\
2070&  0.00&-   & 0.80& 0.70& 0.54& 0.16  &  0.11   & 0.08\\
2087&  0.00&-   & 0.01& 0.01&  -  &  -    &   -     &  -  \\
2094&  0.00&-   & -   &  -  &  -  &  -    &   -     & 0.03\\
\hline
\end{longtable}
\normalsize
\twocolumn

\addtocounter{table}{0}
\onecolumn
\scriptsize
\begin{longtable}{rcccc}
\caption{The output parameters from fitting the SED}\\
\hline
\noalign{\smallskip}
\hline
\noalign{\smallskip}
{\rm VCC}&{\rm stellar fit}& $[F_{6.75}(d+s)/F_{6.75}(s)]$ & c & d\\
(1) & (2) & (3) & (4) & (5) \\
\hline           
1      &   S & 1.15 &     -       &     -     \\
4      &   - &   -  &     -       &     -     \\
17     &   S & 3.48 &     -       &     -     \\
24     &   S & 0.43 &     -       &     -     \\
26     &   - &   -  &     -       &     -     \\
66     &   S & 8.53 &   0.555     &  -1.536   \\
81     &   S &   -  &     -       &     -     \\
87     &   S & 2.77 &     -       &     -     \\
92     &   S & 3.63 &   0.632     &  -1.555   \\
130    &   - &   -  &     -       &     -     \\
152    &   S &10.10 &   1.151     &  -4.827   \\
159    &   S &   -  &     -       &     -     \\
169    &   - &   -  &     -       &     -     \\
171    &   - &   -  &     -       &     -     \\
207    &   - &   -  &     -       &     -     \\
318    &   S & 1.68 &     -       &     -     \\
425    &   - &   -  &     -       &     -     \\
459    &   T & 3.42 &     -       &     -     \\
460    &   S & 1.65 &   0.620     &  -1.951   \\
655    &   S & 5.31 &     -       &     -     \\
664    &   S & 2.29 &     -       &     -     \\
666    &   - &   -  &     -       &     -     \\
692    &   S & 5.17 &     -       &     -     \\
793    &   S &   -  &     -       &     -     \\
802    &   - &   -  &     -       &     -     \\
809    &   S & 4.23 &     -       &     -     \\
836    &   S & 7.30 &   0.622     &  -1.141   \\
848    &   S & 1.70 &     -       &     -     \\
857    &   S & 1.79 &     -       &     -     \\
873    &   S &12.68 &   0.847     &  -2.698   \\
890    &   - &   -  &     -       &     -     \\
912    &   S & 3.75 &     -       &     -     \\
945    &   S &   -  &     -       &     -     \\
950    &   S &   -  &     -       &     -     \\
971    &   S & 2.49 &     -       &     -     \\
984    &   S & 0.57 &     -       &     -     \\
995    &   S & 1.62 &     -       &     -     \\
1001   &   - &   -  &     -       &     -     \\
1002   &   S & 7.30 &   1.133     &  -5.148   \\
1003   &   S & 1.10 &     -       &     -     \\
1043   &   S & 1.46 &   0.569     &  -0.838   \\
1047   &   S &   -  &     -       &     -     \\
1106   &   - &   -  &     -       &     -     \\
1110   &   S & 1.06 &     -       &     -     \\
1121   &   - &   -  &     -       &     -     \\
1158   &   S & 0.69 &     -       &     -     \\
1189   &   S & 4.17 &     -       &     -     \\
1196   &   S & 0.71 &     -       &     -     \\
1200   &   S &   -  &     -       &     -     \\
1217   &   S & 1.67 &     -       &     -     \\
1253   &   S & 0.74 &  0.678      &  -2.555   \\
1257   &   - &   -  &     -       &     -     \\
1287   &   - &   -  &     -       &     -     \\
1313   &   S &   -  &     -       &     -     \\
1326   &   S & 1.66 &     -       &     -     \\
1356   &   S &   -  &     -       &     -     \\
1368   &   S & 0.58 &     -       &     -     \\
1377   &   S &   -  &     -       &     -     \\
1379   &   S & 7.41 &  0.757      &  -3.368   \\
1403   &   - &   -  &     -       &     -     \\
1410   &   S & 5.55 &     -       &     -     \\
1411   &   S &   -  &     -       &     -     \\
1412   &   S & 0.59 &     -       &     -     \\
1419   &   S & 1.55 &     -       &     -     \\
1426   &   S &   -  &     -       &     -     \\
1448   &   S &   -  &     -       &     -     \\
1450   &   S &11.32 &     -       &     -     \\
1486   &   S &   -  &     -       &     -     \\
1552   &   S & 1.01 &     -       &     -     \\
1554   &   S &10.18 &  0.903      &  -2.677   \\
1569   &   S & 1.58 &     -       &     -     \\
1575   &   S &10.38 &     -       &     -     \\
1581   &   S &   -  &     -       &     -     \\
1596   &   - &   -  &     -       &     -     \\
\hline
                                              
\newpage                                                                        
                                              
\caption{continue}\\                                                            
\hline
\noalign{\smallskip}
{\rm VCC}&{\rm stellar fit}& $[F_{6.75}(d+s)/F_{6.75}(s)]$& c & d\\
(1) & (2) & (3) & (4) & (5) \\
\hline  
1644    &   -&    - &      -      &     -  \\
1673    &   S&  6.33&      -      &     -  \\
1675    &   T&    - &      -      &     -  \\
1676    &   S&  8.14&   0.671     &  -1.487\\
1678    &   S&    - &      -      &     -  \\
1686    &   S&  5.75&      -      &     -  \\
1690    &   S&  3.76&   0.461     &  -0.583\\
1699    &   S&  2.87&      -      &     -  \\
1725    &   S&  2.07&      -      &     -  \\
1726    &   S&    - &      -      &     -  \\
1727    &   S&  1.91&   0.075     &   1.601\\
1730    &   S&  2.67&      -      &     -  \\
1750    &   -&  1.78&      -      &     -  \\
1757    &   S&    - &      -      &     -  \\
1758    &   S&  2.94&      -      &     -  \\
1784    &   -&    - &      -      &     -  \\
1789    &   S&  2.49&      -      &     -  \\
1791    &   S&  3.23&      -      &     -  \\
1804    &   S&  1.04&      -      &     -  \\
1811    &   S&  7.28&   0.863     &  -3.643\\
1813    &   S&  0.80&      -      &     -  \\
1822    &   S&  4.70&      -      &     -  \\
1869    &   S&    - &   0.659     &  -3.058\\
1885    &   -&    - &      -      &     -  \\
1918    &   -&  5.73&      -      &     -  \\
1929    &   S&  4.60&      -      &     -  \\
1932    &   S& 16.83&   0.784     &  -2.629\\
1952    &   -&    - &      -      &     -  \\
1970    &   -&    - &      -      &     -  \\
1972    &   S& 12.67&   0.962     &  -3.348\\
1987    &   S& 11.78&   0.659     &  -1.480\\
1992    &   S&  3.03&      -      &     -  \\
1999    &   S&  0.51&      -      &     -  \\
2006    &   -&  0.94&      -      &     -  \\
2007    &   -&  4.23&      -      &     -  \\
2023    &   S&  1.83&      -      &     -  \\
2033    &   S&  1.06&      -      &     -  \\
2034    &   S&  0.51&      -      &     -  \\
2037    &   S&    - &      -      &     -  \\
2058    &   S&  4.90&   0.675     &  -2.473\\
2066    &   S&  3.33&   0.044     &   0.357\\
2070    &   S&  0.67&   0.000     &   0.301\\
2087    &   S&  0.44&      -      &     -  \\
2094    &   -& 17.81&      -      &     -  \\
\hline
\end{longtable}
\normalsize
Column 1: VCC name.

Column 2: S indicates galaxies with a fitted Bruzual \& Charlot spectrum
determined from UV, optical and near-IR spectro-photometry, T for galaxies
whose fit is Bruzual \& Charlot spectrum of the template.

Column 3: The ratio of the total flux (star plus dust) to the stellar flux at 6.75 $\mu$m, 
$[F_{6.75}(d+s)/F_{6.75}(s)]$.

Column 4 and 5: the fit of the radio continuum data, where $a$ and $b$ are the slope and the 
intercept of the relation $log F(\nu) = c \times log \lambda + d$, with $\nu$ in Hz and $\lambda$
in $\mu$m. Given the inconsistency in their radio continuum flux densities, the fitting coefficients of the galaxies
VCC 857, 1110 and 1450 are not given.
\twocolumn

\addtocounter{table}{0}
\onecolumn
\begin{landscape}
\scriptsize
\begin{longtable}{rrrrrrrrrrrrr}
\caption{Template dust extinction corrected SEDs}\\
\hline
\noalign{\smallskip}
\hline
$\lambda$ ($\mu$m) & \multicolumn{12}{c}{$Log(F(\nu)/F(K))$} \\
\hline
       &S0a    &Sa     &Sab-Sb & Sbc-Sc& Scd-Sd& Im    & BCD  &$L_H$$<$8.3   &8.3$\leq$$L_H$$<$9& 9$\leq$$L_H$$<$9.8&9.8$\leq$$L_H$$<$10.5&$L_H$$\geq$10.5\\
\hline
  0.20 & -3.22(2)& -2.46(5) & -1.51(5)& -1.04(15)& -0.72(4)& -0.38(7) & -0.36(4) &    $-$ (1) &      -0.37(11) &  -0.87(9) &  -1.22(12)&  -1.62(10) \\
  0.37 & -1.25(5)& -0.96(10)& -0.92(7)& -0.59(16)& -0.47(6)& -0.26(15)& -0.32(9) &   -0.12(3) &      -0.28(21) &  -0.58(17)&  -0.88(15)&  -1.06(16) \\
  0.44 & -0.78(6)& -0.42(11)& -0.45(7)& -0.26(22)& -0.16(6)&  0.06(21)& -0.10(11)&    0.11(4) &      -0.03(27) &  -0.21(21)&  -0.40(19)&  -0.53(18) \\
  0.55 & -0.48(6)& -0.21(11)& -0.25(7)& -0.13(22)& -0.07(6)&  0.13(21)& -0.04(10)&    0.23(3) &       0.08(27) &  -0.09(21)&  -0.18(19)&  -0.28(18) \\
  1.25 &  0.06(5)&  0.07(9) &  0.05(6)& -0.02(7) &   $-(-)$&  0.19(6) &  0.08(2) &    0.22(2) &       0.19(4)  &   0.09(3) &   0.05(12)&   0.06(16) \\
  1.65 &  0.13(6)&  0.13(10)&  0.11(6)&  0.13(14)&   $-(1)$&  0.15(7) &  0.11(2) &    0.21(2) &       0.15(5)  &   0.14(7) &   0.13(19)&   0.12(16) \\
  2.10 &  0.00(6)&  0.00(11)&  0.00(7)&  0.00(22)&  0.00(6)&  0.00(35)&  0.00(17)&    0.00(20)&       0.00(32) &   0.00(22)&   0.00(19)&   0.00(18) \\
  6.75 & -0.95(5)& -0.96(9) & -0.46(7)& -0.14(21)& -0.37(5)& -0.19(14)& -0.57(11)&   -0.15(4) &      -0.35(18) &  -0.34(18)&  -0.16(18)&  -0.61(17) \\
    12 &   $-$(1)& -0.54(4) & -0.24(7)&  0.26(13)&   $-$(1)&   $-$(1) &   $-$(-) &    $-$ (-) &       $-$ (-)  &   $-$ (1) &   0.13(13)&  -0.33(14) \\
    15 & -1.37(5)& -1.30(10)& -0.47(7)& -0.13(21)& -0.14(4)&  0.06(9) & -0.48(7) &    0.02(2) &      -0.30(10) &  -0.46(18)&  -0.22(19)&  -0.63(17) \\
    25 &   $-$(1)&   $-$(1) & -0.40(7)&  0.46(13)&   $-$(1)&   $-$(1) &   $-$(1) &    $-$ (-) &       $-$ (1)  &   0.72(3) &   0.28(11)&  -0.10(12) \\
    60 & -0.19(2)&  0.19(7) &  0.43(7)&  1.14(17)&  1.29(5)&  1.44(6) &  1.39(6) &    $-$ (-) &       1.48(10) &   1.18(14)&   1.05(14)&   0.43(15) \\
   100 &  0.36(2)&  1.01(6) &  0.97(7)&  1.63(17)&  1.72(5)&  1.78(7) &  1.65(8) &    2.00(3) &       1.70(9)  &   1.67(14)&   1.61(13)&   0.97(15) \\
   170 &  0.36(4)&  1.15(5) &  1.27(7)&  1.78(11)&  1.88(4)&  1.89(10)&  1.93(9) &    2.05(4) &       1.82(12) &   1.83(15)&   1.68(10)&   1.15(11) \\
 28000 &   $-$(1)&   $-$(1) & -1.55(5)& -1.46(8) &   $-(-)$&   $-$(1) &   $-$(-) &    $-$ (-) &       $-$ (-)  &   $-$ (-) &  -1.63(9) &  -1.44(9)  \\
 63000 & -2.46(2)& -2.09(2) & -1.57(5)& -1.08(11)&   $-(-)$&   $-$(1) &   $-$(-) &    $-$ (-) &       $-$ (-)  &   $-$ (1) &  -1.21(9) &  -1.40(12) \\
126000 & -2.00(3)& -1.68(6) & -1.34(5)& -1.01(10)&   $-(1)$&   $-$(1) &   $-$(1) &    $-$ (-) &       $-$ (-)  &   $-$ (1) &  -1.10(13)&  -1.47(14) \\
210000 &   $-$(1)&   $-$(1) & -1.31(5)& -0.81(10)& -0.72(2)&  0.33(4) &   $-$(1) &    $-$ (-) &       0.51(3)  &  -0.72(3) &  -0.95(11)&  -1.06(10) \\
\hline
\end{longtable}
\normalsize
Note: the values in parenthesis give the total number of objects in each Hubble type  
and wavelength bin that were combined to form the templates
\end{landscape}

\addtocounter{table}{0}
\scriptsize
\begin{longtable}{rrrrrrrrrrrrr}
\caption{Example of Template dust extinction corrected B\&C SEDs}\\
\hline
\noalign{\smallskip}
\hline
$\lambda$ ($\mu$m) & \multicolumn{12}{c}{$Log(F(\nu)/F(K))$} \\
\hline
       &S0a    &Sa     &Sab-Sb & Sbc-Sc& Scd-Sd& Im    & BCD  &$L_H$$<$8.3   &8.3$\leq$$L_H$$<$9& 9$\leq$$L_H$$<$9.8&9.8$\leq$$L_H$$<$10.5&$L_H$$\geq$10.5\\
\hline
 0.1005 & -3.26 & -2.88 & -2.04 & -1.03 & -0.81 & -0.58 & -0.40 & -0.36 & -0.51 & -0.97 & -1.26 & -2.08\\
 0.1015 & -3.25 & -2.88 & -2.06 & -1.05 & -0.83 & -0.62 & -0.44 & -0.40 & -0.55 & -1.01 & -1.25 & -2.09\\
 0.1025 & -3.36 & -3.02 & -2.25 & -1.28 & -1.02 & -0.85 & -0.67 & -0.63 & -0.77 & -1.24 & -1.42 & -2.28\\
 0.1035 & -3.28 & -2.91 & -2.10 & -1.09 & -0.87 & -0.65 & -0.46 & -0.43 & -0.57 & -1.04 & -1.30 & -2.13\\
 0.1045 & -3.19 & -2.83 & -1.98 & -0.98 & -0.75 & -0.55 & -0.37 & -0.34 & -0.48 & -0.95 & -1.16 & -2.02\\
 0.1055 & -3.20 & -2.83 & -1.98 & -0.97 & -0.75 & -0.54 & -0.36 & -0.32 & -0.47 & -0.93 & -1.17 & -2.01\\
    ... &   ... &   ... &   ... &   ... &   ... &   ... &   ... &   ... &   ... &   ... &   ... &   ...\\
 1.0025 & -0.02 &  0.14 &  0.08 &  0.08 &  0.15 &  0.05 &  0.22 &  0.32 &  0.18 &  0.12 &  0.11 &  0.06\\
 1.0075 & -0.02 &  0.13 &  0.07 &  0.08 &  0.15 &  0.04 &  0.21 &  0.32 &  0.17 &  0.12 &  0.10 &  0.05\\
 1.0125 & -0.02 &  0.14 &  0.08 &  0.08 &  0.15 &  0.05 &  0.22 &  0.33 &  0.18 &  0.12 &  0.11 &  0.06\\
 1.0175 & -0.02 &  0.14 &  0.08 &  0.08 &  0.15 &  0.05 &  0.22 &  0.32 &  0.18 &  0.12 &  0.10 &  0.06\\
 1.0225 & -0.01 &  0.14 &  0.08 &  0.08 &  0.15 &  0.05 &  0.22 &  0.32 &  0.18 &  0.12 &  0.11 &  0.06\\
    ... &   ... &   ... &   ... &   ... &   ... &   ... &   ... &   ... &   ... &   ... &   ... &   ...\\
 9.7800 & -1.09 & -1.05 & -1.10 & -1.09 & -1.02 & -1.24 & -1.06 & -0.96 & -1.10 & -1.06 & -1.07 & -1.10\\
 9.8200 & -1.09 & -1.05 & -1.10 & -1.09 & -1.02 & -1.24 & -1.07 & -0.96 & -1.11 & -1.06 & -1.07 & -1.11\\
 9.8600 & -1.09 & -1.05 & -1.11 & -1.09 & -1.02 & -1.24 & -1.07 & -0.96 & -1.11 & -1.07 & -1.07 & -1.11\\
 9.9000 & -1.10 & -1.05 & -1.11 & -1.09 & -1.02 & -1.24 & -1.07 & -0.96 & -1.11 & -1.07 & -1.07 & -1.11\\
 9.9400 & -1.10 & -1.06 & -1.11 & -1.10 & -1.03 & -1.25 & -1.08 & -0.97 & -1.12 & -1.07 & -1.08 & -1.11\\
 9.9800 & -1.10 & -1.06 & -1.12 & -1.11 & -1.03 & -1.25 & -1.08 & -0.97 & -1.12 & -1.08 & -1.08 & -1.12\\
10.0200 & -1.11 & -1.07 & -1.12 & -1.11 & -1.04 & -1.26 & -1.09 & -0.98 & -1.12 & -1.08 & -1.09 & -1.12\\
\hline
\end{longtable}
\normalsize
Notes: Table 10, 11 (template dust extinction uncorrected) and 12 (template FIR fit) are available only in electronic format at http://cdsweb.u-strasbg.fr
\twocolumn

\end{document}